%% file: main.tex
\documentclass{sig-alternate-05-2015}
\input{commands}

\usepackage{bm}
\usepackage{balance}  
\usepackage{wrapfig}

\usepackage[hidelinks]{hyperref}

\CopyrightYear{2017}
\setcopyright{acmcopyright}
\conferenceinfo{SIGMOD'17,}{May 14-19, 2017, Chicago, IL, USA}
\isbn{978-1-4503-4197-4/17/05}\acmPrice{\$15.00}
\doi{http://dx.doi.org/10.1145/3035918.3064042}

\begin{document}

\setcopyright{acmcopyright}

\title{ML4all: Democratizing Machine Learning using Database Techniques}
\title{ML4all: Democratizing Scalable Machine Learning}
\title{ML4all: Scalable Machine Learning for Everyone}
\title{A Cost-based Optimizer for Gradient Descent Optimization}

\numberofauthors{1}
\author{
	Zoi Kaoudi$^1$ \hspace{2ex}
	Jorge-Arnulfo Quian\'e-Ruiz$^1$ \hspace{2ex}  
	Saravanan Thirumuruganathan$^1$ \hspace{2ex}\\
	Sanjay Chawla$^1$ \hspace{2ex}
	Divy Agrawal$^2$ \hspace{2ex}
\vspace{1.0mm}\\
\affaddr{
	{$^1$Qatar Computing Research Institute, HBKU}
}
\vspace{1.0mm}\\
\affaddr{
	{$^2$UC Santa Barbara}
}
\vspace{1.6mm}
\\
\fontsize{9}{9}\selectfont\ttfamily\upshape
{
{\tt \{zkaoudi,jquianeruiz,sthirumuruganathan,schawla\}@qf.org.qa, agrawal@cs.ucsb.edu}
}
}

\maketitle

\begin{abstract}
\input{abstract}

\end{abstract}

\section{Introduction}
\label{sec:intro}
\input{introduction}

\section{Gradient Descent Primer}
\label{sec:background}
\input{background3}

\section{GD Optimizer Architecture}
\label{sec:arch}
\input{architecture2}

\section{GD abstraction}
\label{sec:abstraction}
\input{abstraction}



\section{GD Iterations Estimation}
\label{sec:iterations}
\input{iterations-estimator2}

\section{GD Plan Space}
\label{sec:plans}

\input{plans2}

\section{GD Cost Model}
\label{sec:costmodel}
\input{costmodel3}

\section{Experimental Evaluation}
\label{sec:evaluation}
\input{evaluation}

\section{Related Work}
\label{sec:relatedwork}
\input{relatedwork}

\section{Conclusion}
\label{sec:conclusion}
\input{conclusion}

\newpage

\bibliographystyle{abbrv}
\bibliography{da}

\appendix

\section{ML4all Language}
\label{app:language}
\input{language}

\section{GD Operators Listings}
\label{app:codesnippets}
\input{codesnippets}

\section{Accelerating GD algorithms}
\label{app:more-plans}
\input{additional_plans}

\section{Implementation}
\label{sec:implementation}
\input{implementation}

\section{Additional results}
\label{app:experiments}
\input{additional_experiments}

\end{document}

%% file: commands.tex
\newcommand{\eat}[1]{}

\usepackage{latexsym}
\usepackage{amsfonts}
\usepackage{amsmath}
\usepackage{amssymb}
\usepackage{color}
\usepackage{colortbl}
\usepackage{epsfig}
\usepackage{xspace}
\usepackage{graphicx}
\usepackage{subfigure}
\usepackage{enumerate}
\usepackage[table]{xcolor}
\usepackage[all]{xy}
\usepackage{cite}
\usepackage{booktabs}
\usepackage{pdfpages}

\usepackage{epsfig}
\usepackage{multirow}

\newcommand{\myparagraph}[1]{\vspace{0.1cm}\noindent\textbf{{#1}.~}}

\newcommand{\printIfExtVersion}[2]
{
	\ifthenelse{\equal{\extVersion}{true}}{#1}{}
	\ifthenelse{\equal{\extVersion}{false}}{#2}{}
}

\newcommand{\at}[1]{\protect\ensuremath{\mathsf{#1}}\xspace}

\newcommand{\bi}{\begin{itemize}}
\newcommand{\ei}{\end{itemize}}

\newcommand{\be}{\begin{enumerate}}
\newcommand{\ee}{\end{enumerate}}
\newcommand{\beqn}{\begin{eqnarray*}}
\newcommand{\eeqn}{\end{eqnarray*}}

\newcommand{\ie}{i.e.,\xspace}
\newcommand{\eg}{{e.g.,}\xspace}

\newcommand{\cst}{c_{\mathcal{S}}}
\newcommand{\ctr}{c_{\mathcal{T}}}
\newcommand{\cco}{c_{\mathcal{C}}}
\newcommand{\cpd}{c_{\mathcal{U}}}

\newcommand{\cdt}{c_{\mathcal{CV}}}
\newcommand{\clp}{c_{\mathcal{L}}}
\newcommand{\csa}{c_{\mathcal{SP}}}

\newcommand{\pagerio}{pageIO}

\newcommand{\seek}{SK}
\newcommand{\nt}{NT}

\newcommand{\If}{\mbox{\bf if}\ }

\newcommand{\Else}{\mbox{\bf else}\ }

\newcommand{\While}{\mbox{\bf while}\ }

\newcommand{\For}{\mbox{\bf for}\ }

\newcommand{\Return}{\mbox{\bf return}\ }

\newcounter{ccc}

\newcommand{\eop}{\hspace*{\fill}\mbox{$\Box$}\vspace{1ex}}     
\newcounter{example}
\renewcommand{\theexample}{\arabic{example}}

\newcommand{\nthesection}{\arabic{section}}
\newcounter{theorem}
\renewcommand{\thetheorem}{\arabic{theorem}}
\newcounter{prop}[section]

\newcounter{lemma}[section]
\renewcommand{\thelemma}{\nthesection.\arabic{theorem}}
\newcounter{cor}
\renewcommand{\thecor}{\arabic{theorem}}
\newcounter{definition}[section]
\renewcommand{\thedefinition}{\nthesection.\arabic{definition}}

\newcounter{alg}[section]
\renewcommand{\thealg}{\nthesection.\arabic{alg}}

\newcounter{arule}
\renewcommand{\thearule}{\arabic{arule}}

\newcounter{claim}
\renewcommand{\theclaim}{\arabic{claim}}

\newcommand{\w}{\mathbf{w}}

\usepackage{soul}
\newif\ifnip

\newcounter{enum}
\newenvironment{packed_enum}{
\begin{list}{\textbf{(\arabic{enum})}}{
  \setlength{\itemsep}{0pt}
  \setlength{\parskip}{0pt}
  \setlength{\labelwidth}{-5 pt}
  \setlength{\leftmargin}{0 pt}
  \setlength{\itemindent}{0pt}
  \usecounter{enum}}
}{\end{list}}

\newcommand{\rheem}{\textsc{Rheem}\xspace}
\newcommand{\mlall}{{ML4all}\xspace}

\usepackage{listings}
\definecolor{javared}{rgb}{0.6,0,0} 
\definecolor{javagreen}{rgb}{0.25,0.5,0.35} 
\definecolor{javapurple}{rgb}{0.5,0,0.35} 
\definecolor{javadocblue}{rgb}{0.25,0.35,0.75} 

\usepackage[linesnumbered,ruled,vlined]{algorithm2e}
\let\oldnl\nl
\newcommand{\nonl}{\renewcommand{\nl}{\let\nl\oldnl}}

\makeatletter
    \newcommand\figcaption{\def\@captype{figure}\caption}
    \newcommand\tabcaption{\def\@captype{table}\caption}
\makeatother

\newcommand{\add}[1]{{\textcolor{black}{{#1}}}}

%% file: abstract.tex

As the use of machine learning (ML) permeates into diverse application domains, there is
an urgent need to support a declarative framework for ML. Ideally, a user will specify an ML
task in a high-level and easy-to-use language and the framework will invoke the appropriate
algorithms and system configurations to execute it. An important observation
towards designing such a framework is that many ML tasks can be expressed as mathematical
optimization problems, which take a specific form. Furthermore, these optimization problems
can be efficiently solved using variations of the gradient descent (GD) algorithm.
Thus, to decouple a user specification of an ML task from its execution, a key component
is a GD optimizer. We propose a cost-based GD optimizer that selects the best GD plan for a given ML task.
To build our optimizer, we introduce a set of abstract operators for expressing GD algorithms and propose a novel approach to estimate the number of iterations a GD algorithm requires to converge.
Extensive experiments 
on real and synthetic datasets show that our optimizer not only chooses the best GD plan but also allows for optimizations that achieve orders of magnitude performance speed-up.

%% file: introduction.tex

Can we design a Machine Learning (ML) system that can replicate the success of relational
database management systems (RDBMs)? A system where users' needs are decoupled  from 
the underlying algorithmic and system concerns?  The starting point of such an attempt
is the observation that, despite a huge diversity of tasks, many ML problems can
be expressed as  mathematical optimization problems that take a very specific form~\cite{mlbook,sgd-tricks,petuum}.  For example, a classification task can be expressed as 

\vspace*{-0.2cm}
\begin{small}
\begin{equation}
f(\mathbf{w}) = \sum_{i \in \mbox{data}}\ell(\mathbf{x_{i}},y_{i},\mathbf{w}) + \cal{R}(\mathbf{w})
\label{eq:opt}
\vspace{-0.1cm}
\end{equation}
\end{small}

\noindent where $\mathbf{x_i}$ is the assembled feature vector, $y_i$ is the binary
label, $\mathbf{w}$ is the model vector, $\ell$ is the loss function
that we seek to optimize, and $\cal{R}$ is the regularizer that helps guiding
the algorithm to pre-defined parts of the model space. A key (but now well-known) observation is that if $\ell$ and $R$ are convex functions then a gradient descent (GD) algorithm can arrive at the global optimal solution (or  a local optimum for non-convex functions). 
One can apply GD to most of the supervised, semi-supervised, and unsupervised ML problems. For example, we can apply GD to support vector machines (SVM), logistic regression, matrix factorization, conditional random fields, and deep neural networks. 

\begin{figure}[!t]
	\centering
	\includegraphics[scale=0.2]{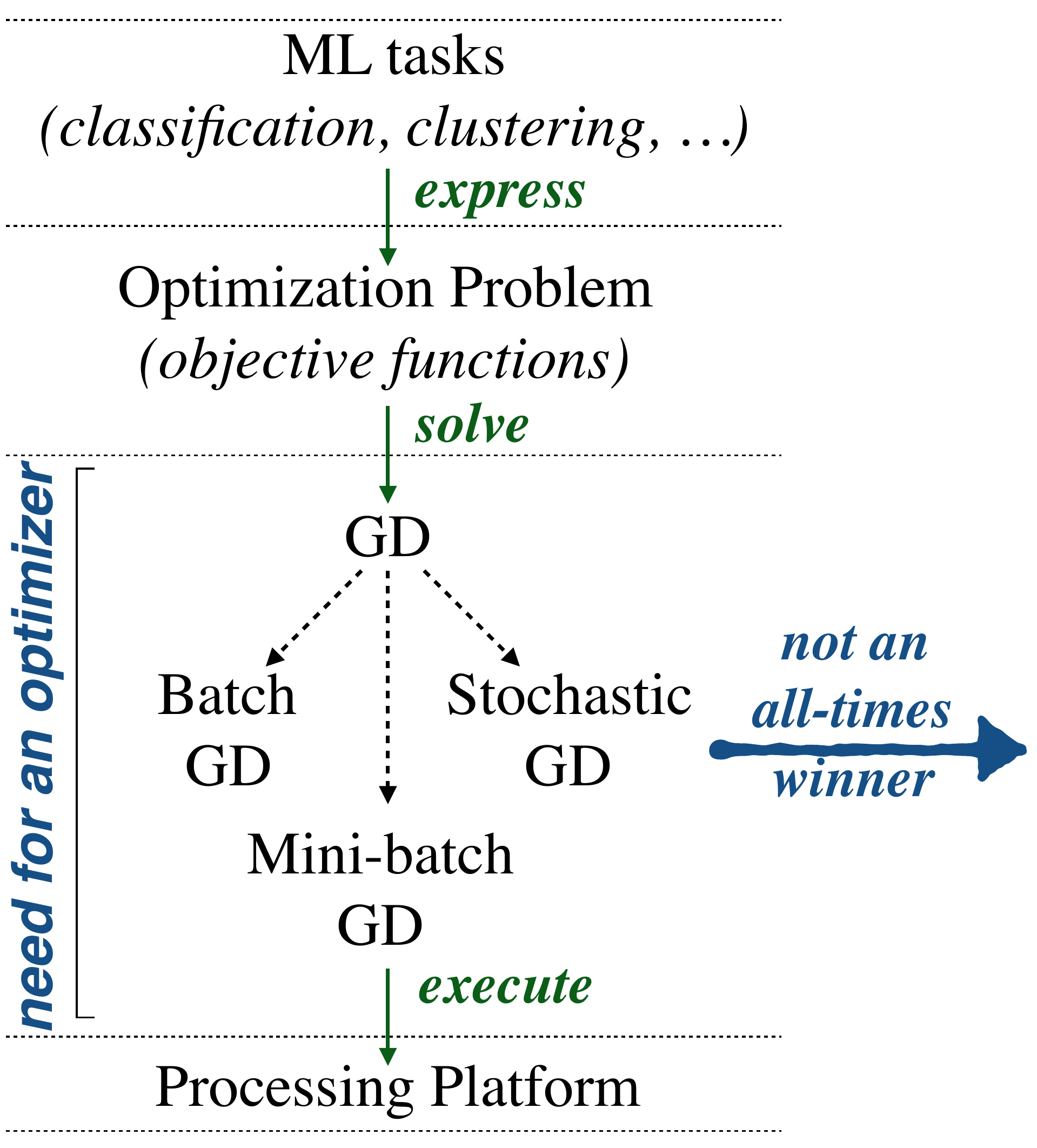}
	\includegraphics[scale=0.33]{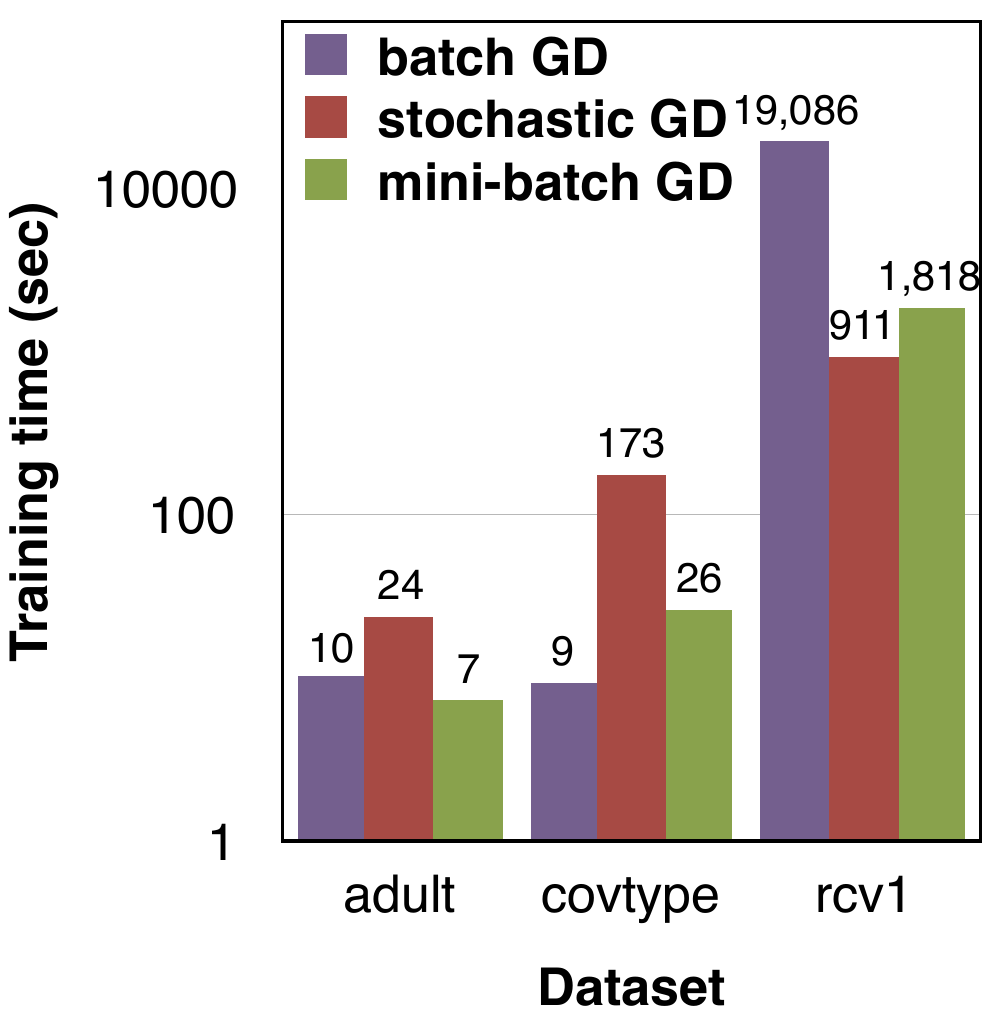}
	\vspace{-0.4cm}
	\caption{Motivation.}
	\label{fig:motivation}
	\vspace{-0.5cm}
\end{figure}

The left-side of Figure~\ref{fig:motivation} shows the general workflow when treating ML tasks as an optimization problem using GD.
Even if one maps an ML task to an optimization problem to solve it with GD, she is still left with the dilemma of {\em which GD algorithm to choose}.
There are different GD algorithms proposed in the literature, with three fundamental ones: batch GD (BGD), stochastic GD (SGD), and mini-batch GD (MGD).
Each of them has its advantages and disadvantages with respect to accuracy and runtime performance. For example, BGD gives the most accurate results, but requires many costly full scans over the entire data. In addition, in contrast to the current understanding (that SGD is always fastest) there is no single algorithm that outperforms the others in runtime. The right-side of Figure 1 shows that indeed: (i) for the \texttt{adult} dataset MGD takes less time to converge \add{to a tolerance value of $0.01$ for SVM}; (ii) for the \texttt{covtype} BGD is faster \add{for SVM and tolerance $0.01$}; and (iii) for the \texttt{rcv1} dataset SGD is the winner \add{for logistic regression to converge to a tolerance of $10^{-4}$}. In particular, we observe that a GD algorithm can be more than one order of magnitude slower than another. This is the case for batch and SGD in Figure~\ref{fig:motivation}. Thus, building an optimizer able to choose among these GD algorithms is a clear need. 

We initiate research towards that goal:
{\em how can we design a cost-based optimizer \add{for ML systems} that takes an ML task (specified in a declarative manner), evaluates different ways of executing the ML task using a GD algorithm, and chooses the optimal GD execution plan?}
We caution that there are substantial differences between query optimizers for RDBMSs and ML systems that make the above goal quite challenging.
Query optimizers used in RDBMs collect statistics about tables and query workload to generate cost estimates for different query execution plans.
In contrast, ML optimizers have a {\em cold start} problem as the best execution plan is often query and data dependent (it also depends on the accuracy required by users). 
Due to the non-uniform convergence nature of ML algorithms, any collected statistics is often rendered useless when query or data changes.
\add{\hypertarget{dbms-idea}{Thus}, the challenge resides on how to bring the cost-based optimization paradigm, which is routinely used in databases, to ML systems.}
A GD optimizer must be nimble enough to identify the cost of different execution plans under very strict user constraints, such as accuracy.
A key ingredient of a cost-based optimizer for iterative-convergent algorithms is to be able to estimate both (i)~the cost per iterations and (ii)~the number of iterations.
Already, the latter is a hard problem by itself that has only been addressed in theory.
However, the theoretical bounds provided in the literature can hardly be used in practice as they are far from reality.

We present a cost-based optimizer that frees users from the burden of GD algorithm selection and low-level implementation details.
In summary, after giving a brief GD primer (Section~\ref{sec:background}) and the architecture of our optimizer (Section~\ref{sec:arch}), we make the following contributions:
\begin{packed_enum}
	\item We propose a concise and flexible GD abstraction. The optimizer leverages this abstraction for parallelization and optimization opportunities.~(Section~\ref{sec:abstraction})
	\item We propose a speculation-based approach to estimate the number of iterations a GD algorithm requires to converge. To the best of our knowledge, this is the first solution proposed for estimating the number of iterations of iterative-convergent algorithms in practical scenarios.~(Section~\ref{sec:iterations})
	\item We show how our abstraction allows for new optimization opportunities to generate different GD plans~(Section~\ref{sec:plans}).
	We then describe an intuitive cost model to estimate the cost per iteration in each GD execution plan (Section~\ref{sec:costmodel}).
	\item We implemented our optimizer in \mlall, an ML system built on top of \rheem~\cite{rheem-vision,rheem-demo}, our in-house cross-platform system. We use Java and Spark as the underlying platforms and compared it against state-of-the-art ML systems on Spark (MLlib~\cite{mllib} and SystemML~\cite{systemml-vldb}). Our optimizer always chooses the best GD plan and achieves performance of up to more than two orders of magnitude than MLlib and SystemML as well as more than one order of magnitude faster than the abstraction proposed in~\cite{bismarck}.~(Section~\ref{sec:evaluation})
\end{packed_enum}

%% file: background3.tex
ML tasks can be reduced to mathematical optimization
problems. This problem entails minimizing
Equation~\ref{eq:opt} to arrive at the optimal solution $\mathbf{w}^{\ast}$.
The algorithm of choice for optimizing $f(\mathbf{w})$ is GD that we now briefly explain.

Suppose $f(\mathbf{w})$ is a sufficiently smooth function. We can use
Taylor's Series to expand $f(\mathbf{w})$ in the $\mathbf{w}$'s neighborhood. 

\vspace{-0.2cm}
\begin{small}
\begin{equation*}
f(\mathbf{w} + \alpha\bm{\epsilon}) \approx f(\mathbf{w}) + \alpha\nabla f(\mathbf{w})^{T}\bm{\epsilon}
\end{equation*}
\end{small}
\vspace{-0.3cm}

\noindent Now, the choice of $\mathbf{\epsilon}$ which will minimize the value in
the neighborhood must be $\epsilon = -\nabla f(\mathbf{w})$. Thus,
starting at an initial value $\mathbf{w}^{0}$, we iterate as follows,
until convergence:

\vspace{-0.1cm}
\begin{small}
\begin{equation}
\label{equation:minimization}
\mathbf{w}^{k+1} = \mathbf{w}^{k} - \alpha_{k}\nabla f(\mathbf{w}^{k})
\end{equation}
\end{small}
\vspace{-0.3cm}

\noindent $\alpha_{k}$ is called the step size and has the property that $\alpha_{k} \rightarrow 0$ as $k \rightarrow \infty$. The following two reasons
explain why GD algorithms are so popular in ML:
\vspace{-0.1cm}
\begin{packed_enum}
\item If $f$ is convex, then, starting from {\em any}
initial value $\mathbf{w^0}$, a GD algorithm guarantees to
converge to \add{the} global \add{optimum}.
\item If $f$ is convex and non-smooth, the gradient
can be replaced by a sub-gradient (a generalization of the gradient
operator) and the convergence guarantee still holds, albeit at a slower rate
of convergence.
\end{packed_enum}
\vspace{-0.1cm}
Not only one can express many ML tasks as convex programs, but also ML tasks
take on a very specific form as specified in Equation~\ref{eq:opt}.
Abstractly, ML tasks reduce to the optimization problem $\sum_{i=1}^{n}f_{i}(\mathbf{w}) + g(\mathbf{w})$.
Due to the linearity of the gradient operator $\nabla$, we have $\nabla(\sum_{i=1}^{n}f_{i}(\mathbf{w}) + g(\mathbf{w})) 
=
\sum_{i=1}^{n}\nabla(f_{i}(\mathbf{w})) + \nabla(g(\mathbf{w}))$.
\noindent Note that data directly appears only in the first term $\sum_{i=1}^{n}\nabla(f_{i}(\mathbf{w}))$. In large data, computating this
term is the main bottleneck that needs to be addressed to
make the system scalable. Basically, there are three GD algorithms to compute
$\sum_{i=1}^{n}\nabla(f_{i}(\mathbf{w}))$: {\em Batch GD}, {\em Stochastic GD}, and {\em Mini-Batch GD}.

\myparagraph{Batch GD (BGD)} This algorithm keeps the term as it is, \ie~ no
approximation is carried out. In which case the cost of computing the
gradient expression is $O(n)$, where $n$ is the number of data points. 
Thus, each iteration of the GD algorithm requires a 
complete pass over the data set.

\myparagraph{Stochastic GD (SGD)} This algorithm takes a single
random sample $r$ from the data set for approximation, \ie $\nabla f_{r}(\mathbf{w}) \approx
\sum_{i=1}^{n}\nabla(f_{i}(\mathbf{w}))$.
Furthermore, by linearity of Expectation: $E_{r}(f_{r}(\mathbf{w}))  =
\sum_{i=1}^{n}\nabla(f_{i}(\mathbf{w}))$.
\noindent Thus, the cost of each iteration is $O(1)$, \ie~completely
{\em independent} of the size of the data. This has made SGD
particularly attracting for large datasets. However, as
at each iteration, the SGD only provides an approximation of
the actual gradient term, the total number of iterations required to
attain a pre-specified convergence guarantee increases. 

\myparagraph{Mini-Batch GD (MGD)} This is a hybrid approach where
 a small sample of size $b$ is randomly selected from the dataset
to estimate the gradient. For example, if $B=\{r_{1},\ldots,r_{b}\}$
is a random sample, the gradient is then estimated as follows: $\sum_{r_{i} \in B}\nabla f_{r_{i}}(\mathbf{w}) \approx
\sum_{i=1}^{n}\nabla(f_{i}(\mathbf{w})) $.
\noindent MGD is also stochastic and independent of the dataset size.

%% file: architecture2.tex

\begin{figure}[!t]
\centering\includegraphics[scale=0.15]{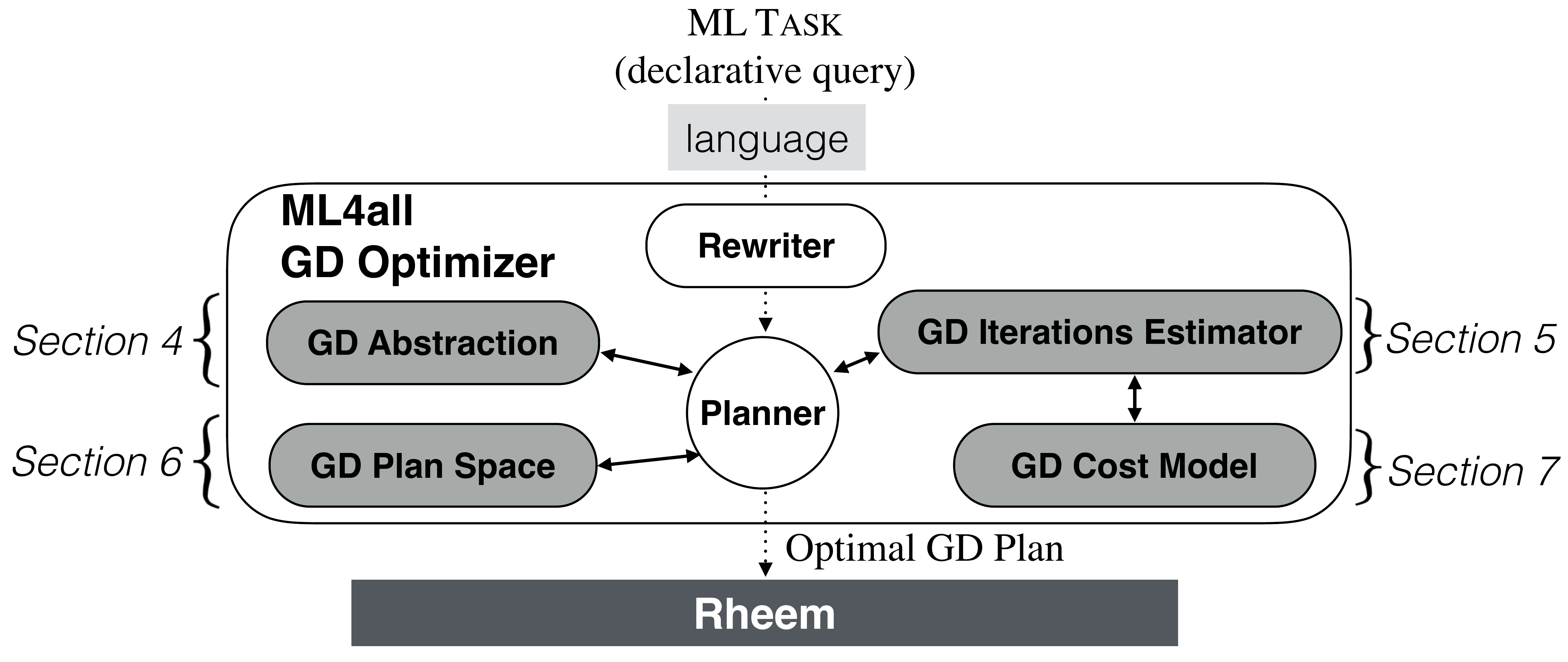}
\vspace{-0.5cm}
\caption{Optimizer Architecture.}
\label{figure:architecture}
\vspace{-0.4cm}
\end{figure}

We are inspired from relational database optimizers to design a cost-based optimizer for gradient descent.
Users send a declarative query and the optimizer outputs the optimal plan that satisfies their requirements.
Figure~\ref{figure:architecture} illustrates the architecture of our cost-based optimizer composed of four main components: a GD {\em abstraction}, an {\em iterations estimator}, a {\em plan search space}, and a {\em cost model}.
Overall, the optimizer first uses the GD abstraction, which contains the set of GD operators, to translate a declarative query into a GD plan.
It then produces an optimized GD plan based on a cost model, which relies on (i)~an iterations estimator to know in how many iterations a GD plan converges and (ii)~a couple of optimizations that define the GD search space.

\myparagraph{Declarative GD Language}
Users can interact with our GD optimizer through a simple declarative language.
\add{We briefly sketch the language with the query below.
Further details can be found in Appendix~\ref{app:language}.}


\vspace{-0.2cm}
\begin{center}
{\sc run} classification {\sc \textcolor{darkgray}{on}} training\_data.txt 
\\{\sc having \textcolor{darkgray}{time}} 1h30m{\sc , \textcolor{darkgray}{epsilon}} 0.01{\sc , \textcolor{darkgray}{max\_iter}} 1000;
\end{center}
\vspace{-0.1cm}

This query states that the user wants to build a classification model for the given dataset \at{training\_data.txt}, where she wants
(i)~her results before one hour and a half,
(ii)~an epsilon value (\ie~tolerance) smaller or equal to $0.01$, and
(iii)~to run until convergence to this epsilon value or for a maximum of $1,000$ iterations.

\myparagraph{Concise GD Abstraction}
Informally, GD-based algorithms exhibit three major phases:
(i)~a preparation phase, where the algorithm parses the dataset and sets the relevant parameters,
(ii)~a processing phase, where the core computations occur, such as parameters update, and
(iii)~a \add{convergence} phase, where the algorithm determines if it should perform another iteration or not.
Observing this pattern allows us to propose seven GD operators that are sufficient to express most of the GD-based algorithms (Section~\ref{sec:abstraction}).


\myparagraph{Speculative GD Iterations Estimator}
As most ML algorithms are iterative, it is crucial to estimate the number of iterations required to converge to a tolerance value.
We propose a novel speculation-based approach to estimate the number of iterations for any GD algorithm.
In a few words, we obtain a sample of the data and run a GD algorithm under a fixed time budget.
Based on the observations, we estimate the iterations required by the algorithm (Section~\ref{sec:iterations}).

\myparagraph{GD Plan Space}
Given an ML task specified using our abstraction, the optimizer needs to explore the space of all possible GD execution plans.
ML tasks could be solved using any of the BGD, MGD, or SGD  algorithms.
Each of these options forms a potential execution plan.
Transforming an abstracted plan to an execution plan enables us to introduce some core optimizations, namely {\em lazy transformation} and {\em efficient data skipping} and, thus, significantly speed up the execution of GD-based algorithms in many cases (Section~\ref{sec:plans}).

\myparagraph{GD Cost Model}
Once all possible GD execution plans are defined, our optimizer uses a cost model to identify the best execution plan in this search space.
\add{Note that, like database optimizers, the main goal of our optimizer is to avoid the worst execution plans.}
We provide a cost analysis model for computing the operator cost, which together with the estimated number of iterations of a GD algorithm enables the cost estimation of an execution plan (Section~\ref{sec:costmodel}).

%% file: abstraction.tex

We aim at providing an abstraction for GD algorithms that allows our optimizer to build plans considering parallelization and optimization opportunities. 
\add{We found that most ML algorithms have three phases: the {\em preparation} phase, the {\em processing} phase, and the {\em convergence} phase.
In the preparation phase, the algorithm parses and prepares the input dataset as well as it sets all required parameters for its core operations.
Then, it enters into the iterative phases of processing and convergence, which interleave each other.
While the processing phase performs its core computations, the convergence phase decides (based on a given number of iterations or a convergence condition) if it has to repeat its core operations.}

Based on this observation, we introduce seven basic operators that abstract the above phases: \at{Transform}, \at{Stage}, \at{Compute}, \at{Update}, \at{Sample}, \at{Converge}, and \at{Loop}.
The system exposes these operators as {\em User-Defined Functions} (UDFs).
While we provide reference implementations for all the common use cases, \add{expert} users could readily customize or override them if necessary.
We aim at providing a small but adequate set of operators that allows GD algorithms to obtain scalability and high performance.
In the following, we formally define these operators, justify their existence, and illustrate examples of them in Figure~\ref{figure:abstraction}(a) using SGD.



\subsection{Preparation Phase}
\label{section:abstraction_preparation}
The reader might believe that a single preparation operator is sufficient to abstract this phase, such as in~\cite{bismarck}. While this is true in theory, in practice this is not efficient. This is because GD algorithms need to transform the entire input dataset, but, to set their global variables, they usually need no (or a small sample of) input data. Therefore, our system provides two basic operators (\at{Transform} and \at{Stage}), rather than a single one, for users to parse input datasets and efficiently set all algorithmic variables, respectively. We discuss these two operators below.

\vspace{-0.2cm}
\begin{packed_enum}

\begin{figure}[!t]
\centering\includegraphics[width=\columnwidth]{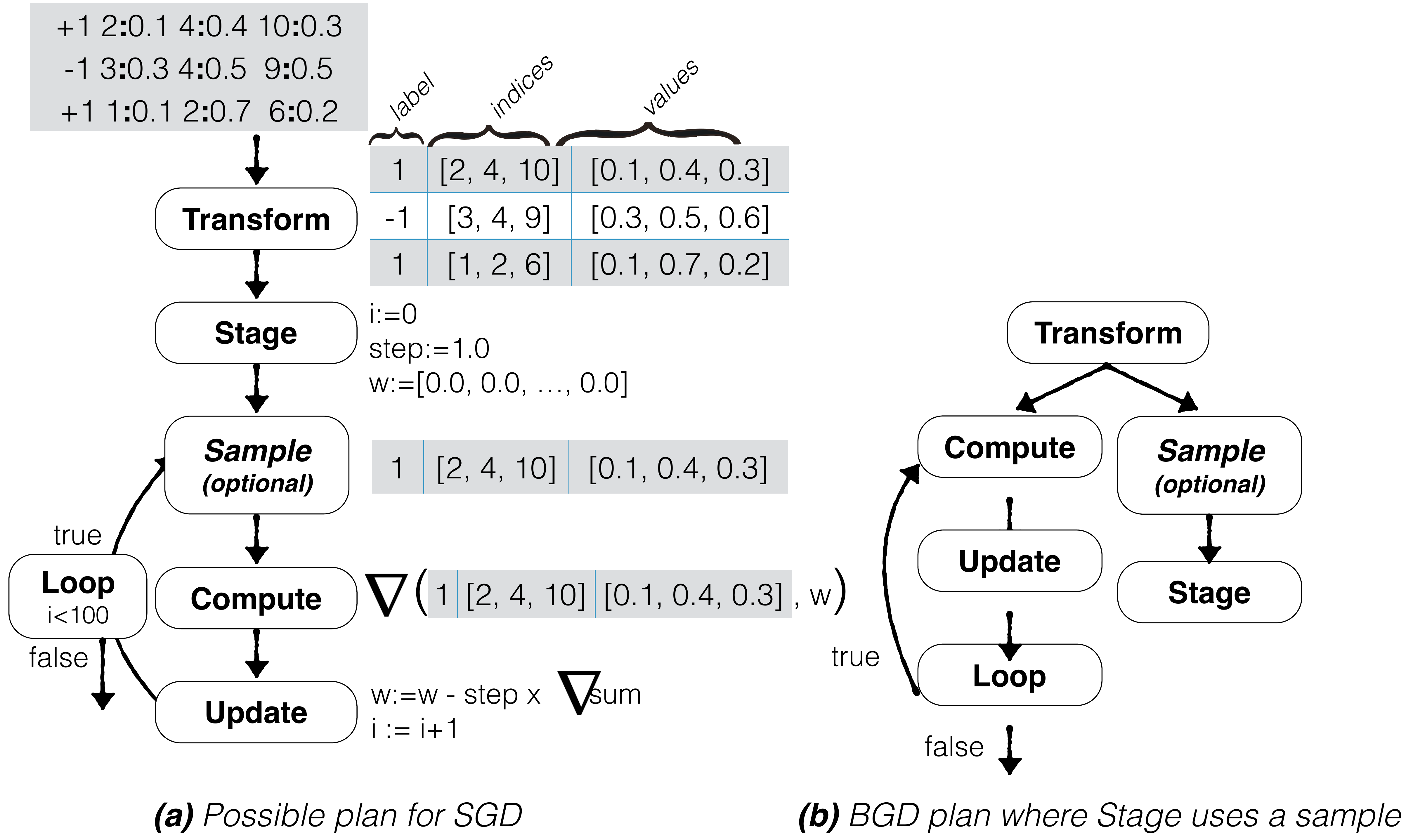}
\vspace{-0.8cm}
\caption{Abstraction.}
\label{figure:abstraction}
\vspace{-0.5cm}
\end{figure}

\item {\bf Transform} prepares input data units\footnote{\add{The system reads the data units from disk using a RecordReader UDF, such as in Hadoop.}} for consequent computation. Basically, it outputs a parsed and potentially normalized data unit ($U_T$) for each input data unit ($U$):
\vspace{-0.1cm}
$$\at{Transform}(U) \rightarrow U_T$$
This operator is important as it allows for a proper computation of the input data units. For example for SGD (Figure~\ref{figure:abstraction}(a)), a possible  \at{Transform} operator identifies the double-type dimensions of each data point as well as its label in the entire sparse input dataset. It outputs a sparse data unit containing a label, a set of indices, and a set of values. Listing~\ref{list:transform} shows the code snippet for a simple \at{Transform} operator. Note that the \texttt{context} contains all global variables.
\begin{lstlisting}[belowcaptionskip=-1.5cm, abovecaptionskip=-1.5cm, aboveskip=-0.2cm, belowskip=-0.2cm, basicstyle=\scriptsize\normalfont\sffamily,float=h!, label=list:transform,caption=Code snippet example of \at{Transform}.]
public double[] transform(String line, Context context) {
1	String[] pointStr = line.trim().split(',');
2	double [] point = new double[pointStr.length];
5	for (int i=0; i<pointStr.length; ++i) {
6		point[i] = Double.parseDouble(pointStr[i]) }
9	return point;}
\end{lstlisting}


\item {\bf Stage} sets the initial values for all algorithm-specific parameters required. Typically, this operator does not require the input dataset for setting the initial parameters.
Still, it may sometimes need a data unit or list of data units to initialize parameters.
For instance, it may use a sample from the input data to initialize the weights (see Figure~\ref{figure:abstraction}(b)).
Thus, we formally define \at{Stage} as follows:
\vspace{-0.1cm}
$$\at{Stage}(\emptyset \, | \, U_T\, | \, \at{list}\langle U_T\rangle) \rightarrow \emptyset \, | \, U_T \, | \,\at{list}\langle U_T\rangle$$
\at{Stage} allows users to ensure the good behavior of the consequent operations. For example in Figure~\ref{figure:abstraction}(a), \at{Stage} sets the initial values for vector $w$ to $0.0$, the step size to $1.0$, and the iteration counter to $0$. It is worth noting that, \at{Stage} is not a data transformation operator and hence it simply outputs any potential data units ($U_T$) it receives.
We show in Listing~\ref{list:stage} of Appendix~\ref{app:codesnippets} the \at{Stage} code for this example.
\end{packed_enum}

\subsection{Processing Phase}
\label{section:abstraction_processing}
As in the preparation phase, the reader might again think that a single operator is sufficient to abstract the main operations of a GD algorithm. This is in fact what is proposed in~\cite{bismarck}.
However, this prevents the parallelization of an algorithm and thus, its performance and scalability.
In a distributed setting, GD algorithms need to know the global state of their operations to update their parameters for the next iteration, \eg~the weights in BGD.
Thus, having a single operator for this phase would lead to centralizing the process phase so that the update can happen and thus would significantly hurt performance.

Therefore, we abstract this phase via two basic operators, that can be used to define the main computations of their algorithms (\at{Compute}) and update the global variables accordingly (\at{Update}).
While these two operators abstract the operations of most batch GD algorithms, some (online) algorithms (such as MGD and SGD) work on a sample of the input data. This is why we introduce a third operator (\at{Sample}) that can be optionally used to narrow the data input for their ML tasks. We detail these operators below.

\vspace{-0.1cm}
\begin{packed_enum}
\item[{\bf (3)}] {\bf Compute} performs the core computations. It takes a data unit ($U_T$) as input and performs a user defined computation over it to output another data unit ($U_C$):
\vspace{-0.1cm}
$$\at{Compute}(U_T) \rightarrow U_C$$
For instance, in Figure~\ref{figure:abstraction}(a), the \at{Compute} operator computes the gradient of a sparse data unit. 
Users can use one of the provided gradient functions or provide their own.
The code of \at{Compute} for our example is in Listing~\ref{list:compute}.
\begin{lstlisting}[belowcaptionskip=0cm, abovecaptionskip=-1.5cm, aboveskip=-0.1cm, belowskip=-0.2cm, basicstyle=\scriptsize\normalfont\sffamily,float=h!,label=list:compute,caption=Code snippet example of \at{Compute}.]
public double[] compute (double[] point, Context context) {
1 	double[] weights = (double[]) context.getByKey("weights");
2 	return this.svmGradient.calculate(weights,point);}
\end{lstlisting}

\vspace{-0.1cm}
\item[{\bf (4)}] {\bf Update} re-sets all global parameters required by the GD algorithm, \eg~vector $w$ for SGD. It outputs a data unit ($U_U$) representing the new global variable value for a given aggregated data unit ($U_{\overline{C}}$). Formally:
\vspace{-0.1cm}
$$\at{Update}(U_{\overline{C}}) \rightarrow U_U$$
Notice that $U_{\overline{C}}$ is the sum of all data units. 
For example, $U_{\overline{C}}$ represents the sum of gradients emitted by \at{Compute} in BGD.
This operator is as important as the \at{Compute} operator as it ensures the good behavior of a GD algorithm by correctly computing its global variables. For example, for SGD (Figure~\ref{figure:abstraction}(a)), \at{Update} computes the new values for vector $w$ as is illustrated in Listing~\ref{list:update}.

\begin{lstlisting}[belowcaptionskip=-1.5cm, abovecaptionskip=-1.5cm, aboveskip=-0.2cm, belowskip=-0.2cm, basicstyle=\scriptsize\normalfont\sffamily,float=h!,label=list:update,caption=Code snippet example of \at{Update}.]
public double[] update (double[] input, Context context) {
1 	double[] weights = (double[]) context.getByKey("weights");
2 	double step = (double) context.getByKey("step");
3 	for (int j=0; j<weights.length; j++) {
4		weights[j] = weights[j] - step * input[j+1]; }
5	return weights;}
\end{lstlisting}
\vspace{-0.1cm}

\item[{\bf (5)}] {\bf Sample} defines the scope of the consequent computations to specific parts of the input dataset. It takes the number of data units in the dataset or a set of data units as input and outputs a list of numbers (no greater than the number of input data units) or a smaller list of data units:
\vspace{-0.1cm}
$$\at{Sample}(n \, | \, \at{list}\langle U\rangle) \rightarrow \at{list}\langle nb\rangle \, | \, \at{list}\langle U\rangle$$
It is via \at{Sample} that users can enable the MGD and SGD methods, by setting the right sample size. Typically, this operator is placed right before \at{Compute} and hence it is called at the beginning of each iteration (Figure~\ref{figure:abstraction}(a)). 
The code of an example sample operator is shown in Appendix~\ref{app:codesnippets}.
\end{packed_enum}


\subsection{Convergence Phase}
\label{section:abstraction_tuning}
In addition to the above five operators, we provide two more operators that allow users to have control on the termination of the algorithm: the \at{Converge} and \at{Loop} operators.

\vspace{-0.1cm}
\begin{packed_enum}
\item[{\bf (6)}] {\bf Converge} specifies how to produce the delta data unit (\ie~the convergence dataset), which is the input of the \at{Loop} operator. 
It takes a data unit from \at{Update} and outputs a delta data unit:
\vspace{-0.1cm}
$$\at{Converge}(U_U) \rightarrow U_{\Delta}$$
For example, it might compute the L2-norm of the difference of the weights from two successive iterations for SGD.
Listing~\ref{list:converge} in Appendix~\ref{app:codesnippets} illustrates the lines of code for \at{Converge} in this example.

\item[{\bf (7)}] {\bf Loop} specifies the stopping condition of a GD algorithm.
For this, we first compute the delta data unit for the stopping condition as defined above.
Then, the \at{Loop} operator decides if the algorithm needs to keep iterating based on this delta data unit ($U_\Delta$). 
Formally:
\vspace{-0.1cm}
$$\at{Loop}(U_\Delta) \rightarrow \at{true} \, | \, \at{false}$$
\end{packed_enum}
In other words, this operator determines the number of iterations a GD algorithm has to perform its main operations, \ie~\at{Compute} and \at{Update}. For instance, the \at{Loop} operator ensures that the algorithm will run for $100$ iterations for our example in Figure~\ref{figure:abstraction}(a).
Listing~\ref{list:loop} in Appendix~\ref{app:codesnippets} illustrates the lines of code for \at{Loop} in this example.

\subsection{GD Plans}
\label{sec:gdplans}
We now demonstrate the power of our abstraction by show how the basic GD algorithms, such as BGD and MGD, are abstracted using our operators. SGD was already shown as the running example.
The implementation of BGD and MGD algorithms using the proposed abstraction is trivial.
We simply have to modify the sample size of the \at{Sample} operator in the SGD plan illustrated (Figure~\ref{figure:abstraction}(a)) to support MGD or to remove the \at{Sample} operator to support BGD (\eg~Figure~\ref{figure:abstraction}(b)). 
\add{We also show how two more complex algorithms, namely line search and SVRG~\cite{srvg}, can be expressed using our abstraction (see Appendix~\ref{app:more-plans}).} \add{Our abstraction allows the implementation of any GD algorithm regardless of the step size and other hyperparameters.}

%% file: iterations-estimator2.tex

One of the biggest challenges of having an effective optimizer is to estimate the
number of iterations that a gradient descent (GD) algorithm requires to reach a pre-specified
tolerance value. A GD algorithm only uses first-order information (the gradient).
However, the rate of convergence of GD depends on second-order information, such as the condition
number of the Hessian. This constitutes a roadblock as the Hessian not only is very expensive to compute, but also changes at every iteration. In addition,
the rate of convergence requires an inversion of a $d \times d$ dense matrix, where $d$ is the dimensionality of the problem. 

However, for classical machine learning models, like logistic regression and SVM with
$\ell_{2}$ regularization, the loss function is convex and smooth. A function is $L$-smooth
if $\|\nabla f(\mathbf{v}) - \nabla f(\mathbf{w}) \| \leq L\|v - w\|$ for all
$\mathbf{v}$ and $\mathbf{w}$ in the domain of $f$~\cite{mlbook}. When a function is convex and
$L$-smooth, it is known that BGD with a step size $\alpha \leq 1/L$~\cite{mlbook}, results in
a sequence $\{w^{k}\}$, which satisfies $|f(w^{k}) - f(w^{\ast})| \leq \frac{\|w^{0} - w^{\ast}\|^{2}_{2}}{2\alpha k}$,
where $w^{\ast}$ is the optimal model vector.



Thus in order to obtain a tolerance $\epsilon$, a sufficient number of iterations ($k$) is
$ k \geq \frac{\|w_{0} - w^{\ast}\|^{2}_{2}}{2\alpha \epsilon} $.
However, note that this is a sufficient, rather than a necessary condition and more
importantly the bound is not practical as $\w^{\ast}$ is not known a priori but only once the GD algorithm has converged.
To obtain practical and accurate estimates, we take a speculation-based approach that we describe below.

\begin{algorithm} [t!]
	\caption{Speculation process\label{algo:speculation}}
	\KwIn{Desired tolerance $\epsilon_d$, speculation tolerance $\epsilon_s$, speculation time budget $B$, dataset $D$}
	\KwOut{Estimated number of iterations $T(\epsilon_d)$}
	\BlankLine
	$D' \leftarrow$ sample on $D$\; 
	initialize $errorSeq$ \tcp{List of \{error, iteration\}} 
	$i = 1$ \tcp{iteration}
	$\epsilon_1 = \infty$\; 
	\While{$\epsilon_i > \epsilon_s~\&~t < B$}{
		Run iteration $i$ of GD algorithm on $D'$\;
		$errorSeq \leftarrow \text{add}(\epsilon_i, i)$\;
		$i$++\;
	}
	$a \leftarrow$ fit $errorSeq$ to the function $T(\epsilon) = \frac{a}{\epsilon}$\;
	compute $T(\epsilon_d) = \frac{a}{\epsilon_d}$\;
	\Return $T(\epsilon_d)$\;
\end{algorithm}

\myparagraph{Speculation-based approach}
Our approach to estimate the number of iterations is based on the observations that (i)~in practical
large scale (batch) settings, the training time is large and (ii)~the ``shape'' of error sequence over a sample is very similar to the one over the entire dataset~\cite{bertsekas1999nonlinear}. We can thus afford a relatively
small speculation time budget $B$ such that we can actually run BGD, MGD, and SGD on
few samples from the dataset for relatively high $\epsilon$ values. 

\add{\hypertarget{iterations}{Prior} research shows that gradient descent based methods on convex functions routinely exhibit only three standard convergence rates -- linear, supra linear (with order p) and quadratic~\cite{bertsekas1999nonlinear}.	Each of these convergence rates can be identified purely through the error sequence. Our iterations estimator leverages this observation by identifying and then parameterizing the error sequence in our speculative stage. Note that this approach works regardless of the dataset, the specific (convex) optimization function, the variant of the gradient descent used and the step size.}
As the rate of convergence is $O(\frac{1}{\epsilon})$ or better~\cite{mlbook}, we can fit the function $\frac{a}{\epsilon}$ using the speculation
output \add{to learn $\alpha$}.
Value $a$ is dependent on the dataset and the form of the loss function (and regularizer). \add{\hypertarget{iterations-parameters}{Thus}, our approach elegantly abstracts from any hyperparameter tuning as parameters are learned purely from the speculative stage, \ie~users do not specify them.}

Algorithm~\ref{algo:speculation} shows the pseudocode of our approach. Assume we want to estimate the number of iterations  $T(\epsilon_d)$ a GD algorithm requires to converge to tolerance value $\epsilon_d$. Given a (large) speculation tolerance $\epsilon_s$ and time budget $B$, our algorithm first takes a sample $D'$ from dataset $D$ and starts running the GD algorithm on it (Lines~1--6). \add{$\epsilon_s$ is set by default to $0.05$ and $B$ to $1$ min. However, the user or system administrator is free to change them.}
In each iteration, the reached tolerance error $\epsilon_i$ together with the iteration $i$ is appended in a list (Line~7). Note that $T(\epsilon_i)=i$.
When the error reaches the speculation tolerance $\epsilon_s$ or the time budget has been consumed, the GD algorithm terminates. Then, we use this list of \{$\epsilon, T(\epsilon)$\} to fit the function $T(\epsilon)=\frac{a}{\epsilon}$ and learn $a$ (Line~9). After $a$ is known for the specific dataset, the output is the number of iterations $T(\epsilon_d)$ (Lines~10 and~11).
We run this algorithm for each GD algorithm, namely BGD, MGD, and SGD, to obtain the estimated number of iterations for each one.
\add{Note that MGD and SGD take their data samples from sample $D'$ and not from the input dataset $D$. BGD runs over the entire $D'$.}

\vspace{0.1cm}
\noindent{\it Sampling effect.}
Our iterations estimator uses a small data sample for running the various GD algorithms.
This is advantageous as the smaller size results in algorithms converging quite fast. 
In this way, we can easily obtain a good fit of the error sequence shape before the time budget is exhausted.
We observed that using a small sample instead of the entire dataset for speculation does not affect the iteration estimation in any major way.
It is known that for many linear and quadratic loss functions, the sample complexity 
(number of training samples needed to successfully learn a function) is finite and depends linearly on its VC-dimension~\cite{mlbook}.
It has also been observed (such as in~\cite{bousquet2008tradeoffs}) that only a small number of training examples 
have a meaningful impact in the computation of gradient. 
Finally, \cite{bousquet2008tradeoffs} also observed that the estimation errors vary between the inverse and the inverse square root of the number of training examples.
Jointly, these observations justify our approach to use a small fraction of the dataset for the iterations estimator.

%% file: plans2.tex
Given an ML task using the proposed abstraction in Section~\ref{sec:abstraction} as input, the GD optimizer produces an optimal GD execution plan.
For this, the GD optimizer exploits the flexibility of the proposed abstraction to come up with several optimized plans for the GD algorithms. 
Basically, it departs from the fact that SGD and MGD work on a data sample in each iteration to introduce two core optimizations: 
{\em lazy transformation}, which allows our optimizer to transform input data units only when required, and;
{\em efficient data skipping}, which allows our optimizer to efficiently read only parts of data a GD algorithm should work on.
The former is possible thanks to the ability to commute the {\tt Transform} and the {\tt Sample} operator, while the latter is thanks to the
decoupling of the {\tt Compute} operator from the {\tt Sample} operator. 
To the best of our knowledge, we are the first to exploit such kind of techniques to boost GD algorithms performance.

\begin{figure}[!t]
	\centering\includegraphics[scale=0.23]{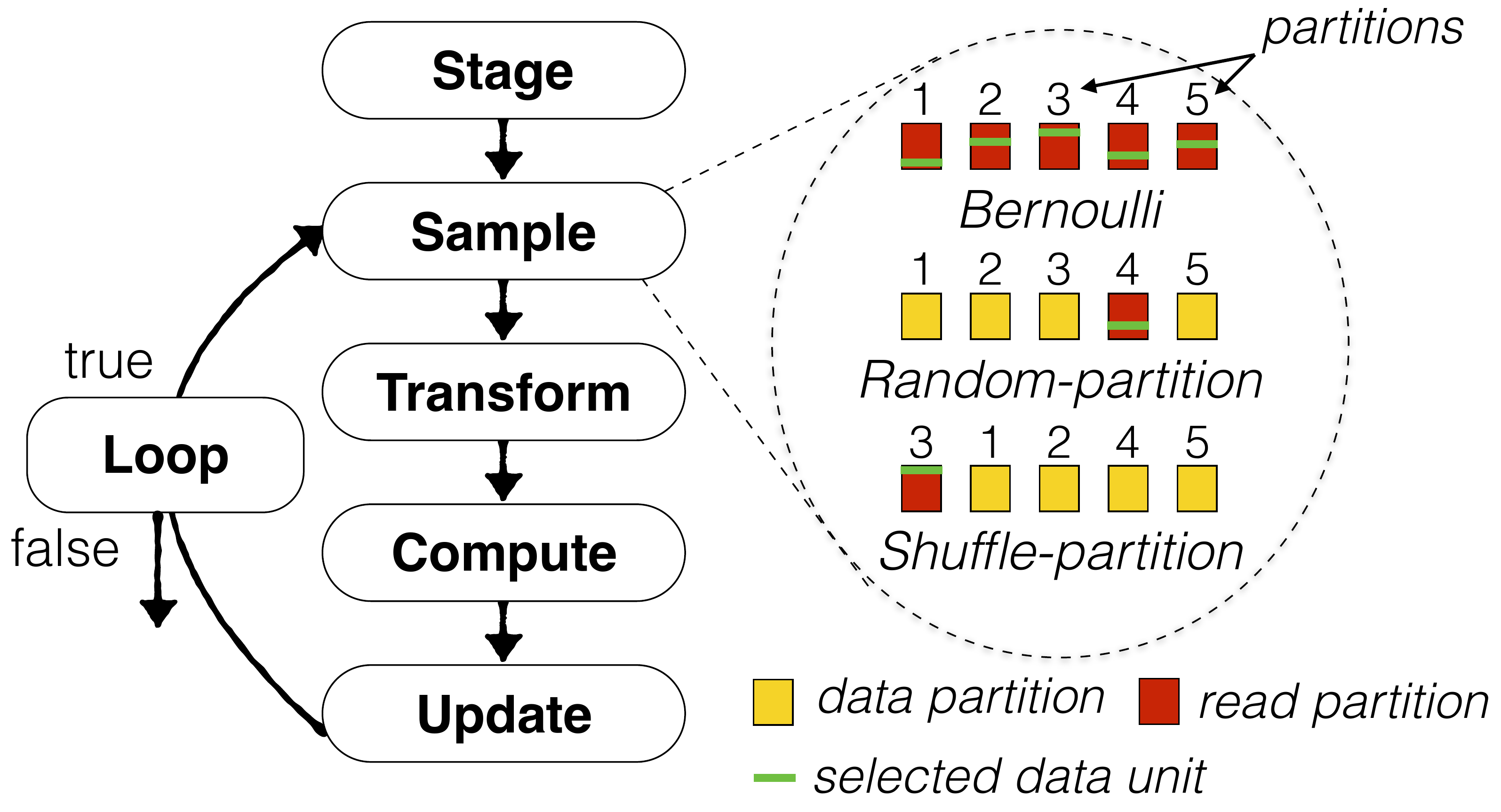}
	\vspace{-0.4cm}
	\caption{Lazy transformation \& data skipping.}
	\label{figure:lazy}
	\vspace{-0.4cm}
\end{figure}

\myparagraph{Lazy-transformation}
Recall that we transform all input data units upfront before all the core operations of an algorithm (see Figure~\ref{figure:abstraction}).
We call this approach {\em eager transformation}.
This approach inherently assumes that all data units are required by GD algorithms.
However, as mentioned above, this is not the case for all algorithms, such as for SGD and MGD.
Thus, our optimizer considers a {\em lazy transformation} approach for those cases where not all data units are required by a GD algorithm.
The main idea is to delay the transformation of data units until they are consumed by the main operations of an algorithm.
We exploit the flexibility of our abstraction in order to move the \at{Transform} operator inside the loop process, right after the \at{Sample} operator.
In this case, when the algorithm runs only few times the transformation cost is alleviated significantly.
\add{Here, the reader might think that our system cannot use this approach whenever the \at{Transform} operator requires any global statistic (such as the mean) of the entire dataset.
However, such possible cases are handled by passing the dataset to the \at{Stage} operator beforehand, which is responsible of obtaining any global data statistics.}
Figure~\ref{figure:lazy} shows the plan for this lazy-transformation approach.

\myparagraph{Efficient data skipping}
Sampling also plays an important role in the performance of stochastic-based  GD algorithms, such as MGD and SGD, especially because these algorithms require a new sample in each iteration.
Thus, apart from changing the order of \at{Transform}, we consider different sampling implementations for SGD and MGD.
The {\em Bernoulli} sampling is a common way to sample data in systems where datasets are chunked into horizontal data partitions.
Then, one has to fetch all data partitions and scan each data unit to decide whether to include it in the sample or not based on some probability. 
MLlib~\cite{mllib} uses this sampling mechanism.
This sampling technique clearly might lead to poor performance as it requires to read the entire input dataset for taking a small sample.
Therefore, our optimizer also considers a {\em random-partition} sampling strategy.
For each sample required, {\em random-partition} first randomly chooses one data partition and then randomly samples a data unit inside this partition (see Figure~\ref{figure:lazy}). 
However, this sampling mechanism might also lead to poor performance due to the large number of random accesses. 
To mitigate this large number of random accesses, we provide an additional sampling strategy: the {\em shuffled-partition}.
With this sampling strategy one randomly-picked data partition is shuffled (see Figure~\ref{figure:lazy}) only once.
Then, at each iteration, the sample operator simply takes the sample in a sequential manner from that shuffled partition.
Whenever there are not enough data units left in the partition to sample, it randomly selects a second partition and shuffles it before taking the sample.
\add{Notice that shuffled-partition might increase the number of iterations that a GD algorithm requires to converge.
However, its cost per iteration is so low that it can still achieve lower training times than the other sampling techniques.}


\begin{figure}[!t]
	\centering\includegraphics[scale=0.2]{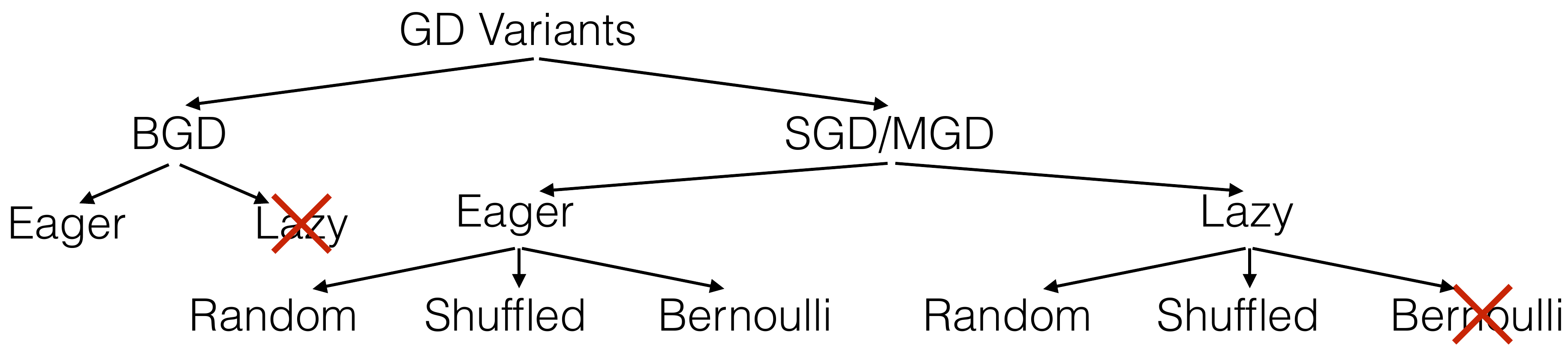}
	\vspace{-0.5cm}
	\caption{Gradient descent plans.}
	\label{figure:variants}
	\vspace*{-0.5cm}
\end{figure}

\myparagraph{Search space}
Taking all possible combinations of the above transformation and sampling techniques leads to potentially six plans for each GD algorithm. However, we consider only one plan for BGD (eager-transformation without sampling) as it requires all input data units at each iteration. Our optimizer also discards the lazy-transformation plan with Bernoulli sampling, because Bernoulli sampling goes through all the data anyways. Thus, our optimizer ends up considering $11$ plans as shown in Figure~\ref{figure:variants}. \add{Even though we consider three GD algorithms in this paper, note that there could be tens of GD algorithms that the user might want to evaluate. In such a case, the search space would increase proportionally. In other words, our search space size  is fully parameterized based on the number of GD algorithms and optimizations that need to be evaluated.}

%% file: costmodel3.tex

As the search space is very small, our optimizer can estimate the cost of all 11 GD plans and pick the cheapest.
To estimate the overall cost of a GD plan, it uses a cost model that is composed of the cost per iteration and the number of iterations of the GD plan.
The latter is obtained by the iteration estimator as explained in Section~\ref{sec:iterations}.
On the other side, the cost per iteration basically depends on the cost of all the operators contained in a GD plan (Section~\ref{sec:opcost}).
The total cost of a GD plan is then simply its cost per iteration times the number of iteration it requires to converge (Section~\ref{sec:plancost}).


\input{operatorcostmodel}
\subsection{GD Plan Cost Model}
\label{sec:plancost}

Now that we have defined the cost per operator we can compose the cost of the different GD algorithms assuming that the algorithm runs for $T$ iterations.

\myparagraph{BGD} 
The cost of running BGD is equal to the cost of \at{Stage}, \at{Transform} for the entire dataset $D$, and plus $T$ times the cost of the \at{Compute}, \at{Update} on the input dataset $D$, \at{Converge}  and \at{Loop}:

\vspace{-0.3cm}
\begin{scriptsize}
	\begin{equation}
	\begin{aligned}
	C_{BGD} (D) = \cst (D) + \ctr(D) + T \times (\cco(D)  + 
	\cpd (D) + \cdt + \clp)
	\end{aligned}
	\end{equation}
\end{scriptsize}
\vspace{-0.3cm}

\myparagraph{MGD with eager transformation} Using the eager transformation, the cost of MGD is the cost of \at{Stage}, \at{Transform} for the entire dataset $D$ plus $T$ times the cost of the \at{Sample} on the entire dataset, \at{Compute}, \at{Reduce}, \at{Update} operators on a sample $m_i$, \at{Converge} and \at{Loop}:

\vspace{-0.3cm}
\begin{scriptsize}
	\begin{equation}
	\begin{aligned}
	C_{MGD}^{eager} (D) = \cst (D) +\ctr(D) +  T \times (\csa (D) + \cco(m_i) \\
	 + \cpd (m_i) + \cdt + \clp)
	\end{aligned}
	\label{eq:mgd-eager}
	\end{equation}
\end{scriptsize}
\vspace{-0.3cm}

\myparagraph{MGD with lazy transformation} For the lazy transformation, the MGD cost is the cost of \at{Stage} for the entire dataset $D$ plus $T$ times the cost of \at{Sample} on $D$, \at{Transform}, \at{Compute}, \at{Update} on a sample $m_i$, \at{Converge} and \at{Loop}:

\vspace{-0.3cm}
\begin{scriptsize}
	\begin{equation}
	\begin{aligned}
	C_{MGD}^{lazy} (D) = \cst(D) +T \times (\csa (D) + \ctr(m_i)  + \\ \cco(m_i) + 
	\cpd (m_i) + \cdt + \clp)
	\end{aligned}
	\label{eq:mgd-lazy}
	\end{equation}
\end{scriptsize}

Formulas~\ref{eq:mgd-eager} and~\ref{eq:mgd-lazy} also apply for SGD and we omit them.


%% file: operatorcostmodel.tex

\subsection{Operator Cost Model}
\label{sec:opcost}
We now provide a cost analysis for the abstraction presented in Section~\ref{sec:abstraction}. We model the cost of an operator in terms of IO (disk or memory), CPU, and network transfer cost (if applicable). In the following, we first define these three costs ({\em IO}, {\em CPU}, and {\em Network}) and then analyze the cost of an operator based on them. Table~\ref{tab:notation} shows the notation of our cost analysis.

\begin{table}[t]
	\center
	\caption{Notation.\label{tab:notation}}
	\scalebox{0.82}{
	\begin{tabular}{l l}
		\hline
		\textbf{Notation} & \textbf{Explanation} \\
		\hline
		$D$			& operator's input dataset \\
		$P$			& data partition \\
		$page$		& data unit for storage access \\
		$packet$		& maximum network data unit \\
		$n$			& \#data units in D \\
		$d$			& \#features in a data unit \\
		$m$			& \#points in a sample \\
		$cap$		&  \#processes able to run in parallel \\
		$\pagerio$	& IO cost for reading/writing a page \\
		$\seek$		& IO cost of a seek\\
		$\nt$			& network cost of 1 byte\\
		$CPU_{u}(op)$	& processing cost for a data unit $U$ \\

		$p(D) = \lceil\frac{|D|_{b}}{|P|_{b}}\rceil$			& \#partitions of D\\
		$w(D) = \frac{p(D)}{cap}$						& \#waves for D \\
		$lwp(D) = \frac{n\,\text{mod}\,(k\times cap\times \lfloor w(D)\rfloor)}{k}$		& \#partitions in the last wave for D \\
		$k = \lceil\frac{n \times |P|_{b}}{|D|_{b}}\rceil$		&  \#data units in one partition \\
		
		\hline
	\end{tabular}
	}
\vspace{-0.6cm}
\end{table}






\myparagraph{IO cost} We consider a disk/memory {\em page} as the minimum unit of data access and we consider a {\em wave} to be the maximum number of parallel processes for an input dataset. For example, consider a compute cluster of $10$ nodes, each being able to process $2$ partitions in parallel. Given this setup, we could parallelize the processing of a given dataset composed of $85$ partitions in $5$ waves: each wave processing $20$ partitions in parallel, except the last wave that processes the remaining $5$ partitions. Thus, we model the cost of reading a dataset $D$ as the cost of reading the pages of a single partition, $\frac{|P|_b}{|page|_b}$, times the number of waves, $w(D)$. But in the last wave, we consider only the remaining data units in case they do not fill an entire partition. Formally:

\vspace{-0.3cm}
\begin{scriptsize}
\begin{equation}
\begin{aligned}
c_{IO} (D) = \lfloor w(D)\rfloor \times (\seek + \frac{|P|_{b}}{|page|_{b}} \times \pagerio)\; + \\
(\seek + \frac{|\min(lwp(D), 1) \times k|_b}{|page|_b} \times \pagerio)
\label{eq:cio1}
\end{aligned}
\end{equation}
\end{scriptsize}
\vspace{-0.3cm}

\myparagraph{CPU cost} Similar to the IO cost, we model the cost of processing a dataset $D$ as the cost of processing the number of data units in one partition times the number of waves. Again, in the last wave, we consider only the remaining data units if they do not fill an entire partition. Formally:

\vspace{-0.2cm}
\begin{scriptsize}
\begin{equation}
\begin{aligned}
c_{CPU} (D, op) = 
\lfloor w(D)\rfloor \times k \times CPU_{u}(op)\; + \\
\lceil\min(lwp(D), 1) \times k\rceil \times CPU_{u}(op)
\end{aligned}
\end{equation}
\end{scriptsize}
\vspace{-0.2cm}


\myparagraph{Network cost} Let a {\em packet} be the maximum network data unit. Notice that the last packet of a dataset can be smaller than the other packets, but its difference in cost is negligible and can be ignored. We thus model the network cost for transferring a dataset $D$ as follows:

\vspace{-0.3cm}
\begin{scriptsize}
\begin{equation}
\begin{aligned}
c_{NT} (D) = \frac{|D|_{b}}{|packet|_b} \times \nt
\end{aligned}
\end{equation}
\end{scriptsize}
\vspace{-0.3cm}

\myparagraph{Operator cost} Given the above costs, we can simply define the cost of any operator $op$ as the sum of its IO, network, and CPU costs. Formally:

\vspace{-0.3cm}
\begin{scriptsize}
\begin{equation}
c_{op}(D) = c_{IO}(D) + c_{NT}(D) + c_{CPU}(D, op) 
\end{equation}
\end{scriptsize}
\vspace{-0.3cm}

\noindent Note that the operators \at{Transform} ($\ctr$), \at{Compute} ($\cco$), \at{Sample} ($\csa$), \at{Converge} ($\cdt$), and \at{Loop} ($\clp$) involve only IO and CPU costs. This is because the data is already partitioned in several nodes and thus \at{Transform}, \at{Sample} and \at{Compute} are performed locally at each node, while \at{Loop} and \at{Converge} are executed in a single node. \at{Stage} ($\cst$) may incur only CPU cost, if it does not receive any data unit as input.
 \at{Update} is the only operator that involves network transfers in its cost ($\cpd$) because all the data units output by the \at{Compute} should be aggregated and thus, sent to a single node where the update will happen.

%% file: evaluation.tex

We designed a suite of experiments to answer the following questions:
(i)~{\em How good is our GD optimizer in estimating the model training time for different
GD algorithms?} This is a key distinguishing feature of our system vis-a-vis all other ML systems (Section~\ref{sec:exp_training});
(ii)~{\em How effective is our optimizer in choosing the correct GD plan for a given dataset?} (Section~\ref{sec:exp_effectiveness})
(iii)~{\em What is the impact of the abstraction in generating GD execution plans?} 
 (Section~\ref{sec:exp_power})
(iv)~{\em Does our sampling techniques affect the accuracy of a model?} (Section~\ref{sec:exp_accuracy})
(v)~{\em What is the impact of each individual optimization that our optimizer offers?} (Section~\ref{sec:exp_indepth})

\subsection{Setup}
\label{sec:exp_setup}

\begin{table}[t]
	\centering
	\caption{Real and synthetic ML datasets.\label{tab:datasets}}
\scalebox{0.75}{
	\begin{tabular}{llrrrr}
		\hline
		\textbf{Name} & \textbf{Task} & \textbf{\#points} & \textbf{\#features} & \textbf{Size} & \textbf{Density}\\
		\hline
		\texttt{adult} & LogR & 100,827 & 123 & 7M & 0.11\\
		\texttt{covtype} & LogR & 581,012 & 54 & 68M & 0.22\\
		\texttt{yearpred} & LinR & 463,715 & 90 & 890M & 1.0\\ 
		\texttt{rcv1} & LogR & 677,399 & 47,236 & 1.2G & $1.5 \times 10^{-3}$\\
		\texttt{higgs} & SVM & 11,000,000 & 28 & 7.4G & 0.92\\
		\texttt{svm1} & SVM & 5,516,800 & 100 & 10GB & 1.0\\
		\texttt{svm2} & SVM & 44,134,400 & 100 & 80GB & 1.0\\
		\add{\texttt{svm3}} & \add{SVM} & \add{88,268,800} & \add{100} & \add{160GB} & \add{1.0}\\
		\hline
	\texttt{SVM A} & SVM & [2.7M-88M] & 100 & [5G-160GB] & 1.0\\
	\texttt{SVM B} & SVM & 10K & [1K-500K] & [180MB-90GB] & 1.0\\
		\hline
	\end{tabular}
}
\vspace{-0.3cm}
\end{table}

\add{We implemented our GD optimizer in \mlall. \mlall is built
	on top of \rheem\footnote{\url{https://github.com/rheem-ecosystem/rheem}}, our in-house cross-platform system~\cite{rheem-vision,rheem-demo}. We used Spark and Java as the underlying platforms and HDFS as the underlying storage. The source code of \mlall's abstraction can be found at \url{https://github.com/rheem-ecosystem/ml4all}.
	Further details about our implementation can be found in Appendix~\ref{sec:implementation}}.

\myparagraph{Cluster} We performed all the experiments on a cluster consisting of four virtual nodes interconnected by a 10Gigabit switch, where each node has: $4\times4$ Intel(R) Xeon(R) CPU E5-2650@2GHz, $30$GB memory, $250$GB disk. 
We used Oracle Java JDK 1.8.0\_25 64bit, HDFS 2.6.2 and Spark 1.6.2. Spark was used in a standalone cluster mode, with four executors each having $20$GB memory and 4 cores. The Spark driver was run in one of the four nodes with the default memory of $1$GB.
We used HDFS with its default settings.

\myparagraph{Datasets} We used a broad range of datasets for different models of supervised learning (SVM, linear regression, logistic regression), of different sizes and different density (\ie~number of non-zeros to total number of values) in order to get comprehensive insights.
The real datasets are from LIBSVM\footnote{\scriptsize{\url{https://www.csie.ntu.edu.tw/~cjlin/libsvmtools/datasets/}}}.
We used eleven synthetic dense datasets for SVM of varying size and dimensionality to stress the scalability of the system.
\add{The datasets of size above $80$GB do not fit entirely into Spark cache memory.}
Table~\ref{tab:datasets} summarizes the datasets along with the tasks that they were used for.

\myparagraph{Baseline systems} 
To the best of our knowledge, there is no other system that uses cost-based optimization to distinguish between
different forms of gradient descent. We thus report the performance of \mlall in absolute terms. 
However,  we do compare our abstraction with the abstraction proposed in Bismarck~\cite{bismarck} (designed to run on a DBMS). For this, we implemented this abstraction on top of Spark. In addition, we compare the plans produced by \mlall with MLlib 1.6.2~\cite{mllib} and SystemML 0.10~\cite{systemml-vldb}, which are state-of-the-art ML systems on top of Spark. \add{MLlib comes with an implementation of the MGD algorithm, and thus, by setting the batch size accordingly we were able to have from BGD to SGD. SystemML provides a declarative R-like language for users to implement their own algorithms. Although it provides scripts for SVM and linear regression, the algorithms used are the native SVM algorithm and the conjugate GD, respectively.
For this reason, we scripted the three GD algorithms we have considered in this paper in their R-like language with appropriate gradient functions.
We then ran these scripts in SystemML with the hybrid execution mode enabled.}
We configured all systems with exactly the same parameters (i.e., step size, maximum number of iterations, initial weights, intercept, regularizer, and convergence condition). \add{ \hypertarget{mllib-stepsize}{In} fact, we use the exact same step size that is hard-coded in MLlib, \ie~$\frac{\beta}{\sqrt{i}}$, where $\beta$ is a user-defined value (set to $1$ in our experiments) and $i$ is the current iteration. As hyperparameter tuning is out of the scope of our paper, we used the same step size not only across the different systems but also across the different GD algorithms.}

\subsection{Estimation of Training Time}
\label{sec:exp_training}

\begin{figure*}[t!]
	\centering
	\mbox{
		\subfigure[\texttt{adult} dataset]{\includegraphics[width=0.58\columnwidth]{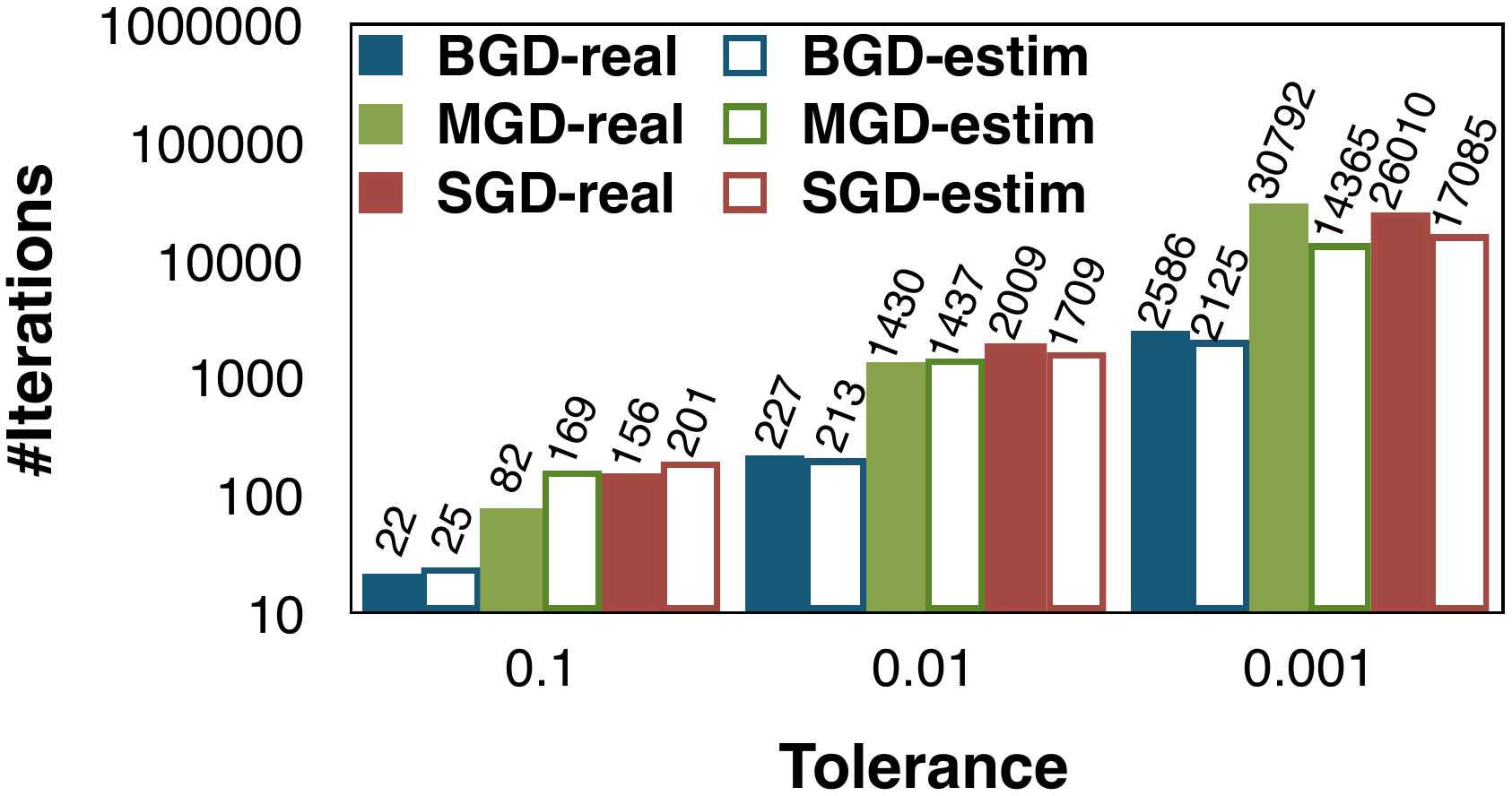}\label{subfig:iter-adult}\hspace{0.4cm}}		
		\subfigure[\texttt{covtype} dataset]{\includegraphics[width=0.58\columnwidth]{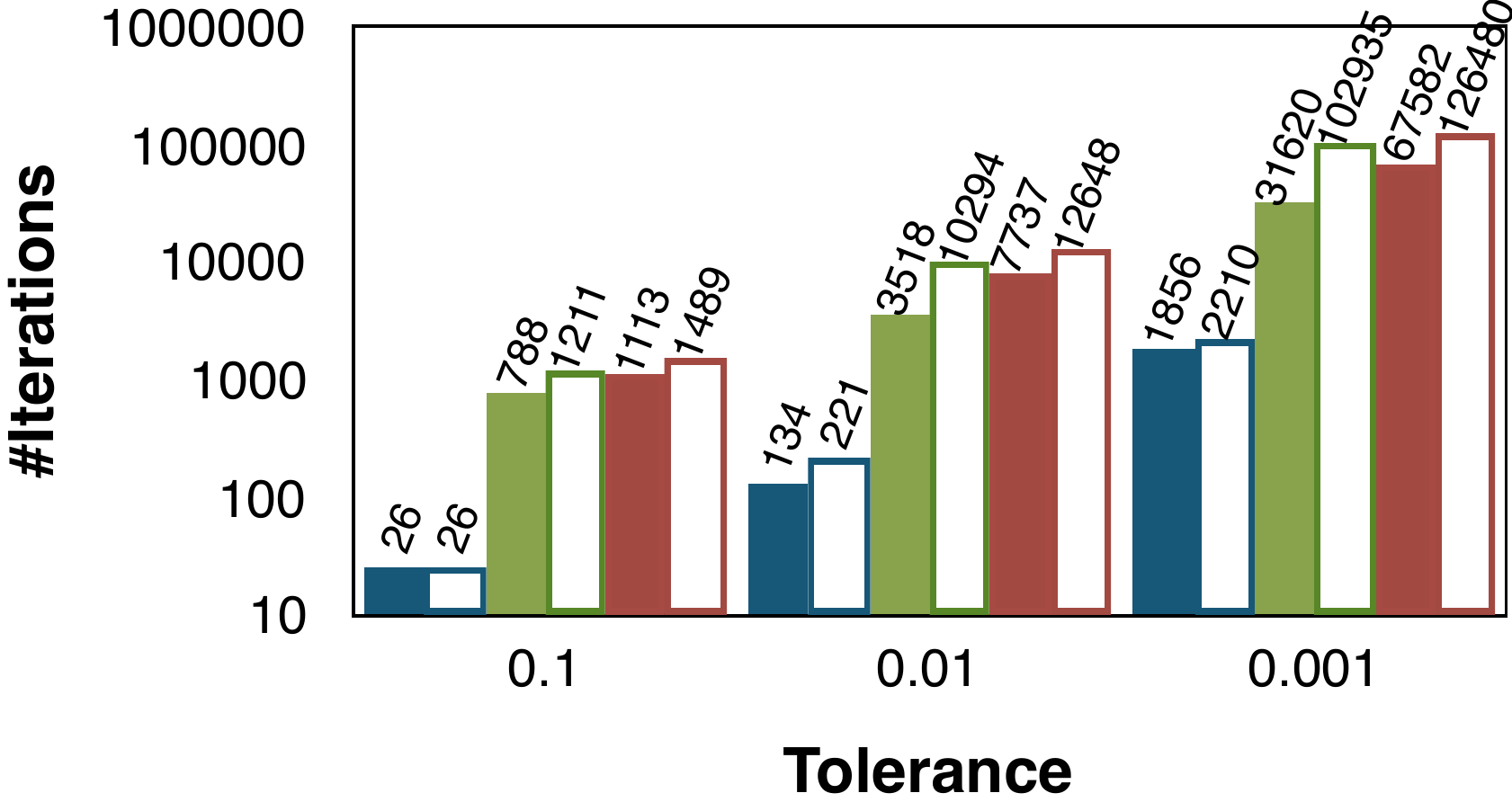}\label{subfig:iter-covtype}\hspace{0.4cm}}
		\subfigure[\texttt{rcv1} dataset]{\includegraphics[width=0.58\columnwidth]{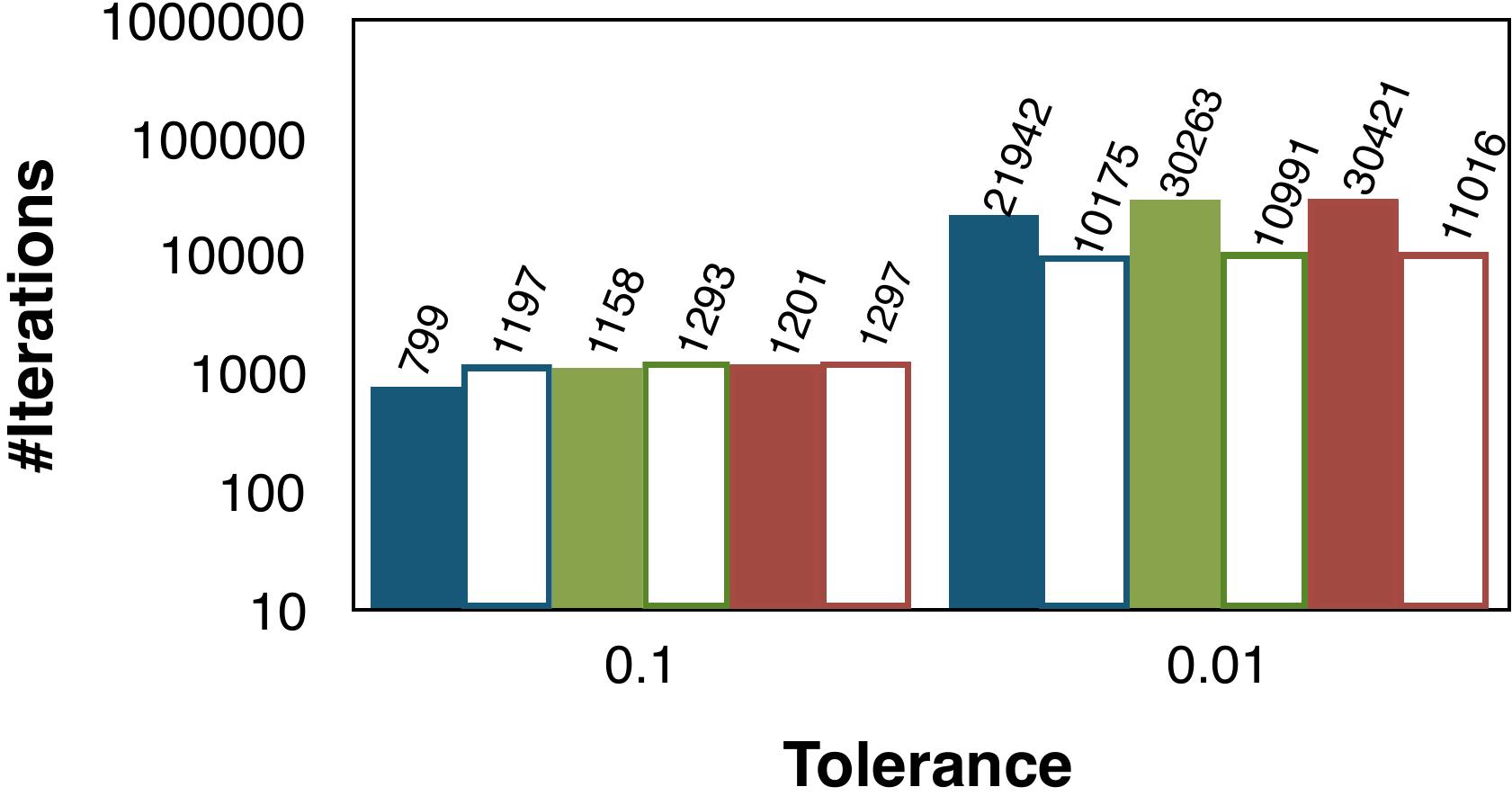}\label{subfig:iter-rcv1}}
	}
	\vspace{-0.3cm}
	\caption{\mlall obtains good estimates for the number of iterations for all GD algorithms. \label{fig:iterations}}
	\vspace{-0.4cm}
\end{figure*}

We first evaluate \mlall on how accurately it can estimate the training time of different plans and thus select the best one. 
We evaluate the estimates of the number of iterations, the cost per iteration, and the combined training time.
For all the experiments below, the speculation tolerance was set to $0.1$, the time budget to $10s$ and the sample size to $1,000$ for the speculative-based iterations estimator.

\subsubsection{Number of iterations estimation}\label{subsec:iterations-eval}
We measure the estimated and the real number of iterations for the three algorithms of GD at different tolerance levels on three real datasets. \add{The results for the other datasets were similar and are omitted due to space limitations.}
Figure~\ref{fig:iterations} shows the results of this experiment, where the full and hollow bars denote the actual and estimated number of iterations, respectively.
Notice that we don't show the results for \texttt{rcv1} with a tolerance of $0.001$ as the GD algorithms did not converge in three hours and we had to stop them.
We observe that the estimated and the actual number of iterations are very close for BGD in all three datasets.
For MGD and SGD, we observe that they are in the same order of magnitude and also very close for a large tolerance.
More importantly, the difference among the estimated number of iterations of BGD, MGD and SGD follows the same trend with the actual number of iterations.
Clearly, as the tolerance decreases all algorithms require more number of iterations to converge.
Even if our estimates are not always very accurate for MGD and SGD, because of stochasticity, they are always in the same order of magnitude with the actual ones.
Especially, we observe that \mlall preserves the same ordering of the estimated number of iterations for all three GD algorithms. \add{ Having the right order is highly desirable in an optimizer as it prevents us from falling into worst cases. In Appendix~\ref{app:experiments}, we demonstrate how our speculation-based approach using curve fitting works well even for different adaptive step sizes.}

\subsubsection{Cost per iteration estimation}
\label{sec:exp_costiter}
To evaluate the cost per iteration, we fixed the number of iterations to $1,000$ and compared the estimated time with the actual time on four real datasets. As the number of iterations is fixed, as expected, \mlall selected SGD for all datasets.
Figure~\ref{fig:time-estimate_iter} reports the results of this experiment.
We observe that \mlall performs remarkably well to estimate the cost per iteration for all datasets.
We see that, in the worst case, \mlall computes a time estimate that is only 17\% away from the actual time.
This shows the correctness and high accuracy of our cost model.
\add{Note that our cost model accurately estimates the cost of any of the GD algorithms.
This is also shown by the following results.}

\subsubsection{Total cost estimation}
\label{sec:exp_costest}
\begin{figure}[t]
	\centering
	\mbox{
		\subfigure[Run of 1,000 iterations]{\includegraphics[width=0.49\columnwidth]{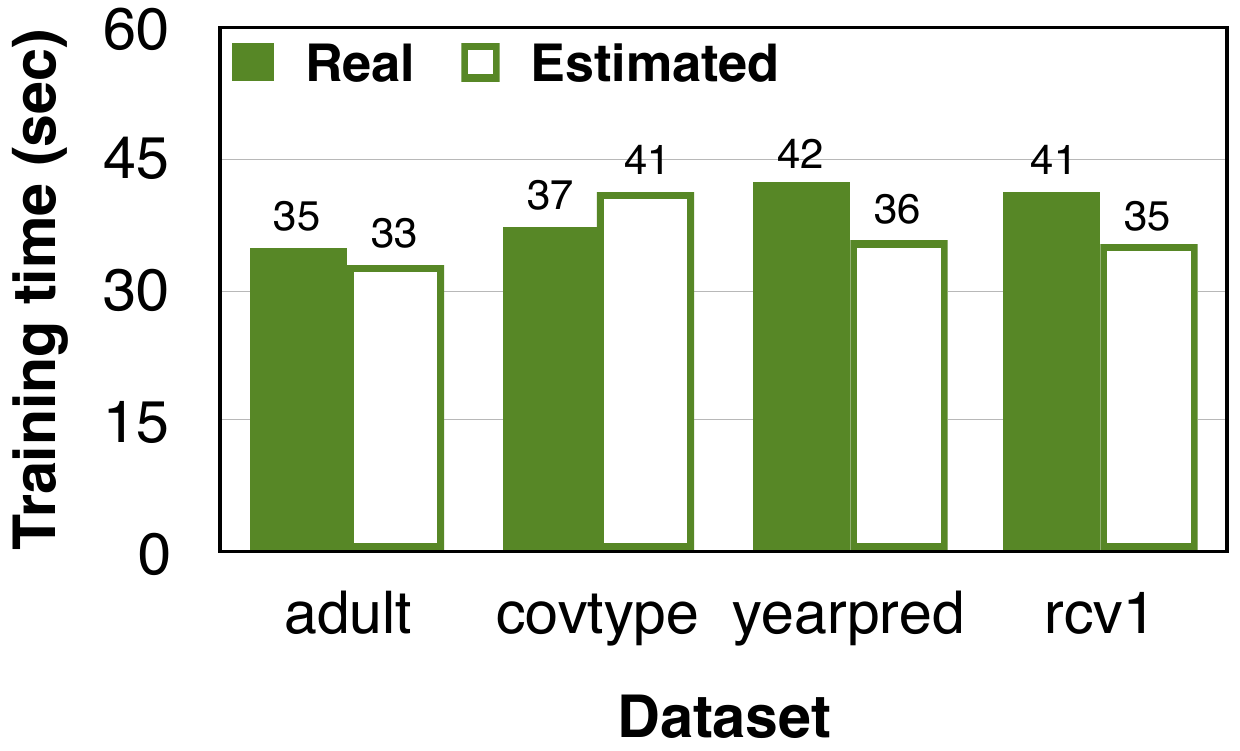}\label{fig:time-estimate_iter}}		
		\subfigure[Run to convergence]{\includegraphics[width=0.53\columnwidth]{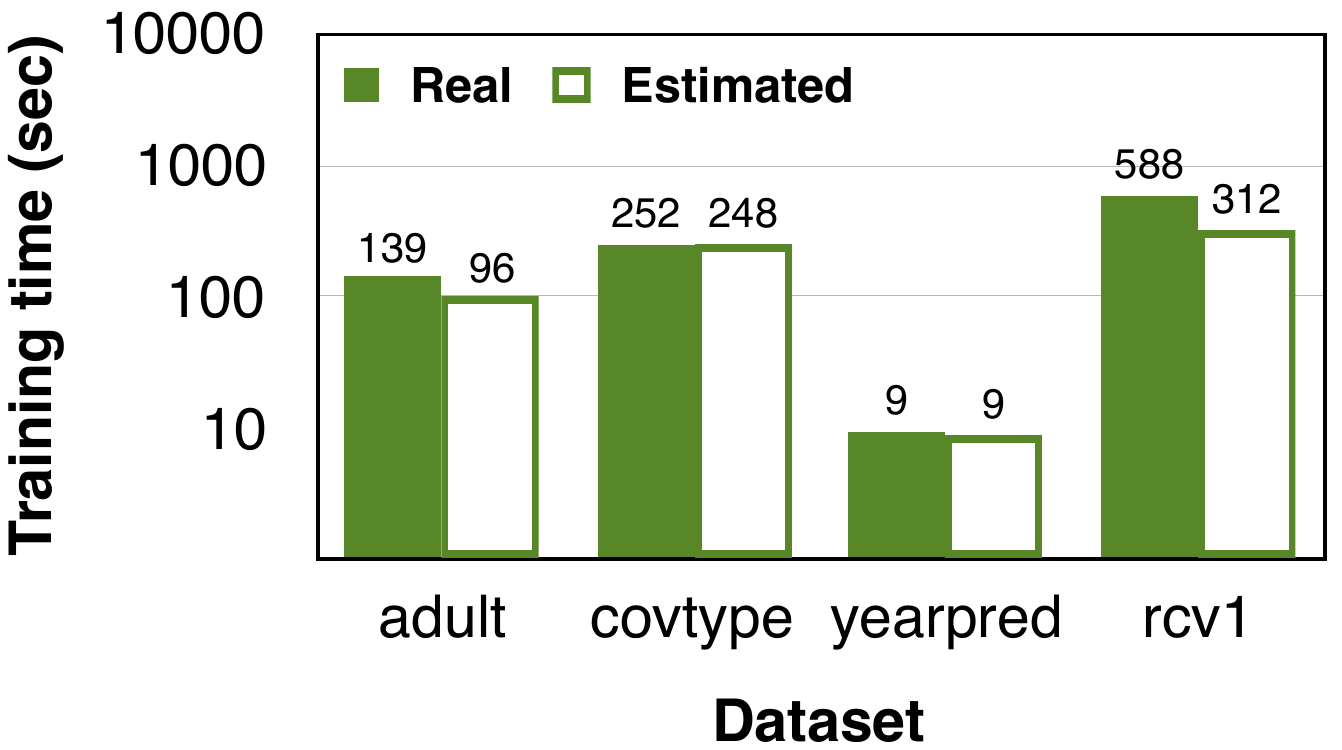}\label{fig:time-estimate_conv}}
	}
	\vspace{-0.5cm}
	\caption{\mlall obtains accurate time estimates. \label{fig:time-estimate}}
	\vspace{-0.2cm}
\end{figure}

We now combine the estimates of number of iterations and cost per iteration to evaluate the overall effectiveness of our optimizer.
For this experiment, we ran all three GD algorithms until convergence.
To get insights on different tolerance values, we set the tolerance to $0.001$ for the datasets \texttt{adult} and \texttt{covtype}, to $0.01$ for \texttt{rcv1}, and to $0.1$ for \texttt{yearpred}. \add{\mlall chose BGD for the first two datasets, and SGD-lazy-shuffle for the last two.}
Figure~\ref{fig:time-estimate_conv} shows the real execution time and the time estimated by \mlall for the algorithm that it decided to be the best choice.
We again observe that the estimated runtimes are very close to the actual ones.
These results confirm the high accuracy of both our cost model and iterations estimator.

\subsection{Effectiveness}
\label{sec:exp_effectiveness}
We now assess the effectiveness of \mlall by evaluating which GD plans it chooses. In addition, we measure the time it takes to choose such plans.
To do so, we exhaustively ran all GD plans until convergence besides the GD plans selected by our optimizer.
For this, we used a larger variety of real and synthetic datasets and measure the training time.

\begin{figure}[t]
	\centering
	\includegraphics[width=0.9 \columnwidth]{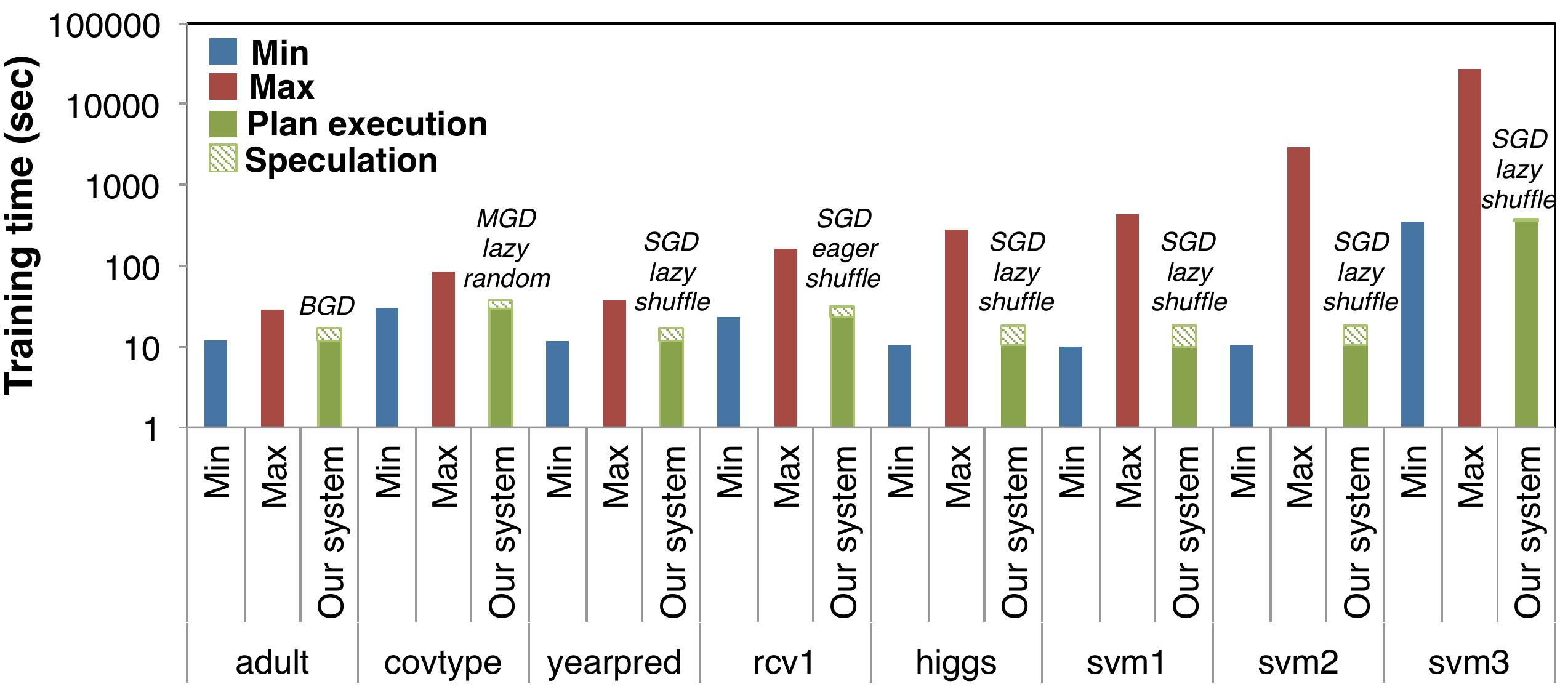}
	\vspace{-0.5cm}
	\caption{\mlall always performs very close to the best plan by choosing it plus a small overhead. \label{fig:optimizer-results}}
	\vspace{-0.4cm}
\end{figure}

Figure~\ref{fig:optimizer-results} illustrates the training times of the best (min) and worst (max) GD plan as well as of the GD plan selected by \mlall for each dataset. Notice that the latter time includes the time taken by our optimizer to choose the GD plan (speculation part) plus the time to execute it. \add{The legend above the green bars indicate which was the GD plan that our optimizer chose. Although for most datasets SGD was the best choice, other GD algorithms can be the winner for different tolerance values and tasks as we showed in the introduction.}
We make two observations from these results.
First, \mlall always selects the fastest GD plan and, 
second, \mlall incurs a very low overhead due to the speculation. Therefore, even with the optimization overhead, \mlall still achieves very low training times - close to the ones a user would achieve if she knew which plan to run.
In fact, the optimization time is between $4.6$ to $8$ seconds for all datasets.
From this overhead time, around $4$ sec is the overhead of Spark's job initialization for collecting the sample.
Given that usually the training time of ML models is in the order of hours, few seconds are negligible.
It is worth noting that we observed an optimization time of less then $100$ msec when just the number of iterations is given.
 
All the above results show the efficiency of our cost model and the accuracy of \mlall to estimate the number of iterations that a GD algorithm requires to converge, while maintaining the optimization cost negligible.


\subsection{The Power of Abstraction}
\label{sec:exp_power}
We proceed to demonstrate the power of the \mlall abstraction. 
 We show how (i)~the commuting of the {\tt Transform} and the {\tt Loop} operator (\ie~lazy vs. eager transformation) can result in rich performance dividends, and
(ii)~decoupling the {\tt Compute} operator with the choice of the sampling method for MGD and SGD can yield substantial performance gains too.
In particular, we show how these optimization techniques allow our system to outperform baseline systems as well as to scale in terms of data points and number of features.
Moreover, we show the benefits and overhead of the proposed GD abstraction.

\begin{figure*}[t!]
	\centering
	\mbox{
		\subfigure[BGD]{\includegraphics[width=0.7 \columnwidth]{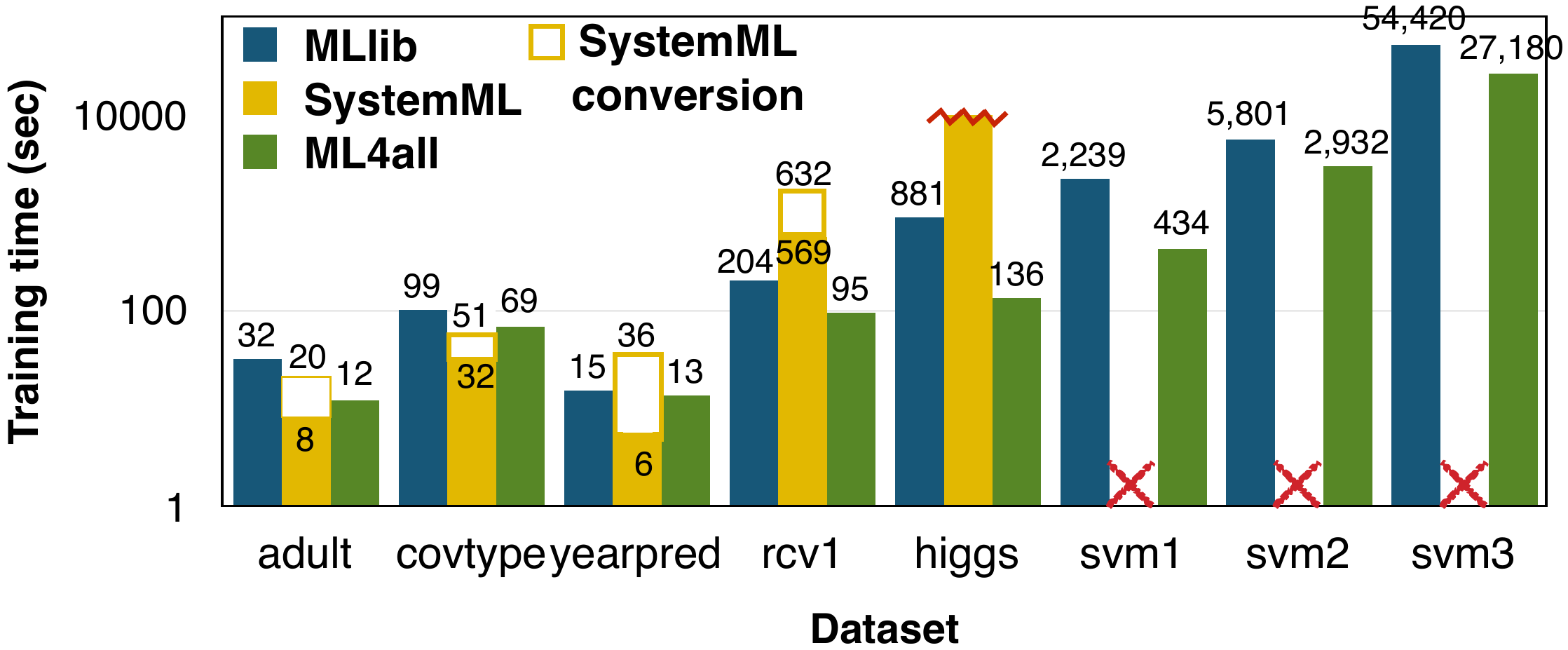}\label{subfig:gd}}
		\subfigure[MGD]{\includegraphics[width=0.7 \columnwidth]{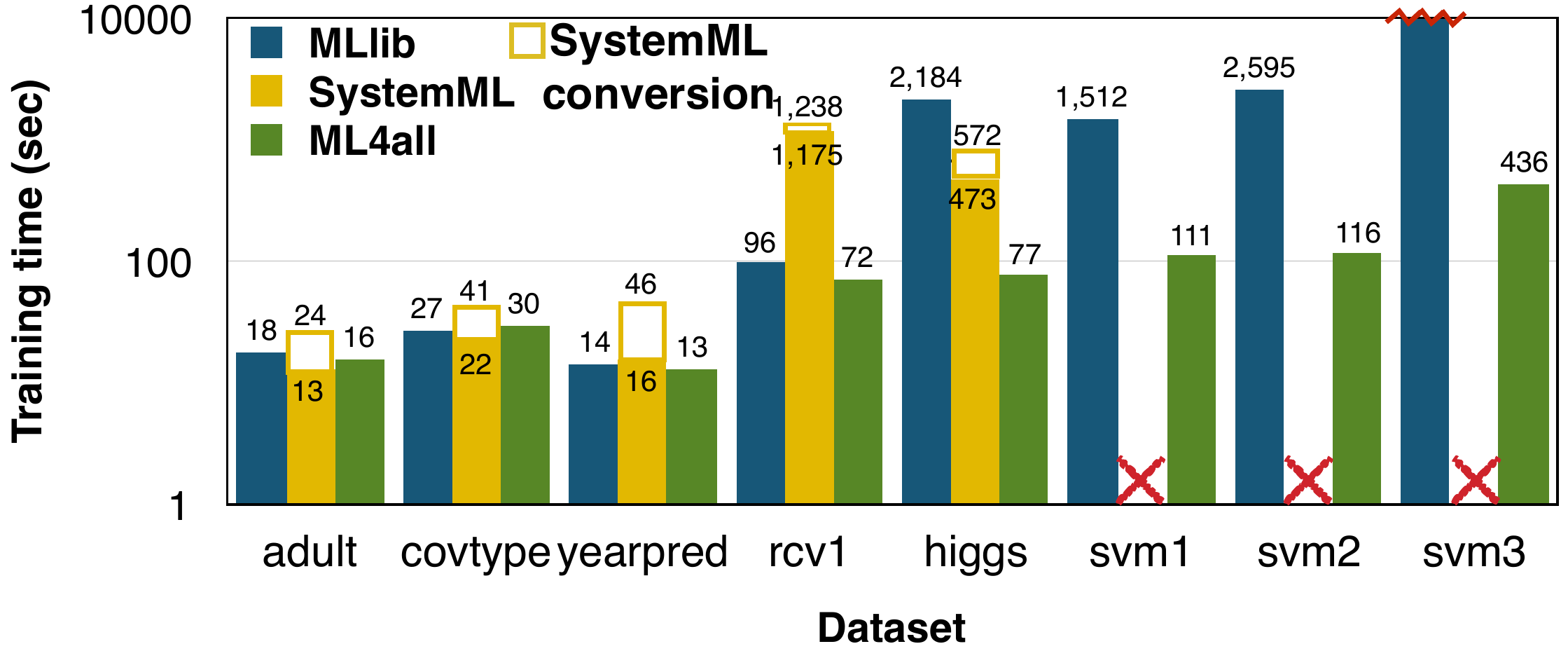}\label{subfig:mgd}}
		\subfigure[SGD]{\includegraphics[width=0.68 \columnwidth]{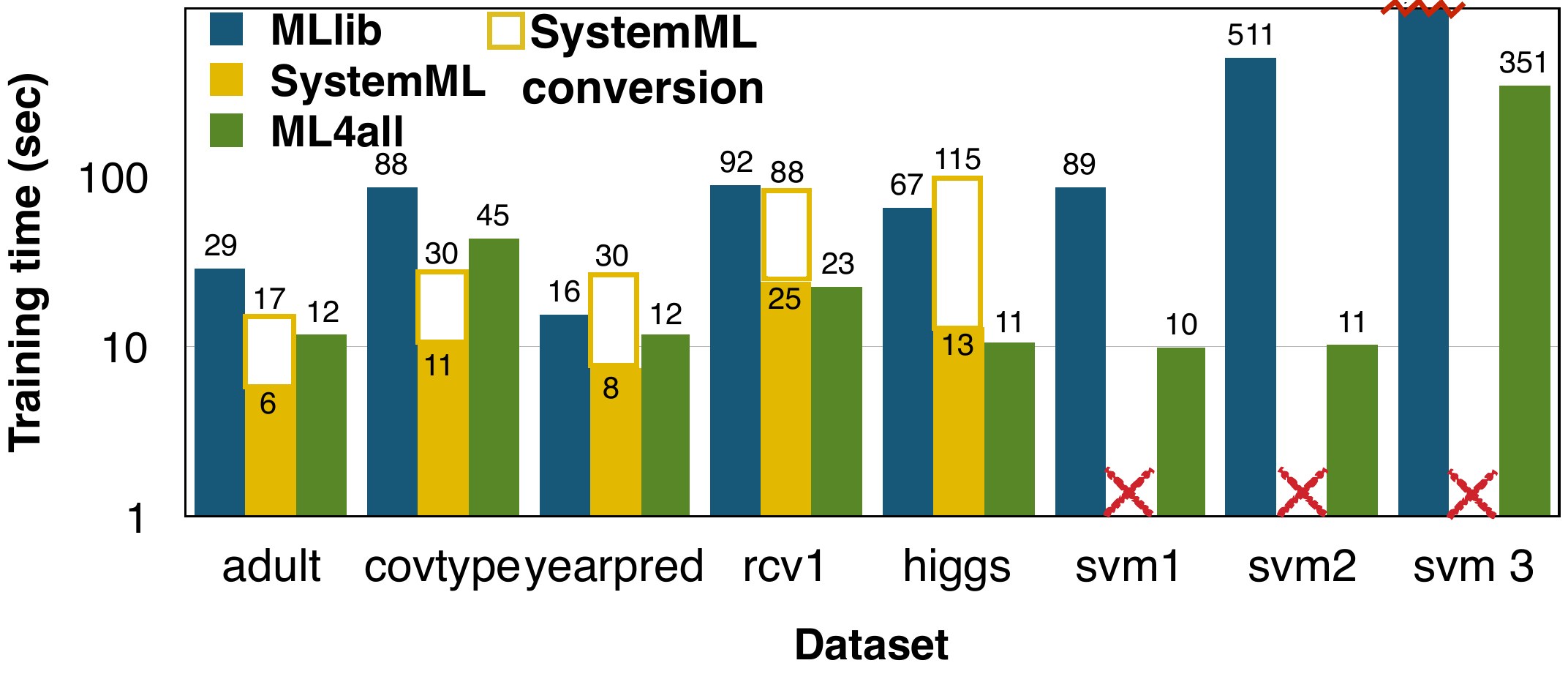}\label{subfig:sgd}}
	}
	\vspace{-0.5cm}
	\caption{Training time (sec). \mlall significantly outperforms both MLlib and SystemML, thanks to its novel sampling mechanisms and its lazy transformation technique. \label{fig:mllib}}
	\vspace{-0.5cm}
\end{figure*}

\subsubsection{System performance}
\label{sec:exp_sysperf}
We compare our system with MLlib and SystemML.
As neither of these systems have an equivalent of a GD optimizer, we ran BGD, MGD and SGD and we used \mlall just to find the best plan given a GD algorithm, \ie~which sampling to use and whether to use lazy transformation or not. We ran BGD, SGD, and MGD with a batch size of $1,000$ in all three systems until convergence. We considered a tolerance of $0.001$ and a maximum of $1,000$ iterations.

Let us now stress three important points.
First, note that the API of MLlib allows users to specify the fraction of the data that will be processed in each iteration. 
Thus, we set this fraction to $1$ for BGD while, for SGD and MGD, we compute the fraction as the batch size over the total size of the dataset.
However, the Bernoulli sample mechanism implemented in Spark (and used in MLlib) does not exactly return the number of sample data requested.
For this reason, for SGD, we set the fraction slightly higher to reduce the chances that the sample will be empty.
We found this to be more efficient than checking if the sample is empty and, in case it is, run the sample process again.
Second, we used the DeveloperApi in order to be able to specify a convergence condition instead of a constant number of iterations.
Third, as SystemML does not support the LIBSVM format, we had to convert all our real datasets into SystemML binary representation.
We used the source code provided to us by the authors of~\cite{systemml}, which first converts the input file into a Spark RDD using the MLlib tools and then converts it into matrix binary blocks.
The performance results for SystemML \add{show the breakdown between the training time and this few seconds conversion time}.

Figure~\ref{fig:mllib} shows the training time in log-scale for different real datasets and three larger synthetic ones.
Note that for our system, the plots of SGD and MGD show the runtime of the best plan for the specific GD algorithm. Details on these plans as well as the number of iterations required to converge can be found in Table~\ref{tab:plans} in Appendix~\ref{app:experiments}.
From these results we can make the following three observations:

\vspace{-0.2cm}
\begin{packed_enum}
\item For BGD (Figure~\ref{subfig:gd}), we observe that even if sampling and \add{ lazy transformation} are not used in BGD, our system is still faster than MLlib.
This is because we used \texttt{mapPartitions} and \texttt{reduce} instead of \texttt{treeAggregate}, which resulted in better data locality and hence better response times for larger datasets.
Notice that SystemML is slightly faster than our system \add {for the small datasets, because it processes them locally}.
The largest bottleneck of SystemML for small datasets is the time to convert the dataset to its binary format.
However, we observe that our system significantly outperforms SystemML for larger datasets, when SystemML runs on Spark.
In fact, we had to stop SystemML after 3 hours for the \texttt{higgs} dataset, while for all the dense synthetic datasets SystemML failed with out of memory exceptions. 

\item For MGD (Figure~\ref{subfig:mgd}), we observe that our system outperforms, on average, both MLib and SystemML:
It has similar performance to MLib \add{and SystemML for small datasets. However, SystemML requires an extra overhead of converting the data to its binary representation.}
It is up to $28$ times faster than MLib and more than $17$ times faster than SystemML for large datasets. \add{Especially, for the dataset \texttt{svm3} that does not fit entirely into Spark's cache, MLlib incurred disk IOs in each iteration resulting in a training time per iteration of $6$ min. Thus, we had to terminate the execution after 3 hours.}
The large benefits of our system come from the shuffle-partition sampling technique, which significantly saves IO costs. 

\begin{figure}[t]
	\centering
	\mbox{			
		\subfigure[Scaling \#points.]{\includegraphics[width=0.5 \columnwidth]{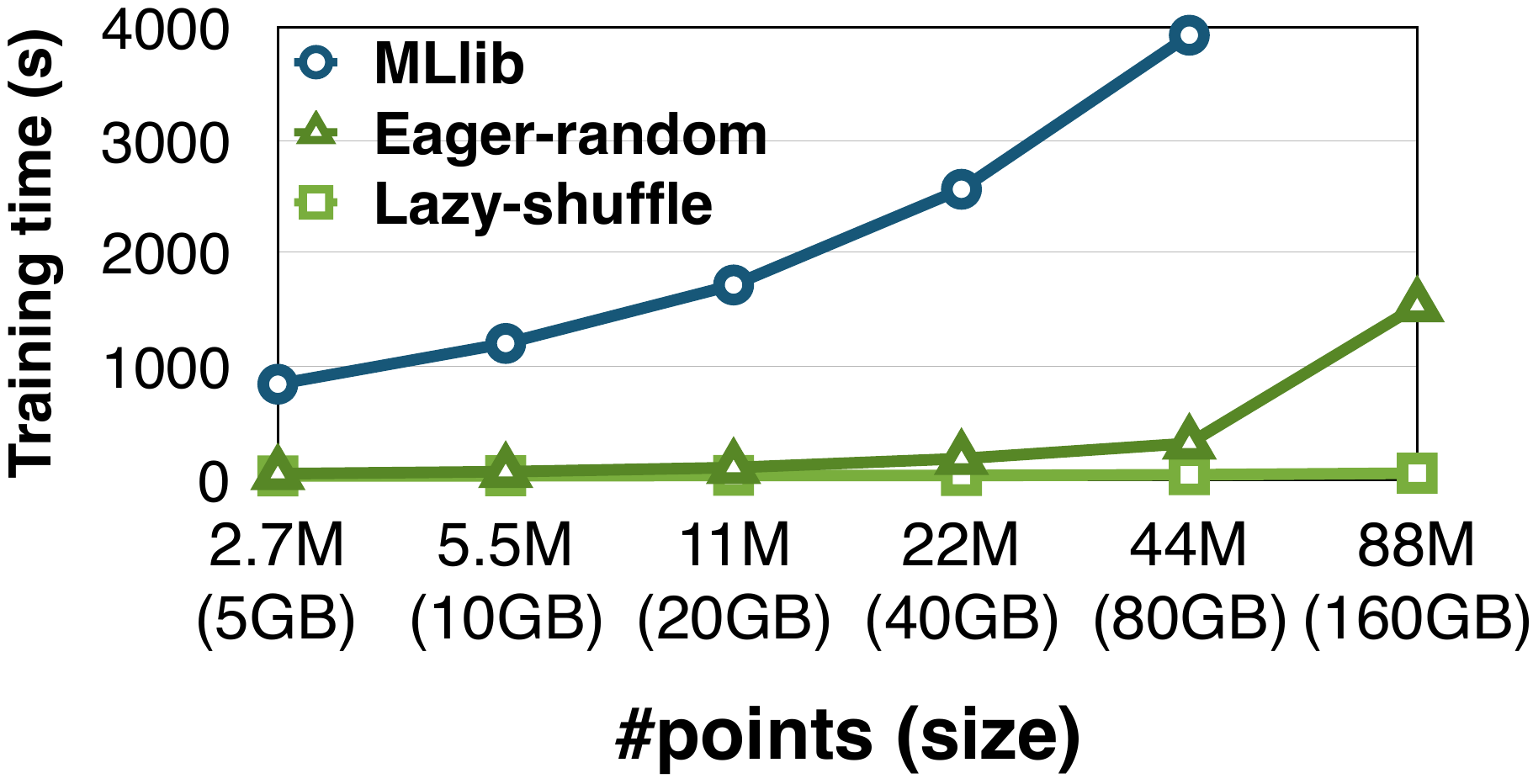}\label{subfig:points}}
		\subfigure[Scaling \#features.]{\includegraphics[width=0.5 \columnwidth]{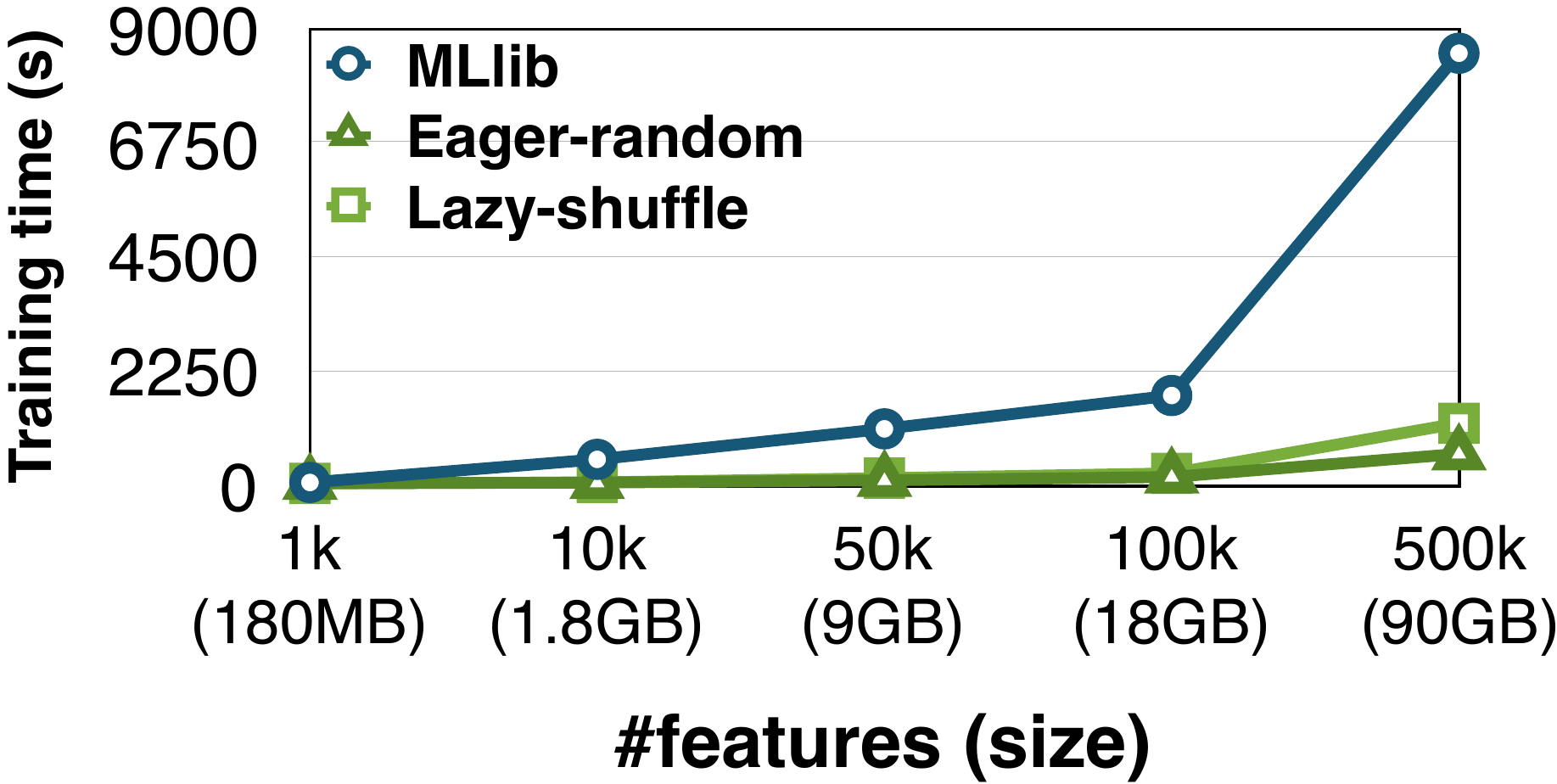}\label{subfig:features}}
	}
	\vspace{-0.5cm}
	\caption{\mlall scalability compared to MLlib. It scales gracefully with both the number of data points and features. \label{fig:comp-scal}}
	\vspace{-0.4cm}
\end{figure}

\item For SGD (Figure~\ref{subfig:sgd}), we observe that our system is significantly superior than \add{MLlib (by a factor from $2$ for small datasets to $46$ for larger datasets).
In fact, similarly to MGD, MLlib incurred many disk IOs for \texttt{svm3}.
We had to stop the execution after 3 hours.
In contrast, SystemML has lower training times for the very small datasets (\texttt{adult}, \texttt{covtype}, and \texttt{yearpred}), thanks to its binary data representation that makes local processing faster.
However, the cost of converting data to its binary data representation is higher than its training time itself, which makes SystemML slower than our system (except for \texttt{covtype}).
Things get worse for SystemML as the data grows and get dense.
Our system is more than one order of magnitude faster than SystemML.}
The benefits of our system on SGD is mainly due to the lazy transformation used by our system.
In fact, as for BGD and MGD, SystemML failed with out of memory exceptions for the \add{three} dense datasets.
Notice that the training time for a larger dataset may be smaller if the number of iterations to converge is smaller.
For example, this is the case for the dataset \texttt{covtype}, which required $923$ iterations to converge using SGD, in contrast to \texttt{rcv1}, which required only $196$.
This resulted in \mlall requiring smaller training time for \texttt{rcv1} than \texttt{covtype}.
\end{packed_enum}
\vspace{-0.2cm}

\begin{figure*}
	\centering
	\mbox{
		\subfigure[\texttt{adult} dataset]{\includegraphics[width=0.6 \columnwidth]{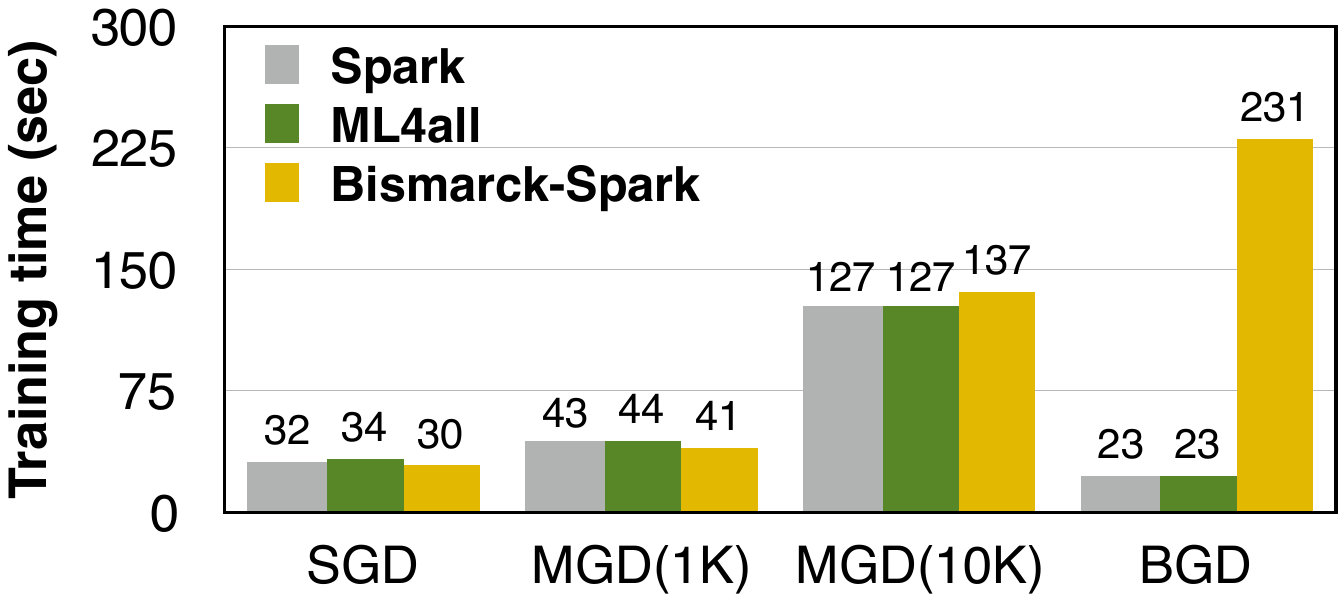}\label{fig:abstraction_adult}\hspace{0.4cm}}
		\subfigure[\texttt{rcv1} dataset]{\includegraphics[width=0.57 \columnwidth]{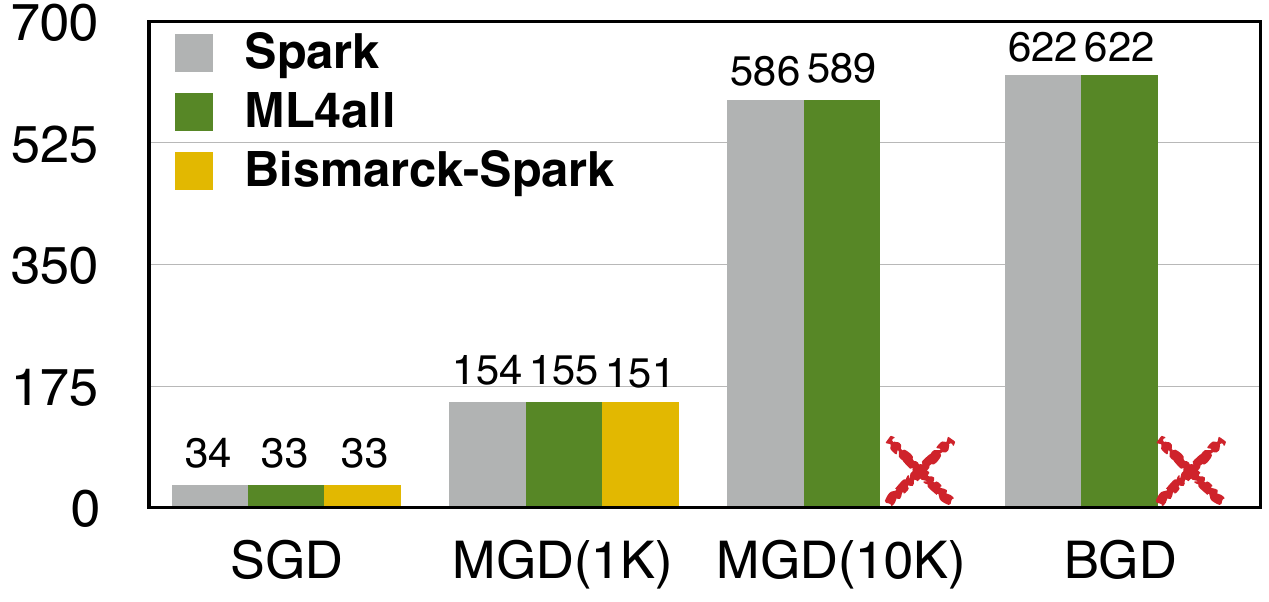}\label{fig:abstraction_rcv}\hspace{0.4cm}}
		\subfigure[\texttt{svm1} dataset]{\includegraphics[width=0.57 \columnwidth]{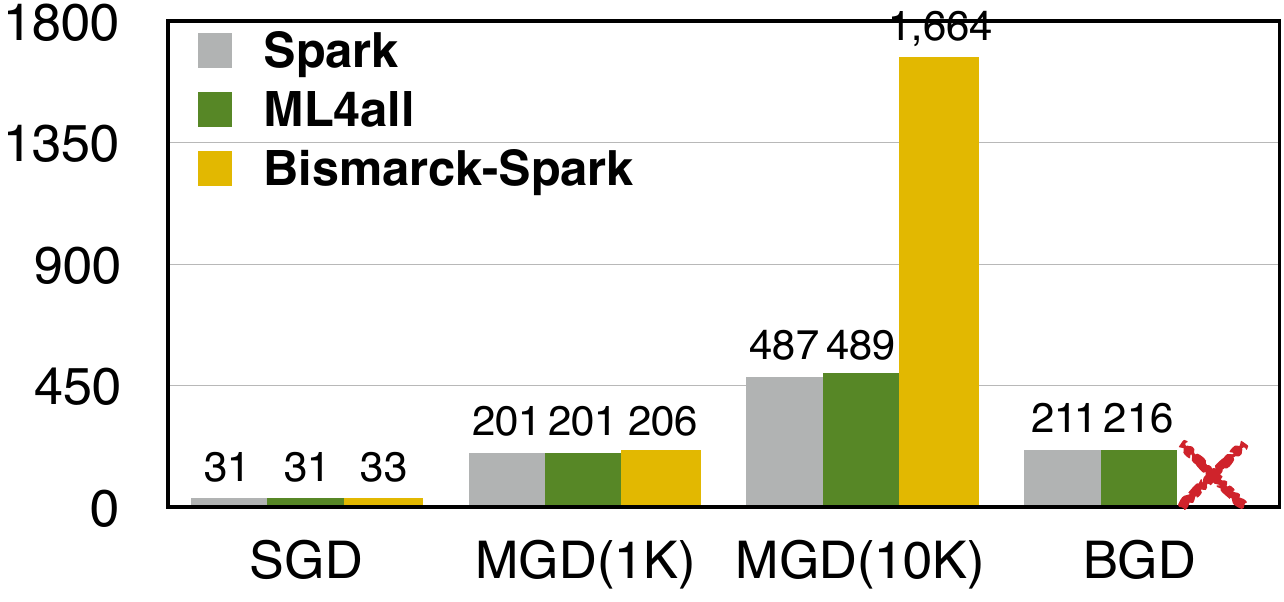}\label{fig:abstraction_syn}}
	}	
	\vspace{-0.5cm}
	\caption{\mlall abstraction benefits and overhead. The proposed abstraction has negligible overhead w.r.t. hard-coded Spark programs while it allows for exhaustive distributed execution.  \label{fig:abstraction}}
	\vspace{-0.0cm}
\end{figure*}

\subsubsection{Scalability}
\label{sec:exp_scalability}
Figure~\ref{fig:comp-scal} shows the scalability results for SGD for the two largest synthetic datasets (\texttt{SVM A} and \texttt{SVM B}), when increasing the number of data points (Figure~\ref{subfig:points}) and the number of features (Figure~\ref{subfig:features}).
Notice that we discarded SystemML as it was not able to run on these dense datasets.
We plot the runtimes of the eager-random and the lazy-shuffle GD plan.
We observe that both plans outperform MLlib by more than one order of magnitude in both cases.
In particular, we observe that our system scales gracefully with both the number of data points and the number of features while MLlib does not. \add{This is even more prominent for the datasets that do not fit in Spark's cache memory.}
Especially, we observe that the lazy-shuffle plan scales better than the eager-random.
This shows the high efficiency of our shuffled-partition sampling mechanism in combination with the lazy transformation.
Note that we had to stop the execution of MLlib after 24 hours for the largest dataset of 88 million points in Figure~\ref{subfig:points}.
MLlib took 4.3 min for each iteration and thus, would require 3 days to complete while our GD plan took only 25 minutes.
This leads to more than $2$ orders of magnitude improvement over MLlib.


\subsubsection{Benefits and overhead of abstraction}
\label{sec:exp_benefits}
We also evaluate the benefits and overhead of using the \mlall abstraction.
For this, we implemented the plan produced by \mlall directly on top of Spark.  
We also implemented the Bismarck abstraction~\cite{bismarck}, \add{which comes with a \at{Prepare} UDF}, while the \at{Compute} and \at{Update} are combined, on Spark. 
Recall that a key advantage of separating {\tt Compute} from {\tt Update} is that the former can
be parallelized where the latter has to be effectively serialized. When these two operators are combined into one, parallelization cannot be leveraged. \add{Its \at{Prepare}  UDF, however, can be parallelized.}

Figure~\ref{fig:abstraction} illustrates the results of these experiments. 
We observe that \mlall adds almost no additional overhead to plan execution as it has very similar runtimes as the pure Spark implementation.
We also observe that our system and Bismarck have similar runtimes for SGD and MGD(1k) and for all three data sets.
This is because our prototype runs in a hybrid mode and parts of the plan are executed in a centralized fashion thus negating the separation of the {\tt Compute} and the {\tt Update} step.
As the dataset cardinality or dimensionality increases, the advantages of \mlall become clear.
Our system is (i)~slightly faster for MGD(10k) for a small dataset (Figure~\ref{fig:abstraction_adult}), (ii)~more than $3$ times faster for MGD(10k) in Figure~\ref{fig:abstraction_syn}, because of the distribution of the gradient computation, and (iii)~able to run MGD(10k) in Figure~\ref{fig:abstraction_rcv} while the Bismarck abstraction fails due to the large number of features of \texttt{rcv1}. 
This is also the reason that the Bismark abstraction fails to run BGD for the same dataset of \texttt{rcv1}, but for \texttt{svm1} the reason it fails is the large number of data points.
This clearly shows that the Bismarck abstraction cannot scale with the dataset size. 
In contrast, our system scales gracefully in all cases as it execute the algorithms in a distributed fashion whenever required.

\subsubsection{Summary}
The high efficiency of our system comes from its (i)~lazy transformation technique, (ii)~novel sampling mechanisms, and (iii)~efficient execution operators.
All these results not only show the high efficiency of our optimizations techniques, but also the power of the \mlall abstraction that allows for such optimizations without adding any overhead.

\subsection{Accuracy}
\label{sec:exp_accuracy}
The reader might think that our system achieves high performance at the cost of sacrificing accuracy.
However, this is far from the truth.
To demonstrate this, we measure the \add{testing} error of each system and each GD algorithm.
\add{We used the test datasets from LIBSVM when available, otherwise we randomly split the initial dataset in training ($80\%$) and testing ($20\%$). We then apply the model (\ie~weights vector) produced on the training dataset to each example in the testing dataset to determine its output label. We plot the mean square error of the output labels compared to the ground truth. }
Recall that we have used the same parameters (\eg~step size) in all systems.

\begin{figure}[t]
\vspace{-0.4cm}
	\centering
	\mbox{
		\subfigure[MGD]{\includegraphics[width=0.5 \columnwidth]{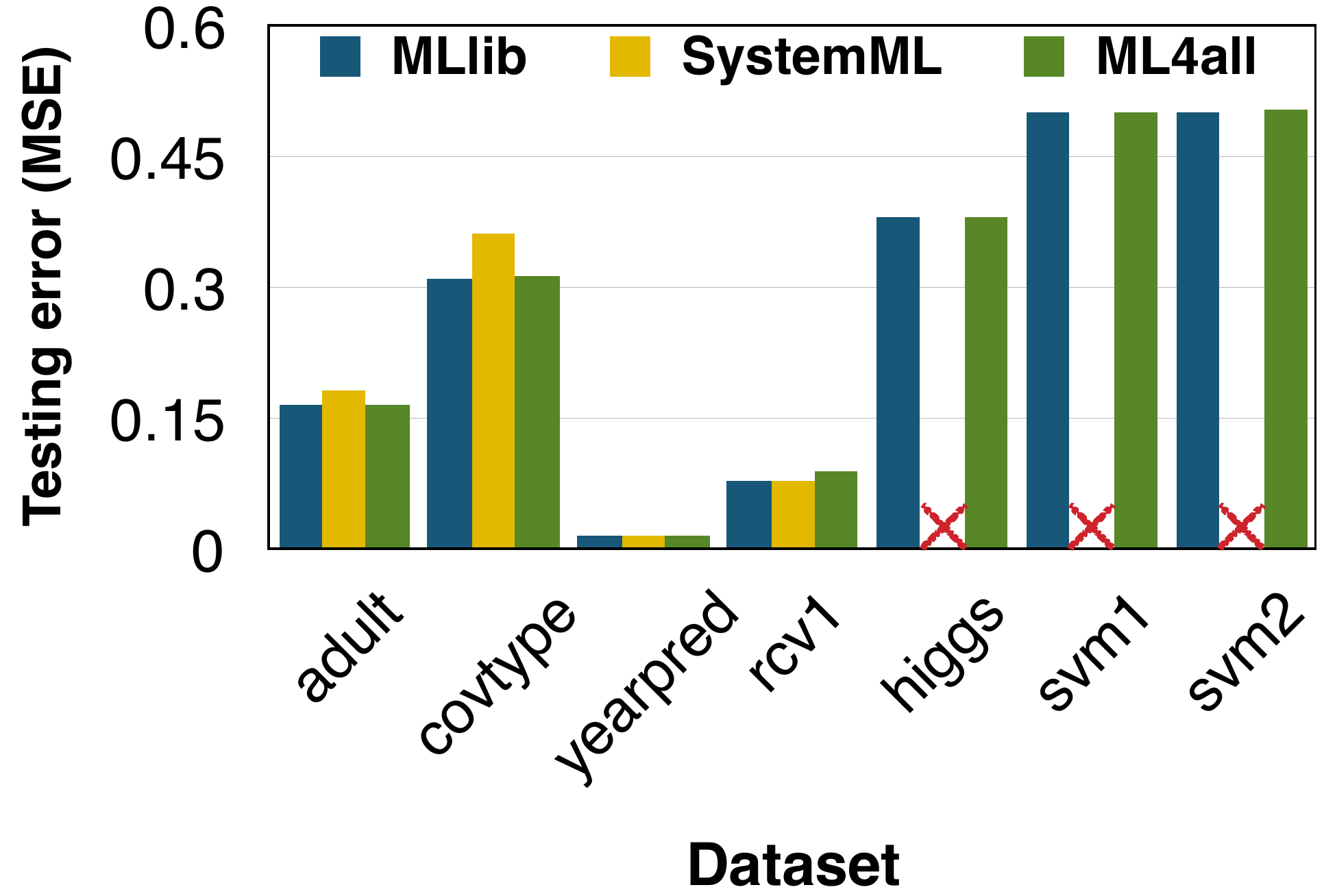}\label{subfig:error-mgd}}
				\subfigure[SGD]{\includegraphics[width=0.5 \columnwidth]{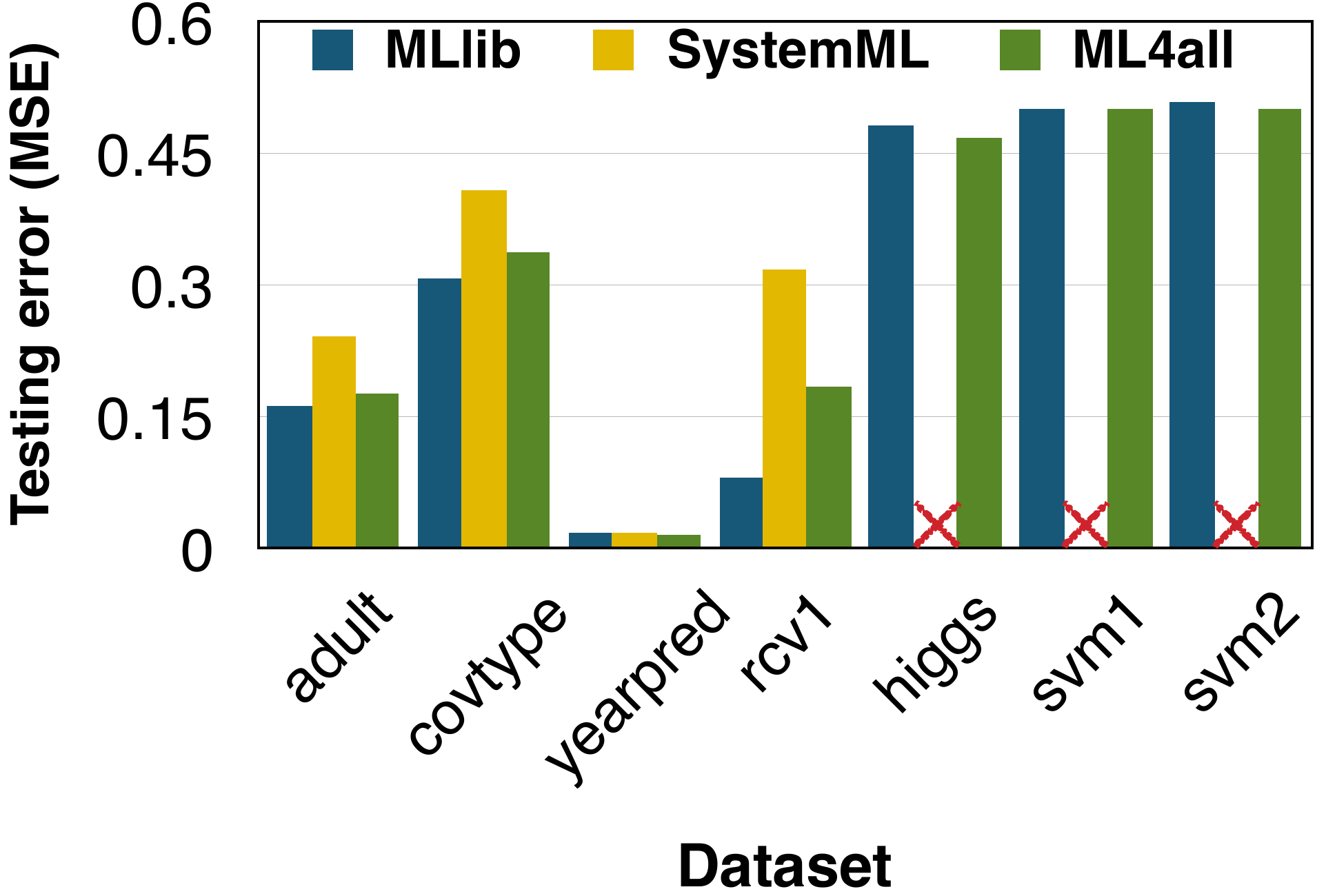}\label{subfig:error-sgd}}
	}
	\vspace{-0.4cm}
	\caption{Testing error (mean square error). For SGD/MGD, \mlall achieves an error close to MLlib even if it uses different sampling methods. \label{fig:error}}
	\vspace{-0.4cm}
\end{figure}

Let us first note that, as expected, all systems return the same model for BGD and hence we omit the graph as the \add{testing} error is exactly the same.
Figure~\ref{fig:error} shows the results for MGD and SGD. \add{We omit the results for \texttt{svm3} as only our system could converge in a reasonable amount of time.}
Although our system uses aggressive sampling techniques in some cases, such as shuffle-partition for the large datasets in MGD\footnote{\add{Table~\ref{tab:plans} in Appendix~\ref{app:experiments} shows the plan chosen in each case.}}, the error is significantly close to the ones of MLlib and SystemML. \add{The only case where shuffle-partition influences the testing error is for \texttt{rcv1} in SGD.
The testing error for MLlib is $0.08$, while in our case it is $0.18$.
This is due to the skewness of the data. SystemML having a testing error of $0.3$ also seems to suffer from this problem. We are currently working to improve this sampling technique for such cases. However, in cases where the data is now skewed our testing error even for SGD is very close to the one of MLlib.}
Thus, we can conclude that \mlall decreases training times without affecting the accuracy of the model.

\subsection{In-Depth}
\label{sec:exp_indepth}

We analyze in detail how the sampling and the transformation techniques affect performance when running MGD with $1,000$ samples and SGD until convergence with the tolerance set to $0.001$ and a maximum of $1,000$ iterations.

\subsubsection{Varying the sampling technique}

\begin{figure}[t]
	\centering
	\mbox{			
		\subfigure[Eager transformation]{\includegraphics[width=0.5 \columnwidth]{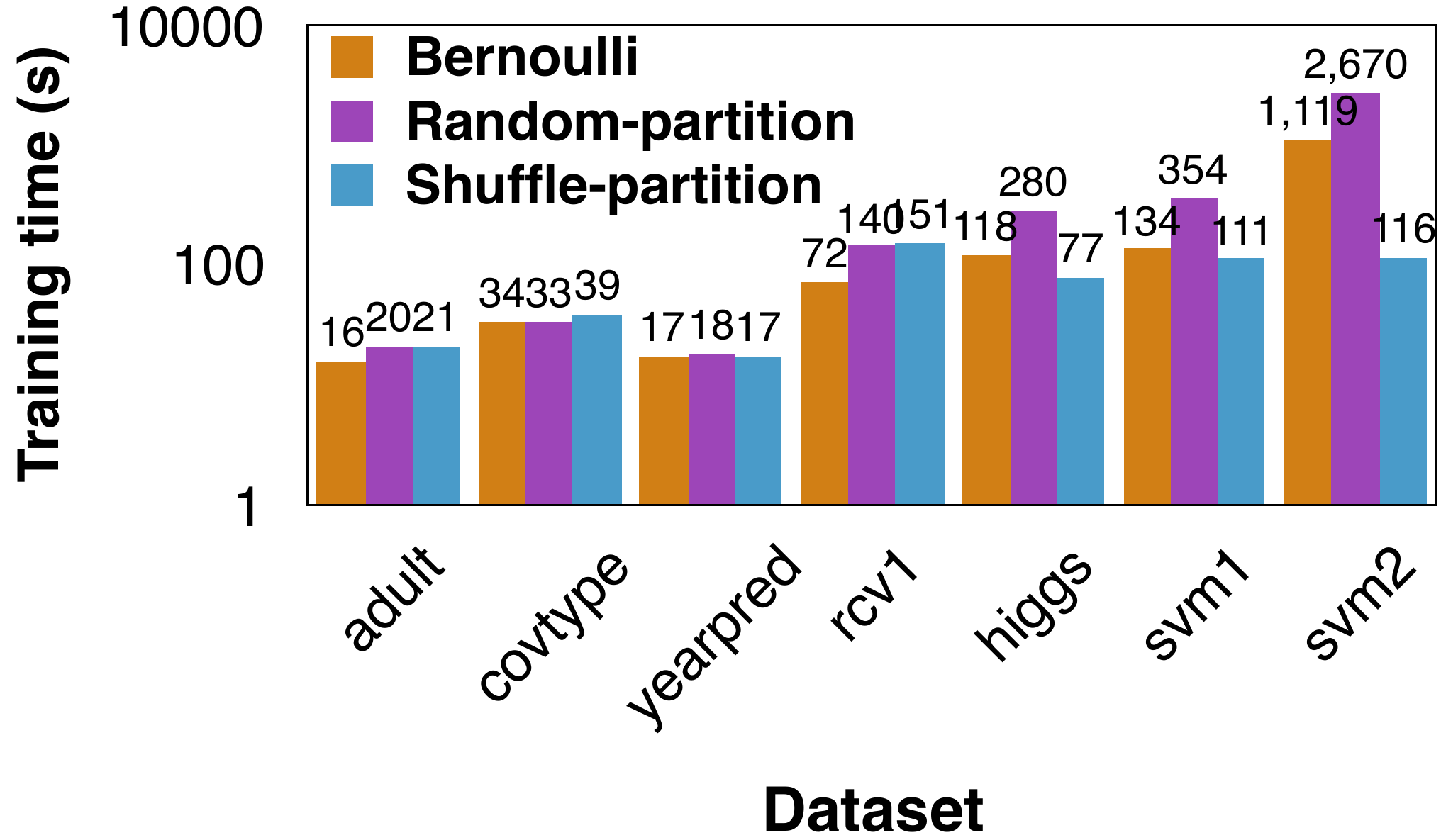}\label{subfig:eagerMGD}}
		\subfigure[Lazy transformation]{\includegraphics[width=0.48 \columnwidth]{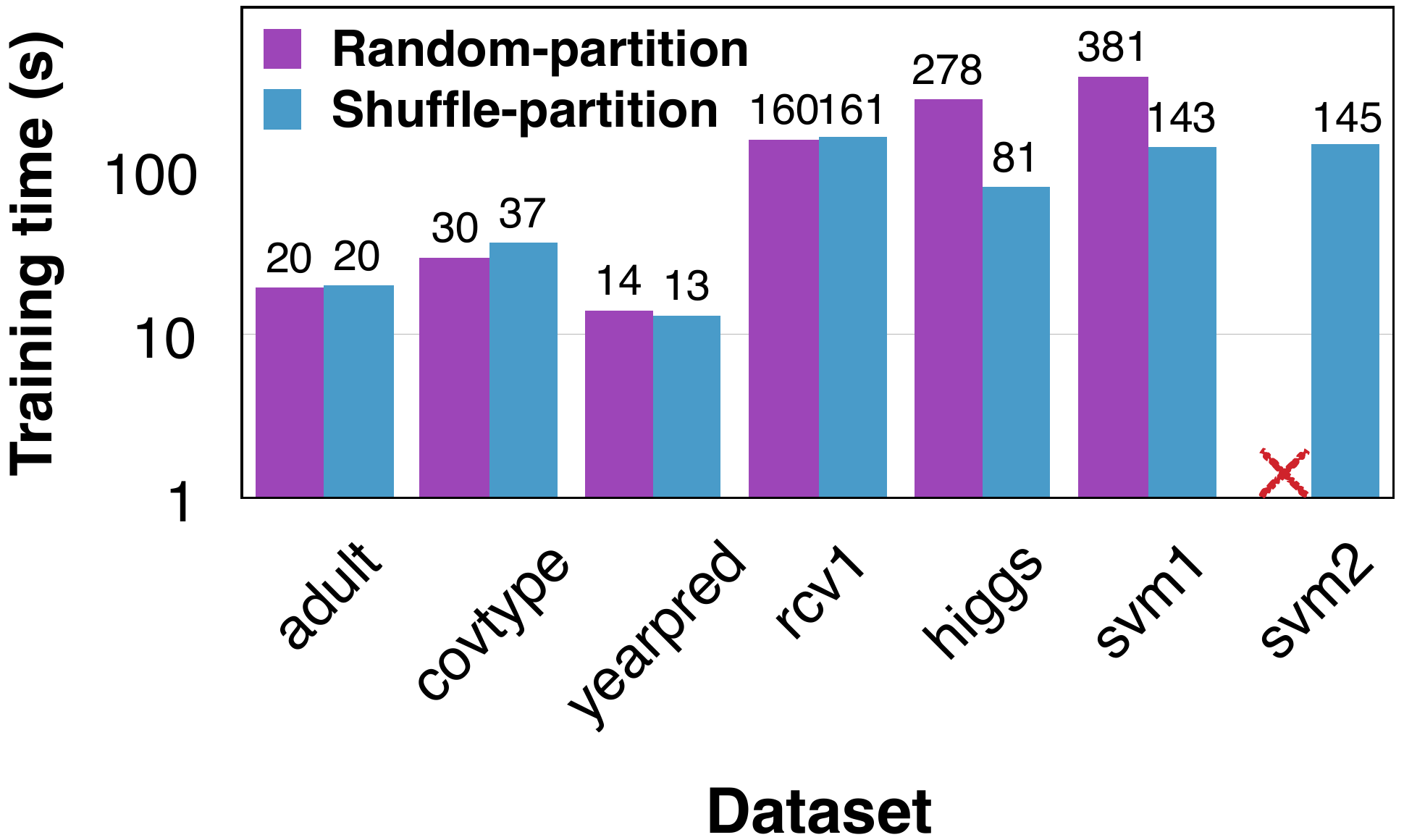}\label{subfig:lazyMGD}}
	}
	\vspace{-0.5cm}
	\caption{Sampling effect in MGD for eager and lazy transformation. \label{fig:mgd}}
	\vspace{-0.4cm}
\end{figure}

We first fix the transformation and vary the sampling technique. Figure~\ref{fig:mgd} shows how the sampling technique affects MGD when using eager and lazy transformation. First, in eager transformation for small datasets, using the Bernoulli sampling is more beneficial (Figure~\ref{subfig:eagerMGD}). This is because MGD needs a thousand samples per iteration and thus, a full scan of the whole dataset per iteration does not penalize the total execution time. However, for larger datasets that consist of more partitions, the shuffle-partition is faster in all cases as it accesses only few partitions.

For the lazy transformation (Figure~\ref{subfig:lazyMGD}), we ran only the random-partition and shuffle-partition sampling techniques.
Using a plan with Bernoulli sampling and lazy transformation is always inefficient as explained in Section~\ref{sec:plans}. We observe that for MGD and the two small datasets of \texttt{adult} and \texttt{covtype}, which consist of only one partition, the random-partition is faster than the shuffle-partition. Again, this is because the re-ordering of the partition does not pay-off in this case. For the rest of the datasets, shuffle-partition shows its benefits as only one partition is now accessed. In fact, for the larger synthetic dataset, we had to stop the execution of the GD plan with lazy transformation and random-partition after one hour and a half.

The results for SGD show that it benefits more from the shuffle-partition technique even for smaller datasets for both eager and lazy transformation (see Appendix~\ref{app:experiments}).

\subsubsection{Varying transformation method}

\begin{figure}[t]
	\centering
	\mbox{			
		\subfigure[SGD]{\includegraphics[width=0.5 \columnwidth]{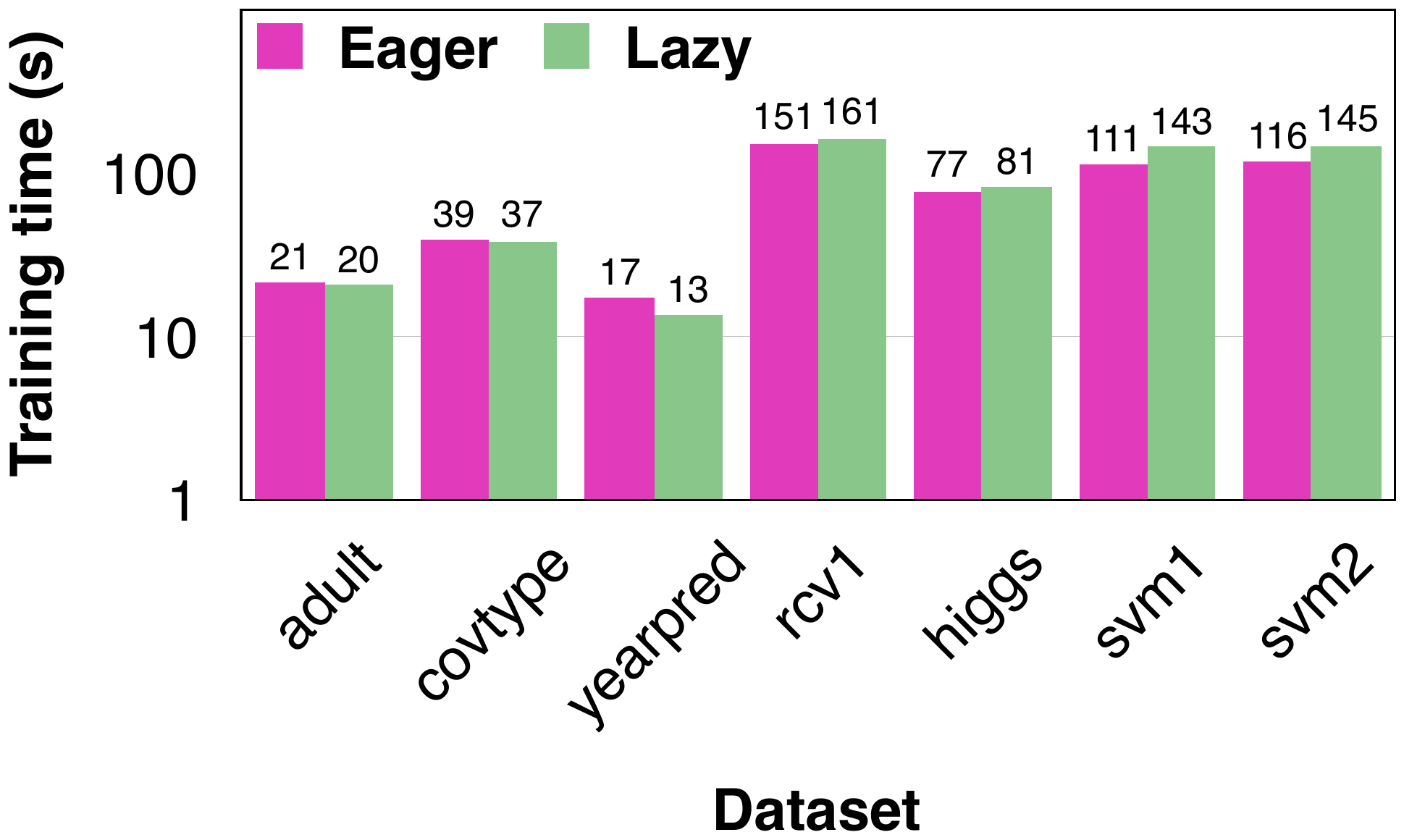}\label{subfig:shuffleSGD}}
		\subfigure[MGD]{\includegraphics[width=0.5 \columnwidth]{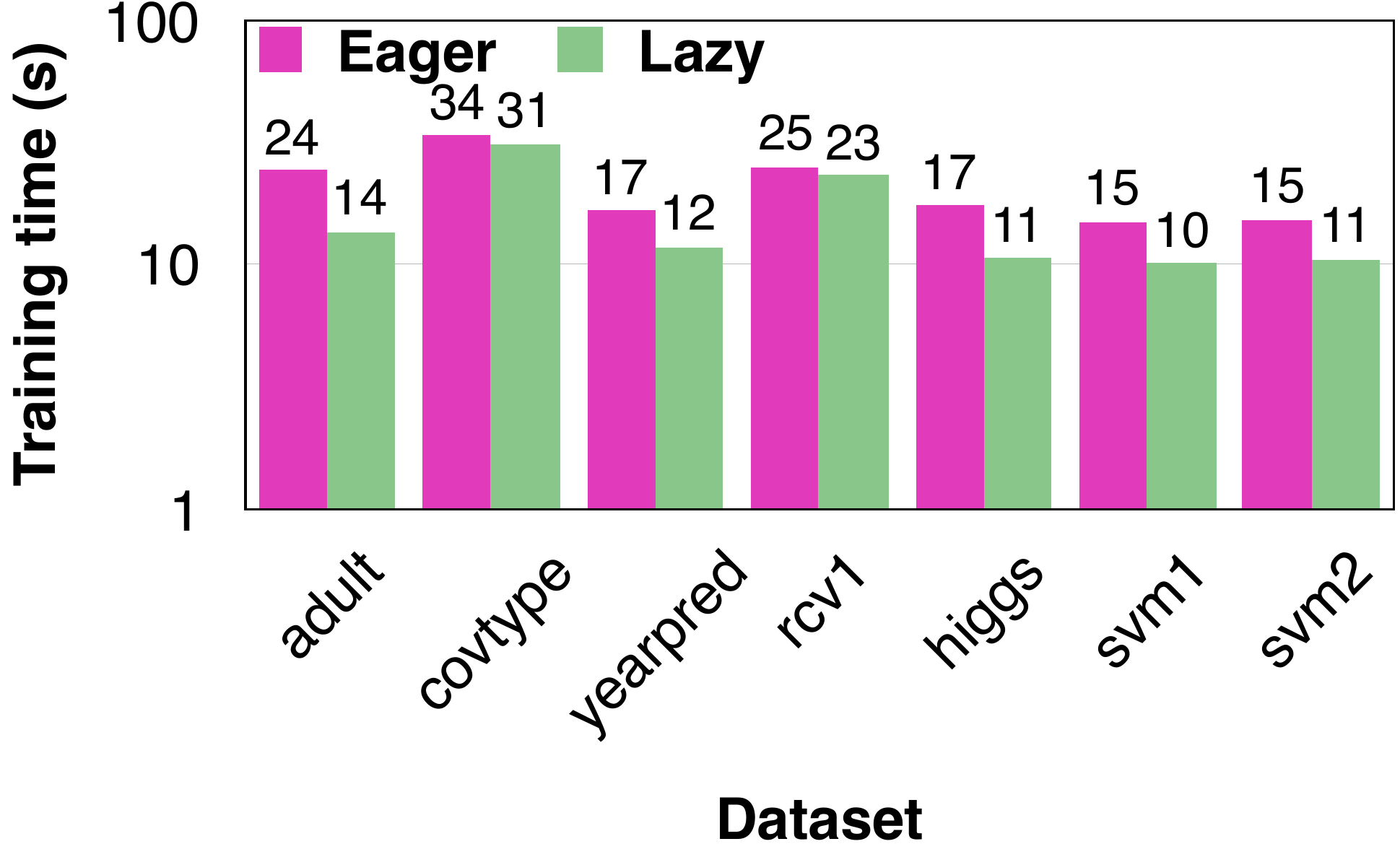}\label{subfig:shuffleMGD}}
	}
	\vspace{-0.5cm}
	\caption{Transformation effect for the shuffle-partition sampling technique. \label{fig:shuffle}}
	\vspace{-0.4cm}
\end{figure}

We now fix the sampling technique to shuffle-partition and vary the transformation. The results are shown in Figure~\ref{fig:shuffle}. Our first observation in Figure~\ref{subfig:shuffleSGD} is that SGD always benefits from the lazy transformation as only one partition is shuffled and only one sample needs to be transformed per iteration. For MGD, eager transformation pays-off more for the larger datasets. This is because the number of iterations required for these datasets arrives to the maximum of $1,000$ and therefore, for a batch size of $1,000$, MGD touches all data units. Thus, it pays off to transform all data units in an eager manner. Appendix~\ref{app:experiments} depicts the results when we fix the sampling technique to random-partition.


To summarize, most of the cases SGD profits from the lazy transformation and the shuffle-partition. For MGD, when the datasets are small then the eager transformation with the Bernoulli sampling is usually a good choice, while for larger datasets the shuffle-partition is usually a better choice. Although, we observed some of these patterns, we still prefer a cost-based optimizer to make such choices as there can be cases of datasets where these rules do not hold.

\subsubsection{Observations}\label{subsec:observations}
\add{In general, we observed some patterns on which our sampling and transformation techniques match best a specific algorithm. For instance, if the dataset size is smaller than one data partition, the eager-bernoulli variant is a good choice.
Another observation is that if the number of iterations to converge is much smaller than the data points of the input dataset, then the lazy transformation is a good choice for SGD. However, even if we observe such patterns, these are not always followed. This confirms the need for a cost-based optimizer instead of a rule-based one.}

%% file: relatedwork.tex

ML has attracted a lot of attention from the database research community over the last years~\cite{madlib,deepdive-vldb15,feature-selection,kumar-sigmod15,dimmwitted}.
The closest work to our GD abstraction is Bismarck~\cite{bismarck}. 
The authors propose to model ML operators as an optimization problem and use SGD to solve it. However, as we have witnessed from our experiments, SGD is not always the best algorithmic choice. 
In contrast, our abstraction covers all GD algorithms.
\add{Yet, similar to the authors of Bismarck, one could leverage the User-Defined Aggregate (UDA) feature in most DBMSes to integrate our ideas in a DBMS.
Another related (but complementary) work is DimmWitted~\cite{dimmwitted}, which also focuses on optimizing the execution of statistical analytics.
However, the authors mainly study the trade-off between statistical 
efficiency (\ie~number of iterations needed to converge) 
and hardware efficiency (\ie~execution time per iteration).
In contrast, we focus on selecting both the best GD plan and execution mode (centralized or distributed) of each GD operator.
Furthermore, the authors exploit their ideas only for NUMA machines and hence for main-memory analytics.
}

There are several distributed ML systems built on top of Hadoop or Spark, such as Mahout~\cite{mahout} and MLlib~\cite{mllib}. 
However, in all these systems adding new algorithms or modifying existing ones requires from users a good understanding of the underlying data processing framework.
In the MLBase vision paper~\cite{mlbase}, 
the authors 
plan to provide all available algorithms for a specific ML task together with an optimizer for choosing among these algorithms. However, building all algorithms for a specific operator is a tedious task where a developer has to deal with adhoc decisions for each algorithm.
 SystemML~\cite{systemml-vldb} also provides a cost-based optimizer but it is more focused on finding plans that parallelize task execution and data processing. 
Another system built on top of Hadoop is Cumulon~\cite{cumulon}. It uses a cost model to estimate time and monetary cost of matrix-based data analysis programs. However, the authors of~\cite{cumulon} do not address the problem of algorithm selection. Finally, Google's deep learning platform, TensorFlow~\cite{tensorflow}, lacks an optimizer.
There have also been other efforts towards optimizing the SGD performance~\cite{hogwild,DBLP:conf/icml/LiuWRBS14}, using SGD in distributed deep learning~\cite{singa-oss}, 
and parallelizing both data and model computations~\cite{petuum}. Nevertheless, all these works are complementary to ours.

Moreover, convergence of GD algorithms and the rate of their convergence has been extensively studied theoretically~\cite{bertsekas1999nonlinear,mlbook,bousquet2008tradeoffs}.
A popular technique, local analysis, studies the behavior of algorithms in the vicinity of the optimal solution. 
Nonetheless, local analysis cannot be directly used in practice, because it requires the knowledge of $w^{*}$, which is only known at the end of the execution of a GD algorithm. 


To the best of our knowledge, our approach is the first one to provide both a general GD abstraction to express most GD algorithms and an optimizer that selects the best GD algorithm for users' declarative queries.

%% file: conclusion.tex

We presented a cost-based optimizer for solving optimization problems using gradient descent algorithms. 
\add{In particular, our cost-based optimizer obviates the need for rules of thumb and justifies standard practices used by ML practitioners when selecting a GD algorithm.}
Our optimizer uses a new abstraction for easy parallelization of GD algorithms and further optimizations that lead to performance speedup.
It is able to choose the best GD plan, while the optimizations of \mlall built on top of \rheem speed up performance by more than two orders of magnitude than state-of-the-art ML systems on top of Spark, \ie~MLlib and SystemML. 
\add{Last but not least, our approach can easily be extended to assist in other design choices in ML systems, such as hyperparameter tuning.}

%% file: language.tex

\add{\mlall exposes to users a simple declarative language to interact with its GD optimizer. 
As this language also targets non-expert users, \ie~with no (or little) knowledge of ML and data processing platforms, it is composed of three main commands only:
{\sc run} for executing an ML operator, {\sc having} for expressing constraints, and {\sc using} for controlling the optimizer to some extent.
Where the {\sc having} and {\sc using} commands are optional.
Notice that a detailed discussion of this language is out of the scope of this paper.}

\myparagraph{Running a  query} \add{Users run a task on a specific dataset using the basic command {\sc run}. This is a mandatory command in any query. For example, a user would write the following query to run a classification on a given dataset:}

\vspace{-0.1cm}
\begin{center}
	\texttt{Q1 =} {\sc run} \texttt{classification} {\sc on} \texttt{training\_data.txt;}
\end{center}
\vspace{-0.1cm}

\noindent \add{This query states that the user wants to build a classification model using the dataset \at{training\_data.txt}. Users can also be more specific and provide a gradient function instead of the ML task. The gradient function can be from the ones provided by the system (\eg~\at{hinge()}) or by users via a Java UDF (see Table~\ref{tab:gradients} for a list of currently supported ML tasks and gradient functions).  
Additionally, users can provide their own parser to read an input dataset. For example, if the dataset \at{traning\_data} is sparse, then the user can utilize the libsvm dataset parser, \at{libsvm(training\_data.txt)}.
Notice that, by default, the system takes the first column as the label and the remaining columns as the features.
Users can also specify the columns for the label and features as shown in \at{Q2}.}

	\begin{table}[h!]
		\centering
		\caption{ML tasks and gradient functions currently supported by our system.\label{tab:gradients}}
		\begin{tabular}{|c|l|}
			\hline
			\textbf{ML task} & \textbf{Gradient function}\\
			\hline
			Linear regression & $g(\mathbf{w},\mathbf{x_i},  y_i) = 2 (\mathbf{w}^{T} \mathbf{x_i} - y_i)  \mathbf{x_i}$ \\
			\hline
			Logistic regression & $g(\mathbf{w},\mathbf{x_i},  y_i) = (\frac{-1}{1 + e^{y_i\mathbf{w}^{T} \mathbf{x_i}}})  y_i \mathbf{x_i}$ \\
			\hline
			SVM & $g(\mathbf{w},\mathbf{x_i},  y_i) = \begin{cases} 
			-y_i  x_i, & y_i  \mathbf{w}^{T}x_i < 1 \\ 
			0, & y_i  \mathbf{w}^{T}x_i \geq 1
			\end{cases}$ \\
			\hline
		\end{tabular}
	\end{table}

\myparagraph{Specifying constraints} \add{Users express their time, accuracy, and iterations constraints as follows:}

\vspace{-0.1cm}
\begin{center}
	\texttt{Q2 =} {\sc run} \texttt{classification} \\ {\sc on} \texttt{input\_data.txt:2, input\_data.txt:4-20,} \\{\sc having time} \texttt{1h30m}{\sc , epsilon} \texttt{0.01}{\sc , max\_iter} \texttt{1000;}
\end{center}
\vspace{-0.1cm}

\noindent \add{With such an extension, the user is now indicating to the system that she wants:
(i)~her results before one hour and half;
(ii)~her results within a tolerance epsilon of $0.01$; and
(iii)~to run the system until convergence or for a maximum of $1000$ iterations. Indeed, any of these constraints are optional and independent from each other.
In case no tolerance is specified, the system uses the value $10^{-3}$ as default.
If the system cannot satisfy any of these constraints, it informs the user which constraint she has to revisit.
Note that, in \at{Q2}, the user is also specifying that column $2$ is the label and attributes $4-20$ are the features.}

\myparagraph{Controlling the optimizer} \add{Advanced users can additionally use the {\sc using} command to control the optimizer to some extend. They can specify the GD algorithm, the convergence function or condition, the step size, and the sampling for SGD and MGD:}

\vspace{-0.1cm}
\begin{center}
	\texttt{Q3 =} {\sc run} \texttt{classification} {\sc on} \texttt{input\_data.txt} \\{\sc using algorithm} \texttt{SGD}{\sc , convergence} \texttt{cnvg()}{\sc , step} \texttt{1}{\sc , sampler} \texttt{my\_sampler();}
\end{center}
\vspace{-0.1cm}

\add{\noindent In contrast to \at{Q1} and \at{Q2}, \at{Q3} tells the system to use: (i)~SGD as algorithm, (ii)~the convergence function \at{cnvg()}, (iii)~the step size 1, and (iv)~the sampling mechanism \at{my\_sampler()}. Similarly to the {\sc having} command, any of these {\sc using} commands are optional and independent from each other.
}

\myparagraph{Storing models and testing data} \add{As explained earlier, once a user sends her query, the system translates it into a GD plan using a cost-model, further optimizes the plan, runs the optimized execution plan, and returns the resulting model. A user can optionally store such a model using the command:}

\vspace{-0.1cm}
\begin{center}
	{\sc persist} \texttt{Q1} {\sc on} \texttt{my\_model.txt}.
\end{center}
\vspace{-0.1cm}

\add{Once a model is obtained, a user can run the test phase over a given dataset: <\texttt{result =} {\sc predict on} \texttt{test\_data} {\sc with} \texttt{my\_model.txt;}>.}

\vspace{-0.1cm}
\begin{center}
	\texttt{result =} {\sc predict on} \texttt{test\_data} {\sc with} \texttt{my\_model.txt;}
\end{center}
\vspace{-0.1cm}

\add{Next, users might indeed use the \at{result} from the test phase to compute the measures they are interested in, such as the precision, recall, and f1\_score.}

%% file: codesnippets.tex

We show in Listing~\ref{list:stage} the code snippet for the \at{Stage}, Listing~\ref{list:update} the \at{Converge}, and Listing~\ref{list:loop} the \at{Loop} operator of the example in Figure~\ref{figure:abstraction}(a). Listing~\ref{list:sample} shows a simple random sampling operator.
\begin{lstlisting}[basicstyle=\scriptsize\normalfont\sffamily,aboveskip=0pt,belowskip=0pt,float=h!,label=list:stage,caption=Code snippet example for \at{Stage}.]
public void stage(Context context) {
1	double[] weights = new double[features];
2	context.put("weights",weights);
3	context.put("step",1.0);
4	context.put("iter",0);
}
\end{lstlisting}

\begin{lstlisting}[basicstyle=\scriptsize\normalfont\sffamily,aboveskip=0pt,belowskip=0pt,float=h!,label=list:converge,caption=Code snippet example for \at{Converge}.]
public double converge (double[] input, Context context) {
1 	double[] weights = (double[]) context.getByKey("weights");
2 	double delta = 0.0;
3	for (int j = 0; j < weights.length; j++) {
4		delta += Math.abs(weights[j] - input[j]);
5	}
}
\end{lstlisting}

\begin{lstlisting}[basicstyle=\scriptsize\normalfont\sffamily,aboveskip=0pt,belowskip=0pt,float=h!,label=list:loop,caption=Code snippet example for \at{Loop}.]
public boolean loop(double input, double tolerance) {
1	boolean stop = input < tolerance;
2   return stop;
}
\end{lstlisting}


\begin{lstlisting}[basicstyle=\scriptsize\normalfont\sffamily,aboveskip=0pt,belowskip=0pt,float=h!,label=list:sample,caption=Code snippet example for \at{Sample}.]
public double[] sample(double[] input, Context context) {
1	double rand = new Random().nextDouble();
2	if (rand < 0.5)
3	   return null;
4   return input;
}
\end{lstlisting}

%% file: additional_plans.tex
\add{We demonstrate how our abstraction can support GD acceleration techniques such as line search or combinations of BGD and SGD.}

\myparagraph{SVRG}
The idea of this algorithm (stochastic variance reduced gradient) is to mix BGD with SGD in order to have fast convergence and fast computation.
It performs SGD by reducing its variance using BGD every $m$ iterations~\cite{srvg}.
This requires a nested loop operation where the outer loop requires the gradient calculation for all input data points (BGD) and the inner loop computes the gradient of a single data point (SGD). \add{In other words, SVRG computes the gradient of all the data points every $m$ iterations, while for the rest it just computes the gradient of one point.  SVRG has also a different update formula in order to reduce the variance in each iteration. We can ``flatten" the nested loops by using an {\it if-else} condition in the \at{Sample}, \at{Compute}, and \at{Update} operators in order to capture the computations of every $m$ iteration. }
Thus, we can express SVRG in our abstraction using the same plan as for SGD (Figure~\ref{figure:abstraction}(a)) but with different implementations of the GD operators.
\add{The pseudocode of the algorithm written to fit our abstraction is shown in Algorithm~\ref{algo:srvg}.}

\begin{algorithm}
	\DontPrintSemicolon 
	\KwIn{Update frequency $m$}
	\textbf{Initialize} $w_0$, $\tilde{w}$\;
	\For{t=1,2,...}{
		\If{(t mod $m$) - 1=0}{
			\If{$t > 1$}{
				$\tilde{w}:=w_{t}$\;
			}
			$\mu:=\frac{1}{n}\sum\limits_{i=1}^{n}\nabla f_i(\tilde{w})$\;
			$w_t := w_{t-1} - \alpha \mu$\;
		}\Else{
		Randomly pick $i_t \in \{1,2,...,n\}$ and update $w_t$\;
		$w_t := w_{t-1} - \alpha (\nabla f_{i_t}(w_{t-1}) -\nabla f_{i_t}(\tilde{w}) + \mu)$\;
	}
	
}
\caption{SVRG}
\label{algo:srvg}
\end{algorithm}

\add{Listing~\ref{list:compute-srvg} shows the modified code snippet for the \at{Compute} operator to implement the SVRG algorithm. Similarly the rest of the operators can be modified. This shows that our template is general enough to capture even algorithms that do not seem to match with a first glance.}
\begin{lstlisting}[basicstyle=\scriptsize\normalfont\sffamily,aboveskip=0pt,belowskip=0pt,float=h!,label=list:compute-srvg,caption=Code snippet for the \at{Compute} of SVRG.]
public Pair<double[], double[]> compute (SparseVector point, Context context) {
1 	int iteration = (int) context.getByKey("iter");
2 	double[] w = context.getByKey("weights");
3 	int m = (int) context.getByKey("m");
4	if ((iteration % m) - 1 == 0) 
5 		return new Pair(this.gradient.calculate(w, point), null);
6	else {
7		grad = this.gradient.calculate(w, point);
8		double[] w_bar = context.getByKey("weightsBar");
9		fullGrad = this.gradient.calculate(w_bar, point);
10		return new Pair(grad, fullGrad);
11	}
}
\end{lstlisting}

\myparagraph{GD with backtracking line-search}
\add{
There has been extensive research on accelerating gradient descent based methods through the step size.
A non-exhaustive list of tricks include step size-based approaches including fixed step size, adaptive step size, optimal step sizes including line search, BB methods, etc. Here we show how our abstraction can be applied to implement BGD\footnote{Usually line search is not used in stochastic algorithms because the correct direction of the gradient is required.} using backtracking line search. Backtracking line search chooses the step size in each iteration of GD as follows: $\alpha_{k_i} = \beta * \alpha_{k_{i-1}}$, where $k$ is the iteration step of BGD and $i$ is the iteration step of the line search. The iterations of the line search repeat until $f(\mathbf{w}^{k}) - f(\mathbf{w}^{k} - \alpha_{k_i} \nabla f(\mathbf{w}^{k})) < \alpha_{k_i} * i$. However, to compute the $f$ function the entire dataset is required. Thus, we need to modify the \at{Compute} and \at{Update}operator to support backtracking line-search step size. Similarly with SVRG we can emulate the nested loops of line search by adding an {\it if-else} condition. Listings~\ref{list:compute-ls} and~\ref{list:update-ls} show the pseudocode for \at{Compute} and \at{Update}, respectively. Similarly other methods, such as the Barzilai-Borwein, can be plugged in our system.
}

\begin{lstlisting}[basicstyle=\scriptsize\normalfont\sffamily,aboveskip=0pt,belowskip=0pt,float=h!,label=list:compute-ls,caption=Code snippet for the \at{Compute} of BGD with backtracking line search.]
public Pair<double, double[]> compute (SparseVector point, Context context) {
1	double[] w = (double[]) context.getByKey("weights");
2	double step = (double) context.getByKey("step");
3	double[] grad = this.svmGradient.calculate(w, point);
4	 boolean isStepSizeIter = (boolean) context.getByKey("isStepSize");
5	if (isStepSizeIter) {
6		double diff = this.objFunction.calculate(w, point) - this.objFunction.calculate(w - step * grad, point);
7 		return new Pair(diff, grad);
8	 }
9	else
10		return new Pair(Inf, grad);
}
\end{lstlisting}

\begin{lstlisting}[belowcaptionskip=-1.5cm, abovecaptionskip=-1.5cm, aboveskip=-0.2cm, belowskip=-0.2cm, basicstyle=\scriptsize\normalfont\sffamily,float=h!,label=list:update-ls,caption=Code snippet example of \at{Update} of BGD with backtracking line search.]
public double[] update (Pair<double,double[]> input, Context context) {
1 	 double[] weights = (double[]) context.getByKey("weights");
2 	 double beta = (double) context.getByKey("beta");
3	 double step = (double) context.getByKey("step");
4	 double diff = input.field0;
5	 int i = (int) context.getByKey("step_iteration");
6	 if (diff >= step * i) {
7		step = beta * step; 
8		 context.put("step", step);
9		 context.put("step_iteration", ++i);
10		return null;
11	 }
12  else {
13		context.put("isStepSizeIter", false);
14 		for (int j=0; j<weights.length; j++)
15			   weights[j] = weights[j] - step * input.field1[j+1];
16		return weights;
17	  }
}
\end{lstlisting}

%% file: implementation.tex

\add{
We implemented our GD optimizer in \mlall. \mlall is an ML system built on top of \rheem\footnote{\url{https://github.com/rheem-ecosystem/rheem}}, our in-house cross-platform system~\cite{rheem-vision,rheem-demo}. 
\rheem is a cross-platform system which abstracts from the underlying platforms and allows not only for platform independence but also for automatic selection of the right platform for a given job.
In \mlall, we use Spark and Java as the underlying platforms of \rheem and HDFS as the underlying storage. The source code of \mlall's abstraction can be found at \url{https://github.com/rheem-ecosystem/ml4all}.
We utilize \rheem's platform independence and map each operator of a GD plan to either Java code (for a centralized execution) or Spark code (for a distributed execution), transparently to users.
The reader may think that running some operators on a centralized mode might be a bottleneck.
However, our optimizer maps an operator to Java only if its input data fits in a single data partition (\ie~one HDFS partition). 
Running an operator on distributed mode for such small input data would just adds a processing overhead.
Thus, it maps an operator to a Spark operator only when its input data spans to multiple data partitions -- having each available processing core executing this operator over a single data partition.
More interestingly, \mlall can produce a GD plan as a mixture of Java and Spark (\ie~a Mix-based GD plan).
This is beneficial when the input size for some operators is large, but the input for other operators in the same plan is much smaller.
For instance, the \at{Transform} and \at{Sample} operators in SGD usually have several orders of magnitude larger input than the operators \at{Compute} and \at{Update}.
In fact, \mlall indeed produces a mix-based plan for SGD.
}

%% file: additional_experiments.tex

\myparagraph{Chosen plans}
Table~\ref{tab:plans} shows for each GD algorithm which was the best GD plan chosen by our optimizer and how many iterations were required for each plan to converge.

\begin{table}
	\centering
	\vspace{-0.3cm}
	\caption{Chosen plan for each GD algorithm.\label{tab:plans}}
	\scalebox{0.8}{
		\begin{tabular}{l|cc|cc|c}
			\hline
			\textbf{Dataset} & \multicolumn{2}{c|}{\textbf{SGD}} & \multicolumn{2}{c|}{\textbf{MGD}}  & \textbf{BGD}\\
			& \textbf{\#iter} & \textbf{plan}  & \textbf{\#iter} & \textbf{plan} & \textbf{\#iter} \\
			\hline
			\texttt{adult} & $433$ & lazy-random & $482$ & eager-bernoulli & $224$ \\
			\texttt{covtype} & $923$ & eager-bernoulli  &  $404$ & lazy-random & $381$\\
			\texttt{yearpred} & $26$ & lazy-shuffle & $14$& lazy-shuffle & $5$\\
			\texttt{rcv1} & $196$ & eager-shuffle & $773$ & eager-bernoulli & $515$\\
			\texttt{higgs} & $6$ & lazy-shuffle  & $1000$  & eager-shuffle  & $264$\\
			\texttt{svm1} & $4$ & lazy-shuffle & $1000$ & eager-shuffle &  $145$\\
			\texttt{svm2} & $5$ & lazy-shuffle & $1000$ & eager-shuffle & $145$\\
			\add{\texttt{svm3}} & \add{$8$} & \add{lazy-shuffle} & \add{$1000$} & \add{eager-shuffle} & \add{$145$}\\
			\hline
		\end{tabular}
	}
\end{table}

\begin{figure*}[t]
	\centering
	\mbox{			
		\subfigure[Step size $1/\sqrt{i}$]{\includegraphics[width=0.25\linewidth]{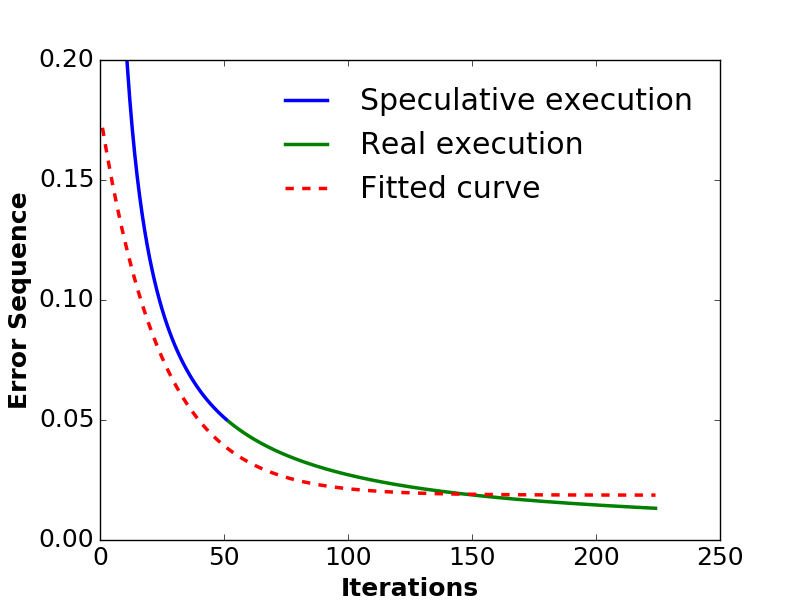}\label{subfig:adult_bgd_sublinear}}
		\subfigure[Step size $1/i$]{\includegraphics[width=0.25 \linewidth]{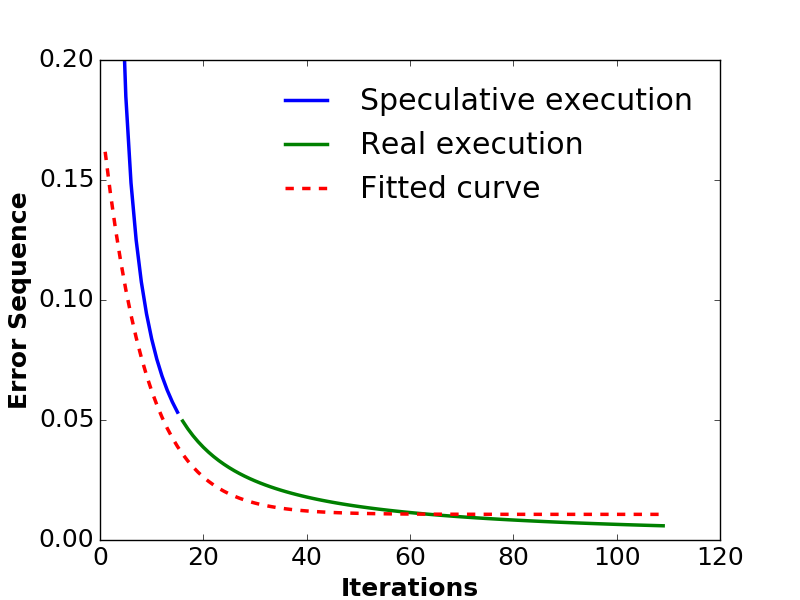}\label{subfig:adult_bgd_linear}}
		\subfigure[Step size $1/i^2$]{\includegraphics[width=0.25 \linewidth]{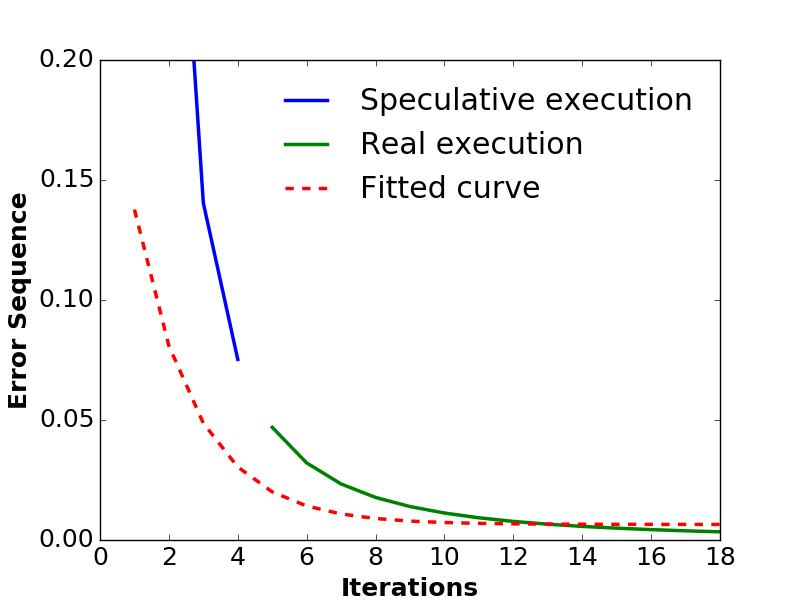}\label{subfig:adult_bgd_superlinear}}
	}
	\caption{\add{Curve fitting using different adaptive step sizes for \texttt{adult} dataset for BGD.} \label{fig:stepsize-adult}}
\end{figure*}

\begin{figure*}[t]
	\centering
	\mbox{			
		\subfigure[\texttt{covtype} dataset]{\includegraphics[width=0.25 \linewidth]{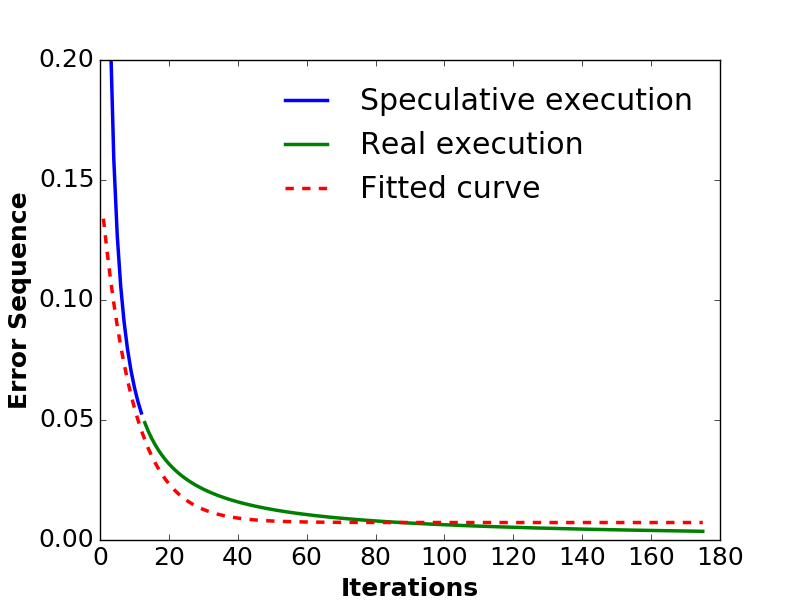}\label{subfig:covtype_bgd_sublinear}}
		\subfigure[\texttt{rcv1}]{\includegraphics[width=0.25 \linewidth]{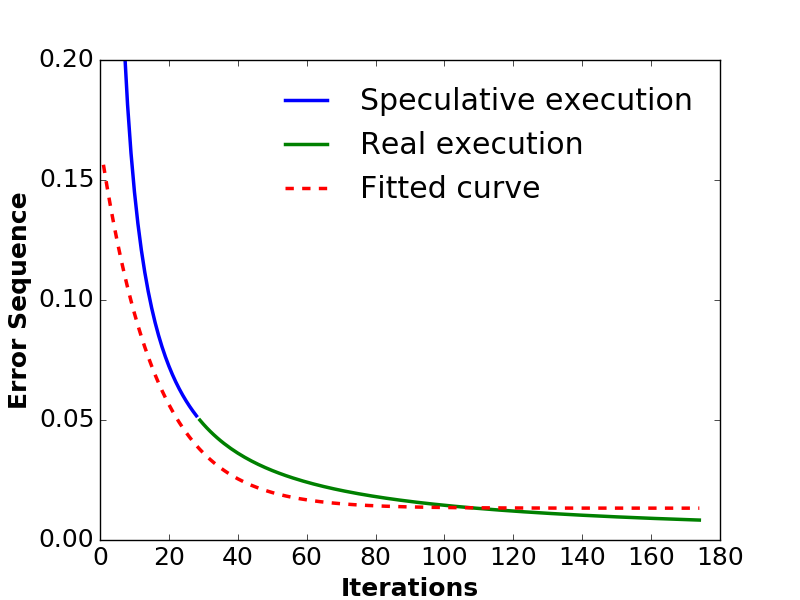}\label{subfig:rcv1_bgd_sublinear}}
		\subfigure[\texttt{higgs} dataset]{\includegraphics[width=0.25 \linewidth]{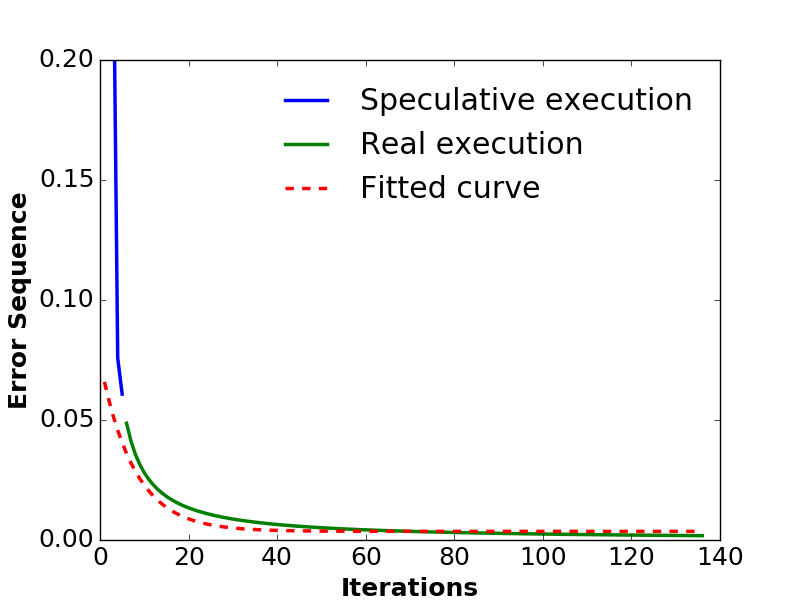}\label{subfig:higgs_bgd_sublinear}}
	}
	\caption{\add{Curve fitting using adaptive step size $1/i$ for BGD for various datasets.} \label{fig:stepsize-rest}}
\end{figure*}

\myparagraph{Adaptive step sizes in iterations estimator}
\add{We demonstrate how using different adaptive step size affects the iteration estimator. For this experiment, we ran the speculation of our iterations estimator on a sample of $1000$ data points until a tolerance value of $0.05$ and collected pairs of <$iteration, error$>. We use these pairs to fit the curve. Our goal is to estimate the number of iterations required to arrive to an error of $0.001$.
In addition, when ran the real execution until a tolerance value of $0.001$ and also collected pairs of <$iteration, error$> in order to compare with the fitted curve.
Figure~\ref{fig:stepsize-adult} shows the results for the \texttt{adult} dataset for BGD and two different step sizes. The $y$-axis shows the error sequence (tolerance values) at each iteration $i$ ($x$-axis). The blue line denotes the execution during speculation on a sample, while the green line depicts the real execution. The red dotted line is the fitted curve from which we can infer the number of iterations that the real execution will require to converge. Notice that the red dotted line arrives at an error of $0.001$ in almost the same number of iterations that the real execution terminates (green line). Similar results are observed for the other datasets as well. Figure~\ref{fig:stepsize-rest} shows some of these results.}

\begin{figure}[t!]
	\centering
	\mbox{			
		\subfigure[Eager transformation]{\includegraphics[width=0.5 \columnwidth]{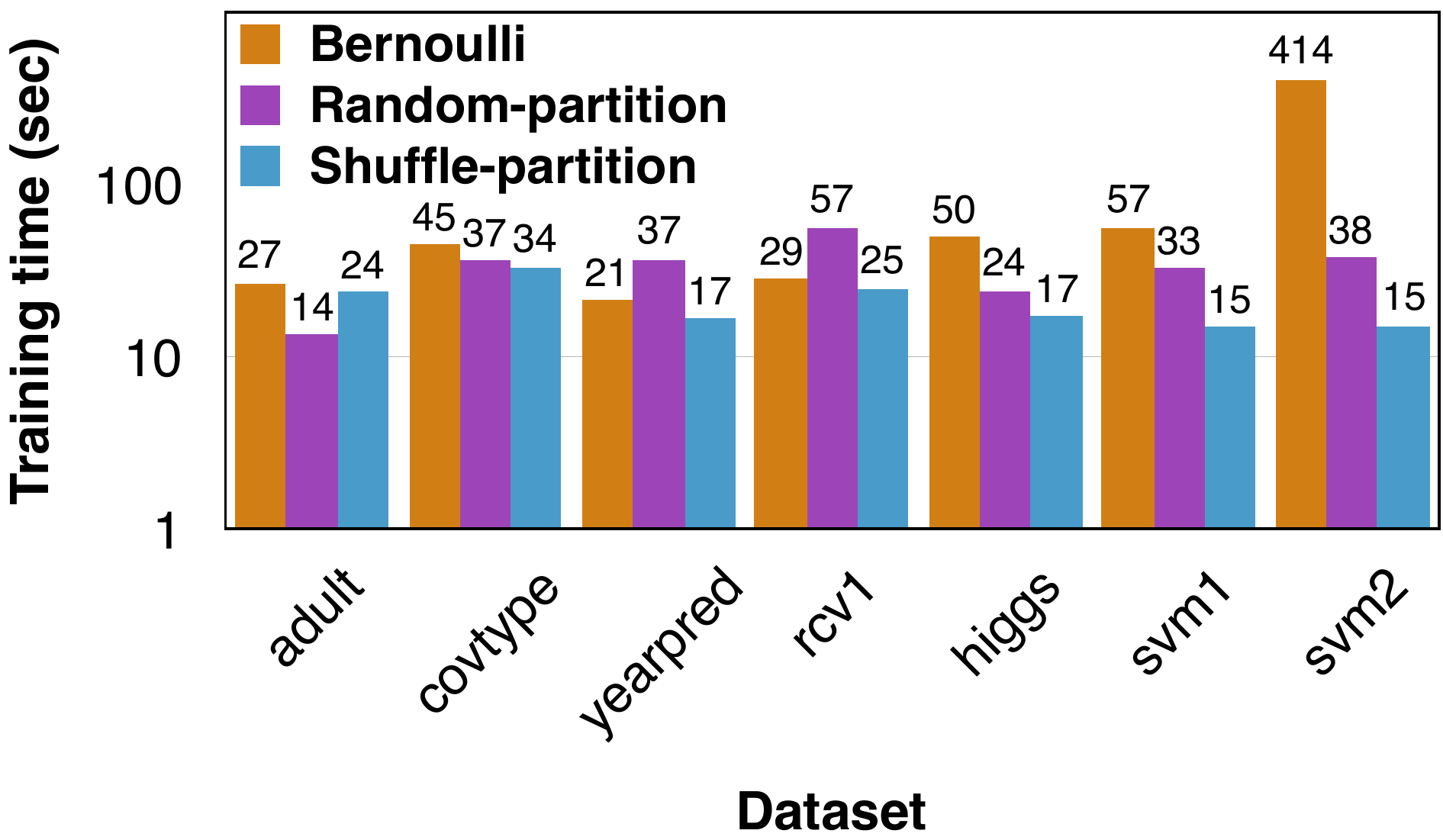}\label{subfig:eagerSGD}}
		\subfigure[Lazy transformation]{\includegraphics[width=0.48 \columnwidth]{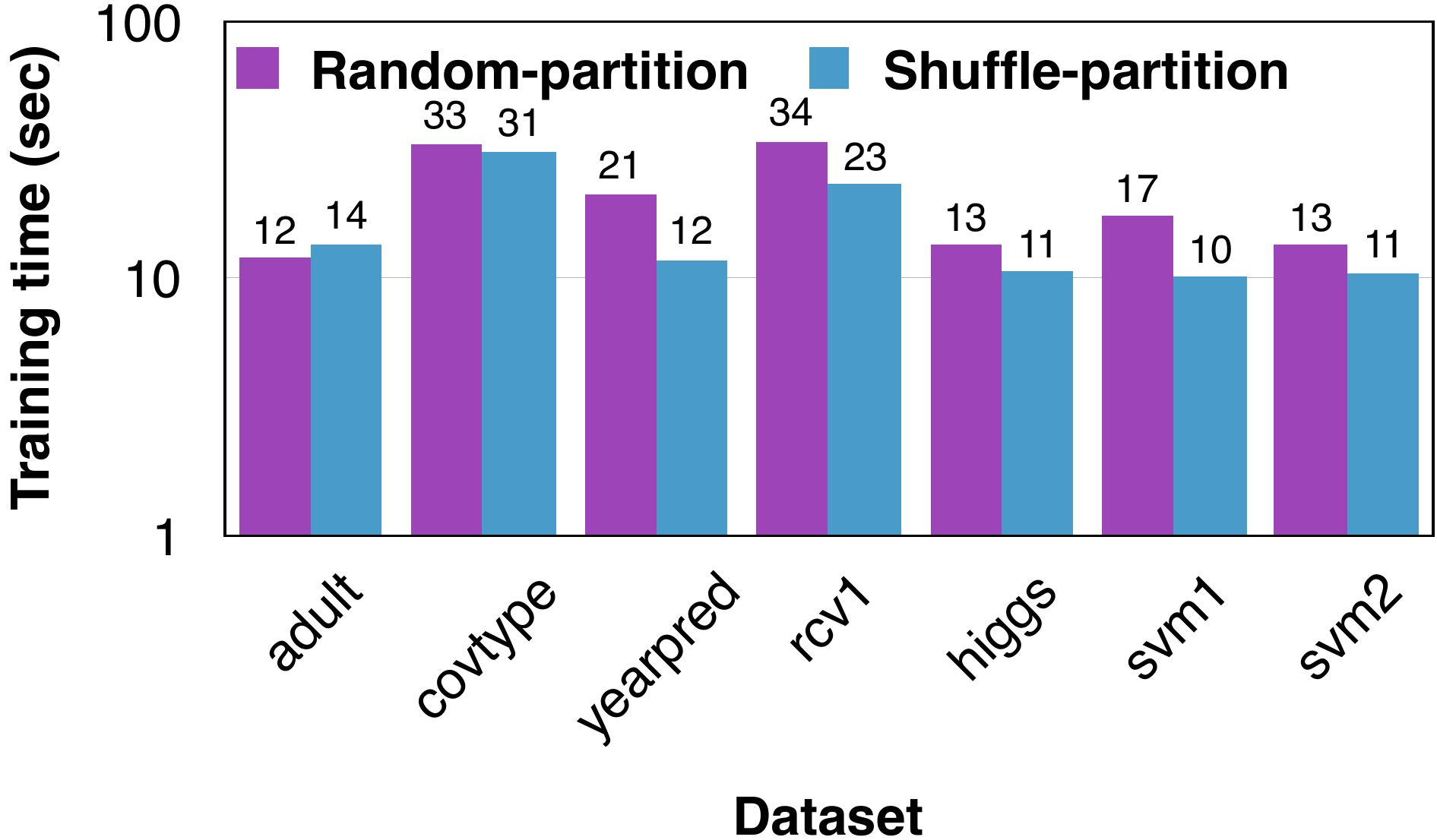}\label{subfig:lazySGD}}
	}
	\caption{Sampling effect in SGD for eager and lazy transformation. \label{fig:gd}}
\end{figure}

\myparagraph{In-Depth}
We now discuss how sampling affects performance when fixing the transformation in SGD to eager or lazy. We observe from Figure~\ref{subfig:eagerSGD} that the shuffle-partition is faster in all datasets but \texttt{adult}. This is because the \texttt{adult} dataset consists of a single partition (HDFS block size is $128$MB) and the cost of re-ordering the entire partition does not pay-off in comparison to the random accesses of the  random-partition sampling for a small number of iterations that are required for convergence.  When using the lazy transformation (Figure~\ref{subfig:lazySGD}), the training times between using random or shuffle-partition are very close with the shuffle-partition being slightly faster in most of the cases.

\begin{figure}[t!]
	\centering
	\mbox{			
		\subfigure[MGD]{\includegraphics[width=0.5 \columnwidth]{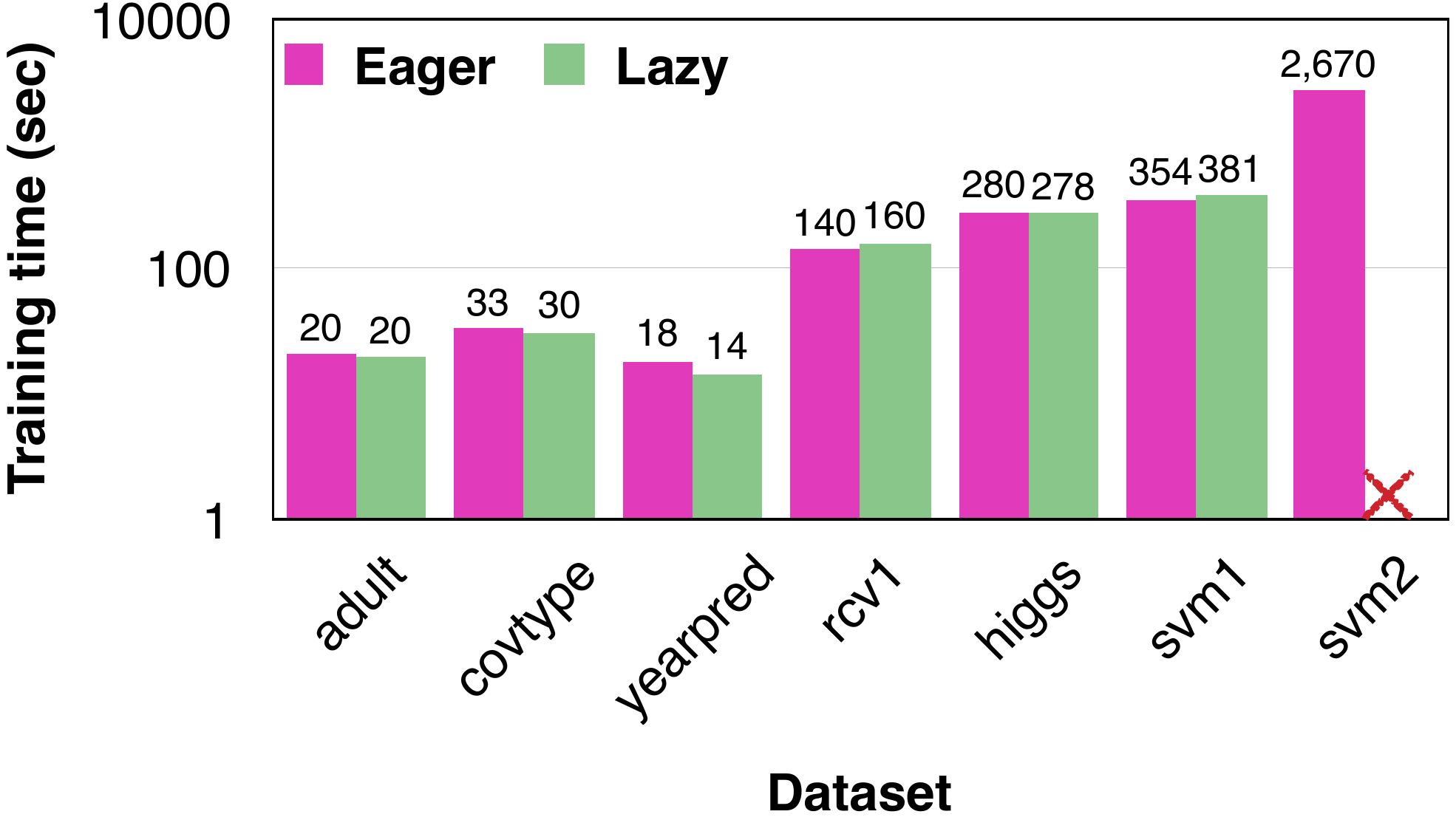}\label{subfig:randomMGD}}
		\subfigure[SGD]{\includegraphics[width=0.5 \columnwidth]{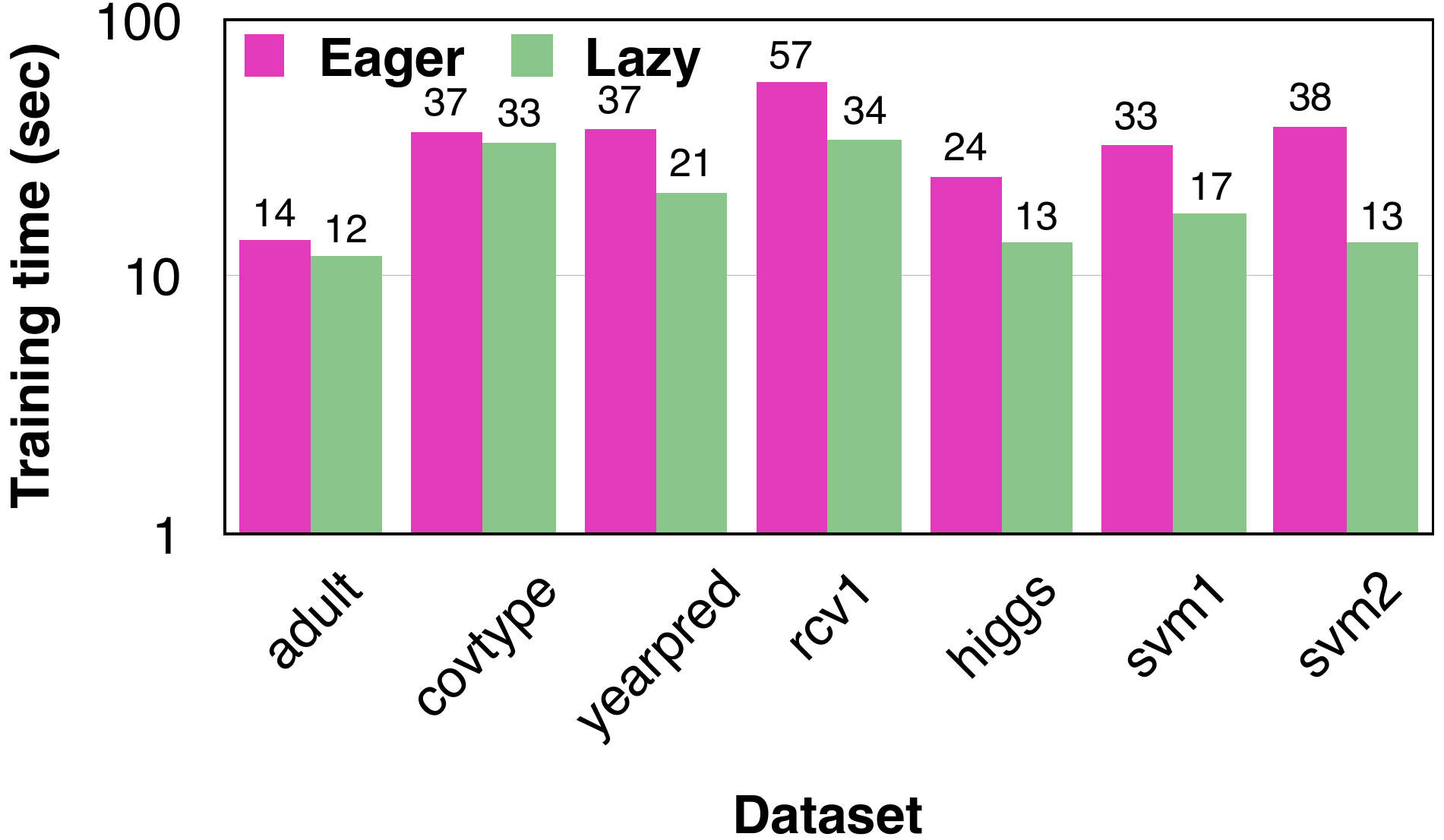}\label{subfig:randomSGD}}
	}
	\caption{Transformation effect for the random-partition sampling technique. \label{fig:random}}
\end{figure}

We now show the impact of the transformation method when we fix the sampling to the random-partition technique. 
We observe from Figure~\ref{subfig:randomSGD} is that there is no significant difference between eager and lazy transformation for MGD, except of the case of \texttt{svm2} where we had to terminate the execution for the lazy transformation. On the other hand, SGD seems to always benefit from the lazy transformation when the random-partition is used (Figure~\ref{subfig:randomSGD}).